\documentclass[12pt,a4paper]{article}
\usepackage[longnamesfirst, authoryear, round]{natbib}

\let\counterwithin\relax
\usepackage[T1]{fontenc}
\usepackage[utf8]{inputenc}
\usepackage[width=1\textwidth, labelfont=bf]{caption}

\usepackage{rotating}

\usepackage{enumitem}
\usepackage{booktabs}
\usepackage[all]{nowidow} 
\usepackage{amsmath}
\usepackage{amsthm} 
\usepackage{dsfont}
\usepackage{amsfonts}
\usepackage{tikz} 
\usepackage{calc}
\usepackage{siunitx}

\usepackage{tikz-qtree} 
\usepackage{pgfplots}
\pgfplotsset{compat = 1.14}
\usepackage{mathtools,amssymb}
\usepackage{xcolor}
\usepackage{cooltooltips}
\usepackage{MnSymbol}
\usepackage{graphicx}
\usepackage{microtype}
\newcommand{\bigCI}{\mathrel{\text{\scalebox{1.07}{$\perp\mkern-10mu\perp$}}}} 
\usepackage{subcaption}
\usepackage[flushleft]{threeparttable}
\usepackage{multirow}

\usepackage{placeins} 
\usepackage{indentfirst} 
\usepackage{lscape}
\usepackage{rotating}
\usepackage[pdfborderstyle={/S/U/W 1},breaklinks,colorlinks]{hyperref}
\AtBeginDocument{%
  \hypersetup{
    citecolor=blue,
    urlcolor =blue,
    menucolor = black,
    linkcolor = black}}
    
\usepackage{tocloft}

\usepackage{chngcntr} 
\counterwithin{figure}{section}
\counterwithin{table}{section}

\usepackage{tcolorbox}
\tcbset{colframe=black, colback=white,size=fbox}

\usepackage{algpseudocode}
\usepackage[linesnumbered,ruled]{algorithm2e}

\usepackage[left=2.5cm,right=2.5cm,top=3.5cm,bottom=3.5cm]{geometry}

\usepackage{listings}
\usepackage{xcolor}
\makeatletter
\def\thanks#1{\protected@xdef\@thanks{\@thanks
        \protect\footnotetext{#1}}}
\makeatother

\title{\vspace{-2.0cm}  Variable Selection for Causal Inference via Outcome-Adaptive Random Forest\thanks{\scriptsize{Financial support from the Deutsche Forschungsgemeinschaft via the IRTG 1792 “High Dimensional Non Stationary Time Series”, Humboldt-Universität zu Berlin, is gratefully acknowledged.} }\\ }

\date{\today}

\author{Daniel Jacob \\ \href{mailto:daniel.jacob@hu-berlin.de}{daniel.jacob@hu-berlin.de}}

\begin{document}

    \maketitle
\thispagestyle{empty}

\begin{abstract}
\footnotesize

Estimating a causal effect from observational data can be biased if we do not control for self-selection. This selection is based on confounding variables that affect the treatment assignment and the outcome. Propensity score methods aim to correct for confounding. However, not all covariates are confounders.  We propose the outcome-adaptive random forest (OARF)  that only includes desirable variables for estimating the propensity score to decrease bias and variance. Our approach works in high-dimensional datasets and if the outcome and propensity score model are non-linear and potentially complicated. The OARF excludes covariates that are not associated with the outcome, even in the presence of a large number of spurious variables. Simulation results suggest that the OARF produces unbiased estimates, has a smaller variance and is superior in variable selection compared to other approaches. The results from two empirical examples, the effect of right heart catheterization on mortality and the effect of maternal smoking during pregnancy on birth weight, show comparable treatment effects to previous findings but tighter confidence intervals and more plausible selected variables.




\noindent

\noindent
\textbf{Keywords:}
\textit{average treatment effect; causal inference; random forest;  variable selection}

\end{abstract}

\newpage
\setcounter{page}{1}

\section{Introduction}

In the causal inference literature, we can classify data into two categories. The one is data from a randomized controlled trial where the researcher or practitioner has full control of the selection process. The counterexample is data from a so-called observational study. In such a setting there are confounding variables that influence both, the outcome and the probability of treatment. To construct unbiased treatment effect estimates from observational studies, propensity score (PS) methods are an increasingly popular tool to control for confounding \citep{rosenbaum1983central}. One model-based approach, the inverse probability of treatment weighting (IPTW), to directly adjust for the confounding bias using propensity scores was proposed by \citet{hirano2001estimation}. Estimating the propensity score can be seen as a classification problem where one seeks to have a good prediction of the assignment probability given covariates, regardless of the functional form of the distribution of the probabilities. Besides logistic regression, non-parametric methods such as random forests \citep{lee2010improving, westreich2010propensity, zhao2016propensity}, neural-networks, and support vector machines \citep{westreich2010propensity} have been proposed to estimate the propensity score. An interesting question is which variables should be included to estimate the propensity score. Common suggestions are, to include all pre-treatment variables that influence the treatment while excluding variables that do not affect the treatment (this can be variables that do not influence any dependent variable - they are spurious, but also variables that are only predictive on the outcome). Following this rule, we would include confounding variables since they influence the treatment and the outcome as well as variables that only predict the treatment and not the outcome but exclude variables that are only predictive of the outcome. 

\begin{figure}
\centering
\begin{tikzpicture}[]
        \draw[->] (0,0) node[left] (D) {$D$} -- (4,0) node[right] (Y) {$Y$}; 
        \path (D) -- coordinate (middle) (Y);
        \node[above of=middle] (Xc) {$X_c$};
        \node[above of=Y] (Xp) {$X_o$};
        \node[above of=D] (Xi) {$X_t$};
        \node[right of=Xp] (Xs) {$X_s$};
        \draw[->] (Xc) -- (D);
        \draw[->] (Xc) -- (Y);
        \draw[->] (Xi) -- (D);
        \draw[->] (Xp) -- (Y);
\end{tikzpicture}%
\caption{Dependencies of the covariates}
\label{fig:DAG}
\end{figure}
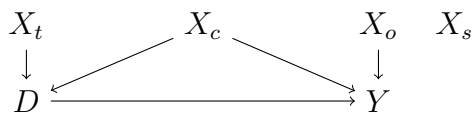

We give an example of the different relationships between the covariates, the outcome, and the treatment in Figure \ref{fig:DAG}. We denote $X_t$ for covariates that predict only treatment but not the outcome, and $X_o$ that predict the outcome but not the treatment probability. Of special interest are the confounding covariates $X_c$ that we need to take into account to get an unbiased estimate of the average treatment effect (ATE) while $X_s$ are spurious covariates that are uncorrelated to both, treatment and outcome. Let us illustrate the role of the variables using the vaccination against COVID 19 as an example. The treatment variable ($D$) is whether a person is vaccinated or not while the outcome variable ($Y$) is the individual probability of severe symptoms. Variable $X_t$ could be the industry sector a person is working in since it influences the probability of being vaccinated but has no influence on symptoms. $X_o$ could be whether a person smokes or not, which has an influence on the degree of symptoms but does not determine the vaccination probability. $X_c$ might be the age of a person, which is associated with vaccination and symptoms while the variable body height might be unrelated and is classified as $X_s$. 

The rational behind the classification of covariates is to perform variable selection when estimating the propensity score. First, and most important to ensure unbiased treatment effects in observational studies is to achieve unconfoundedness. In theory, it is sufficient to only use covariates $X_c$ for the propensity score since the main analytic goal is to eliminate bias. This implies that we do not aim to explain the treatment assignment mechanism with high accuracy. If this would be the goal, variables $X_t$ would need to be included. However, doing so could limit the overlap assumption. These are the two main reasons why we want to exclude covariates $X_t$ in the propensity score model. Last, even if the dependency of $X_o$ and $D$ is zero in the true data generating process, the finite sample bias can be reduced if variables $X_o$ are included in the propensity score model. The finite sample bias arises due to random confounding when the sample size is small.

In a recent paper by \citet{shortreed2017outcome}, the authors suggest a different approach to get unbiased treatment effects from observational studies with a focus on decreasing the variance. Their proposed outcome-adaptive lasso (OAL) approach only selects features in the propensity score estimation that have a relationship with the outcome ($X_c, X_o$) but exclude variables that are predictive on the treatment ($X_t$) as well as spurious variables ($X_s$). To do so, they first find covariates that predict the outcome by regressing the outcome on the treatment and all covariates using a linear model. In the second step, the estimation of the propensity score, they use the lasso with an additional penalty term. The penalty contains individual weights for each covariate based on the coefficients of the covariates from the first step. The result is that the lasso excludes covariates that predict the treatment but are not related to the outcome as well as spurious covariates. Consider $p$ covariates, denoted $X_j$ for $j=1,...,p$, then the OAL estimator is defined as: 

\begin{align}
\begin{split}
\widehat{\alpha}(O A L)=\underset{\beta}{\operatorname{argmin}} &\left[\sum_{i=1}^{N}\left\{-a_{i}\left(\mathbf{x}_{\mathbf{i}}^{\top} \alpha\right)+\log (1+e^{\mathbf{x}_{i}^{\top} \alpha})\right\}\right. \\
&\left.+\lambda_{n} \sum_{j=1}^{p} \widehat{\omega}_{j}|\alpha_{j}|\right]
\end{split}
\end{align}

where $\omega_j = |\tilde{\beta}_j|^{-\gamma}$ s.t. $\gamma>1$. The vector $\tilde{\beta}_j$ refers to the coefficient estimates from regressing the covariates on the outcome, conditioning on the treatment. 

An important limitation of the outcome-adaptive lasso is that this approach is restricted to parametric models. Both, the outcome model and the propensity score model have to be correctly specified. However, to control for selection bias we would like to have as many pre-treatment characteristics as possible to condition on them. This requirement lets new datasets easily become high-dimensional. In such settings, we face two potential challenges. First, the outcome and the treatment variable might not dependent linearly on the covariates. There can be interactions between variables and complex structures. Second, even if we could include such interactions, facing many covariates leads to the problem of which covariates to include in the outcome and propensity score model? It might be the case that only a few characteristics are important - the question is which are they? Such settings call for a non-parametric method that uses regularization. Keeping the idea of outcome-adaptive regularization but accounting for non-linearity and high-dimensionality, we propose the outcome-adaptive random forest (OARF). First, we estimate a modified and standardized variable importance score using a random forest used as the penalty weight. Second, we use the modified variable importance to regularize a random forest that learns the propensity score. To do so we penalize the gain at each split and propose the use of an initial feature space using the modified variable importance. This approach allows replacing both linear models, the regression (which we will refer to as OLS), and the lasso, with a random forest to estimate the ATE when the functions are non-linear or high-dimensional. We also make use of sample splitting to avoid overfitting and apply cross-fitting to restore efficiency. Our approach of a regularized random forest further selects only those covariates in the propensity score that are predictive of the outcome and excludes spurious variables. The OARF is designed to allow for categorical variables and is robust to different amounts of levels between categorical or continuous variables. The result is a flexible non-parametric method that allows unbiased estimation of the ATE while decreasing the variance. Our approach is also fast in computation (about 30 seconds for 2000 observations and 20 covariates).


\section{Illustrating the outcome-adaptive estimation}

\begin{figure}[ht]
\begin{center}
\includegraphics[width=0.8\textwidth]{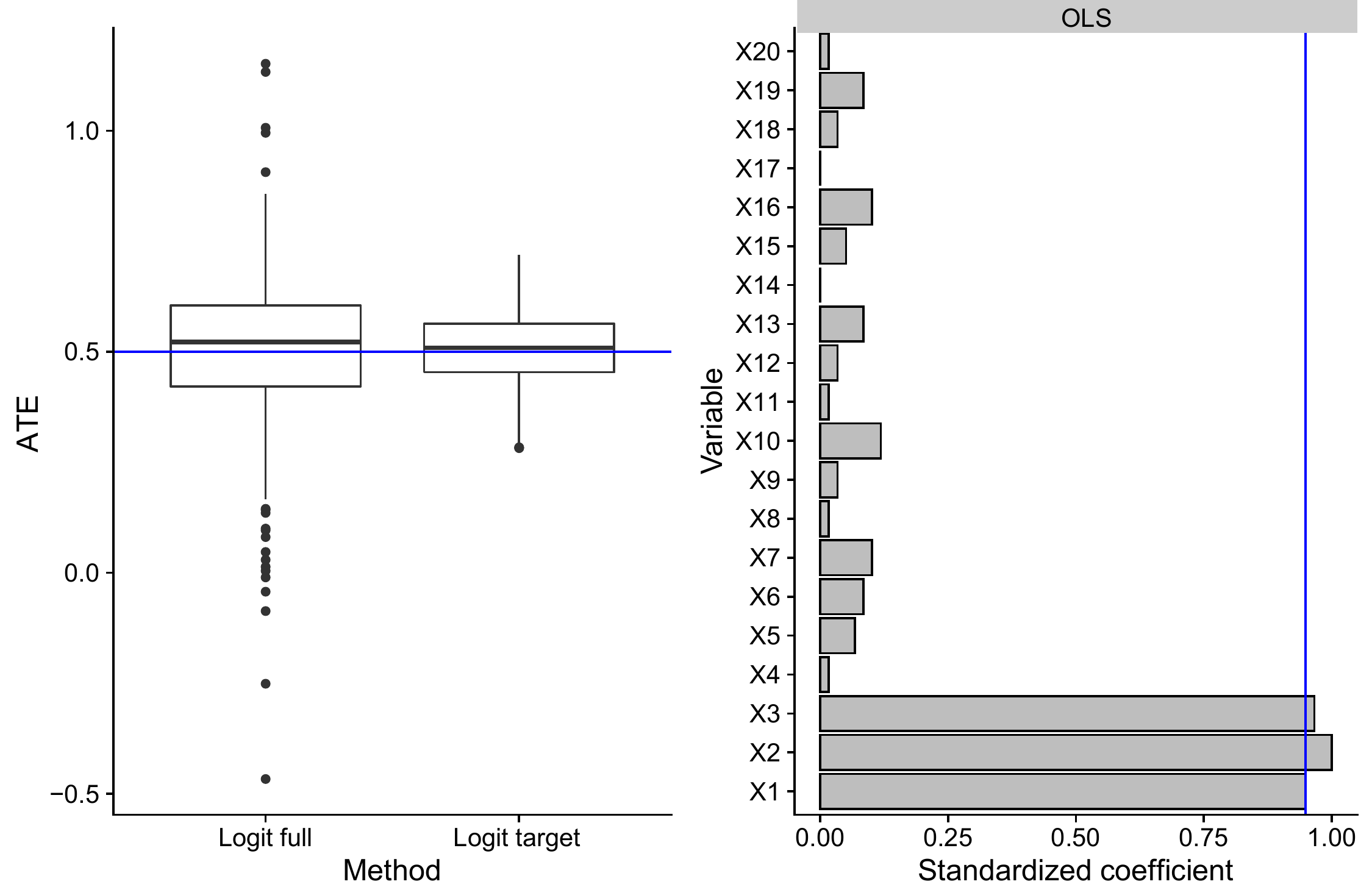}
\caption{Left: IPTW estimates using full and oracle propensity score. Right: Selected variables from OLS.}
\label{fig:logit_ols_coef}
\end{center}
\end{figure}

Let us demonstrate the outcome-adaptive approach in a simple example. Assume the true outcome model as in equation \ref{equ:out_lin}. $Y$ depends on the treatment $D$ and linearly on three covariates ($X_1, X_2, X_3$). We also generate a propensity score model that includes variables $X_1, X_2, X_5$, and $X_6$ (see the Appendix for how the function is generated). The covariates are generated from a multivariate normal distribution and are independent. We set the sample size to $N = 1000$ and generate $p=20$ covariates where only the first three are dependent on the outcome $Y$. Using the notation from Figure \ref{fig:DAG} we have the following structure: $X_1, X_2 \in X_c$, $X_3 \in X_o$ and $X_5, X_6 \in X_t$. Using a linear model (OLS), as in \cite{shortreed2017outcome}, we want variables $X_1$ to $X_3$ to have the highest coefficients and penalize all others. Hence they should have a very small estimated coefficient. We run 500 Monte Carlo simulations and predict the treatment effect using the inverse probability of treatment weighted (IPTW) estimator \citep{lunceford2004stratification, hernan2006estimating}. We use a logit model to estimate the propensity score, using all covariates (the full model) and only using covariates that are confounding or predict the outcome (target model). Figure \ref{fig:logit_ols_coef} shows the estimated treatment effect using both models as well as the standardized and absolute coefficients from the OLS model. The target model (which can be seen as an oracle model) shows a smaller variance around the true treatment effect of $0.5$ (indicated with the horizontal line). We also see that the OLS correctly assigns the highest coefficients to the variables that are predictors of the outcome while all other variables get smaller coefficients.

\begin{align}
Y = 0.5D + 0.6X_1 + 0.6X_2 + 0.6X_3 + \varepsilon, \quad \varepsilon  \sim N(0,1). \label{equ:out_lin}
\end{align}

The suggested approach by \cite{shortreed2017outcome} to select only variables for the propensity score model that are predictive for the outcome works as expected if the underlying outcome and propensity score functions are linear. As soon as we introduce a more complicated function, the selection process in step 1 is biased. To demonstrate this, let us now assume the following true outcome model:

\begin{align}
Y = 0.5D + 0.5X_1 + 0.8X_2 \otimes X_3 + \varepsilon, \quad \varepsilon \sim N(0,1). \label{equ:out_int}
\end{align}

\begin{figure}[ht]
\begin{center}
\includegraphics[width=0.8\textwidth]{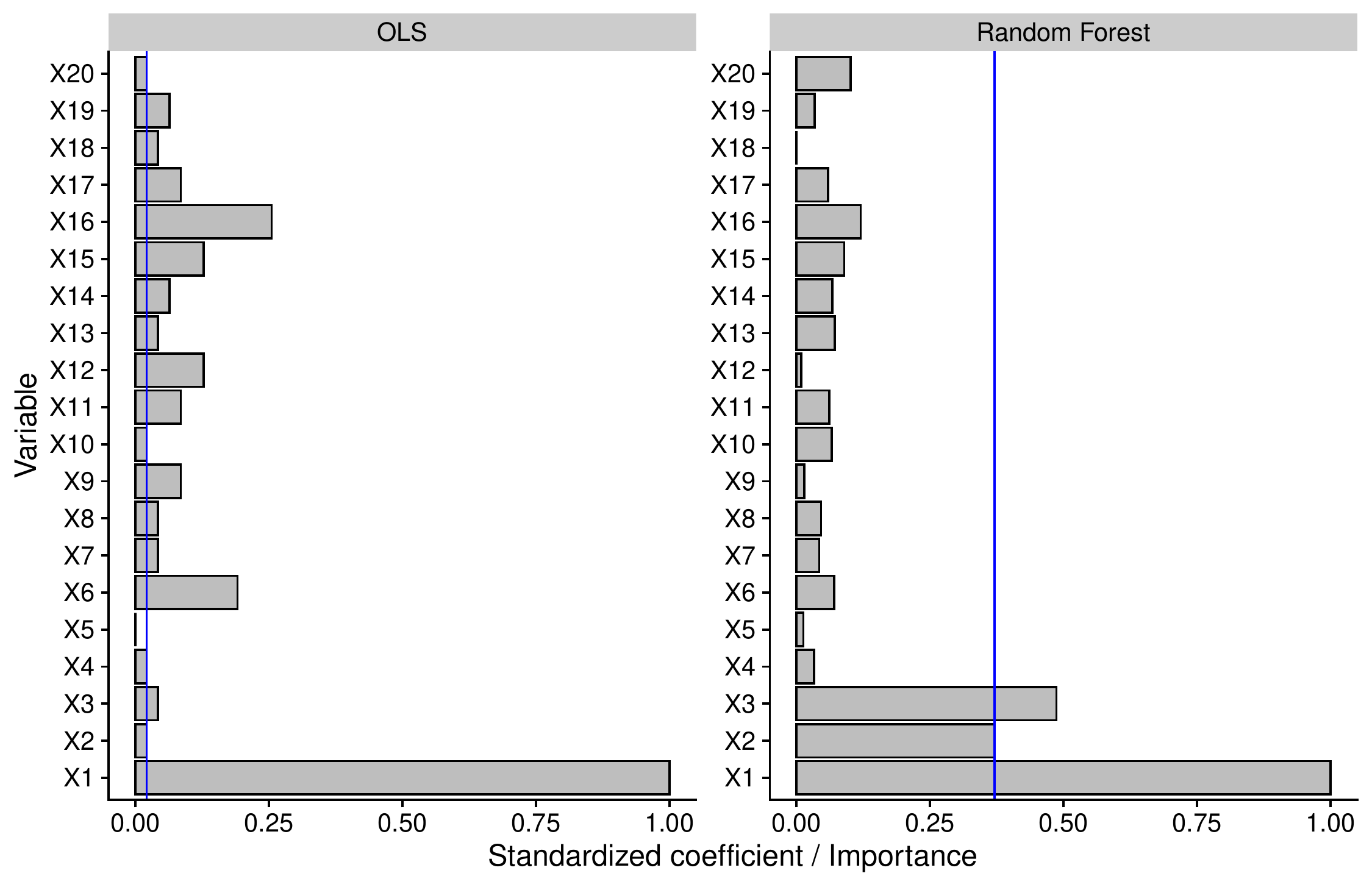}
\caption{Standardized and absolute coefficient values for all covariates.}
\label{fig:OLS_vs_RF_outcome}
\end{center}
\end{figure}

Figure \ref{fig:OLS_vs_RF_outcome} shows the coefficients estimated by the OLS and the random forest, respectively. Values from the linear model are absolute coefficients while values from the random forest are based on the impurity measure which we explain below. The values for both methods are standardized between zero and one to make them comparable.
Say we know that the model should select variables one to three. We draw a line indicating the lowest value from the variables that should be selected. If this would be a threshold (for selection or penalization), clearly the OLS estimates higher coefficients for at least 12 additional variables (besides the two with the highest coefficients). In contrast, the random forest does find the correct variables and assigns much lower importance to all other variables. Using the same threshold method on the random forest, all spurious variables would not be selected or at least heavier penalized compared to the OLS. The problem gets more severe if we not only allow for interaction terms but non-linear structures (say, by including quadratic and trigonometric functions into the data generating process).

\section{Method}
Selecting variables using non-parametric models is not straightforward since we do not directly estimate coefficients for each feature like in the OLS. Using a random forest, we can instead create a variable importance measure to find the most predictive variables from the outcome model. Since we want to use a random forest for both steps, the outcome model and the propensity score, we have to replace the adaptive lasso by penalizing the tree-building mechanism. In this section, we show how to best estimate the variable importance measure from the outcome model and state the importance of the initial feature space. We describe in detail how the information about the covariates is then used to penalize the random forest that estimates the propensity score. This leads to a regularized version of the random forest that shrinks penalized variables to zero. 

To ensure unbiased effects from a causal parameter the following assumptions from the potential outcome framework are required: Each observation has two potential outcomes, $Y^1$ if treated and $Y^0$ if not. We denote treatment by the binary indicator $D \in \{0;1\}$ and  observed covariates $X \in \mathbb{R}^{^p}$. See, for example \cite{rubin1980randomization}. First, ignorability: $\left(Y_{i}^{1}, Y_{i}^{0}\right) \bigCI D_{i}|X_{i}$. It states that the treatment assignment is independent of the two potential outcomes. Second, the stable unit treatment value assumption (SUTVA), $Y_i = Y_i^0 + D_i (Y_i^1-Y_i^0)$, ensures that there is no interference, no spillover effects, and no hidden variation. This means that the treatment status for individual $i$ does not affect the potential outcomes of individual $j$. The third assumptions describes the propensity score: $P(D_i=1|X_i=x) \overset{\mathrm{def}}{=} e(x)$, which needs to be bounded away from zero and one: $\forall x \in supp(X_i), \quad 0 < P(D_i=1|X_i=x) < 1,$. The last assumption states that the covariates are not affected by the treatment: $X_i^1 = X_i^0$.

We are interested in estimating the ATE ($\theta$), assuming a partially linear model that takes the following form: 

\begin{align}
Y = \theta D + g(X) + \varepsilon, \quad \mathsf{E}[\varepsilon |D,X] = 0, \\
D = m(X) + \nu, \quad  \mathsf{E}[\nu |X] = 0.
\end{align}

\subsection{Variable Importance in Random Forest}

Building a tree based on the classification and regression tree (CART) algorithm works as follows: A subset of the feature space $\mathcal{X}$ is selected at each node $t$ in the tree. Internal nodes $t$ are labelled with a split $s_t = (X_j < c)$, which creates at least two further subsets or children $t_L$ and $t_R$ ($L$ and $R$ refer to left and right, as in a tree). This procedure is repeated until we reach terminal nodes (or leaves) that are labeled with the best prediction of the outcome variable. In the regression case, this would be the mean value $\bar{y}_t$. The predicted output for a new instance is the mean value of the leaf reached by the instance when it is propagated through the tree. Using a recursive procedure at every note $t$, a tree identifies the split $s_t = s^*$ for which the partition of the sample into $t_L$ and $t_R$ maximizes the reduction of 

\begin{align*}
\begin{split}
\Delta(s, t)=\operatorname{cost}(\mathbb{D}_t)-\left(\frac{\left|\mathbb{D}_{t_L}\right|}{|\mathbb{D}_t|} \operatorname{cost}\left(\mathbb{D}_{t_L}\right) \right. \\
 \left. +  \frac{\left|\mathbb{D}_{t_R}\right|}{|\mathbb{D}_t|} \operatorname{cost}\left(\mathbb{D}_{t_R}\right)\right).
\end{split}
\end{align*}

Let $\mathbb{D}$ be the set of observations for a specific decision or split. For the first step regression setting, we define the cost function as $\operatorname{cost}(\mathbb{D}) = \sum_{i \in \mathbb{D}}(y_i - \bar{y})^2$, where $\bar{y} = |\mathbb{D}|^{-1}\sum_{i \in \mathbb{D}}y_i$ is the mean of the outcome variable in the specified set or region. Maximising the decrease of the variance within each leaf can be seen as making the leafs pure in terms of the outcome values. Hence, the cost function is called the impurity measure.

The impurity change (or gain) through a split can be used to estimate the importance of a variable by evaluating each cost function given a specific feature $j$ on which the split $s^*$ is based. Hence, we can define the gain in terms of a feature $j$ instead of split $s$: $\Delta(j,t) \overset{\mathrm{def}}= \Delta(s,t)$. The global importance value is given by accumulating the gain over a feature, $\Delta(j)=\sum_{t \in \mathbb{S}_{j}} \Delta(j, t)$ where $\mathbb{S}_j$ represents all the splitting points used in a single tree for the $j$-th feature. This is because a feature $X_j$ can be used multiple times during the recursive growth of a tree. Since a random forest consists of $B$ such trees, the importance value ($Imp_j$) is just an average over all trees:

\begin{align*}
I m p_{j}=\frac{1}{B} \sum_{b=1}^{B} \Delta(j)_{b}.
\end{align*}

\cite{nembrini2018revival} find that the impurity importance can be explained by two parts: The impurity reduction directly related to the true importance and a part of impurity reduction that is highly based on the structure of the feature (i.e. a dummy variable with only two levels vs. a continuous variable). The latter component introduces a bias in the impurity measure. To correct for the structure of different features, the authors propose to extend the feature space dimension $p$ to $2p$. If the selected variable is within $\{1,..., p\}$ then the variable is used as usual. If the variable is instead in the $\{p+1,..., 2p\}$ set, the variable values are permuted. This means that the levels for each observation are reordered such that each observation has a different level as before. If the feature is reordered, the importance value adds negatively to the final measure and positively if the feature is untouched. This procedure, called actual impurity reduction (AIR), allows to de-bias the total importance value by controlling for the structure of each feature. 

There are other measures for variable importance like permutation importance. To calculate the permutation importance of a feature, the prediction performance is calculated for observations that are not included in the bootstrap data (the so-called out-of-bag (OOB) observations). The values of the variable are then randomly permuted for each observation. Calculating the OOB error again with the permuted feature indicates the importance of the variable. The more the prediction error changes, the more important is the feature. The permutation performance is calculated for each feature and averaged over all trees. See \cite{nembrini2018revival} for an overview and comparison of different importance measures. We note that the permutation method is computationally expensive when compared with the AIR method. This is because the former method relies on OOB predictions for each feature. In terms of robustness concerning the different amounts of levels between variables, they perform similarly. 

In our setting, we have at least one binary variable, the treatment assignment, which has a limited range of splitting values compared to continuous variables. Therefore, it is important to take the different structures into account by applying the AIR measure (available in the \texttt{ranger} package as \texttt{'impurity corrected'}).

\subsection{Penalization parameter}

The penalization parameter $\lambda_j$ should depend on the predictive power from the covariates on the outcome. Our measure of predictive power is the importance score $Imp^*_j$ which can be included as proposed by \cite{deng2013gene}: 

\begin{align}
\lambda_{j}=(1-\gamma) \lambda_{0}+\gamma Imp^*_j, \label{equ:reg_gain_deng}
\end{align}

where $\lambda_0$ is a general penalization parameter and $\gamma$ is a weight parameter that determines the proportion of general and specific feature penalization.
Next, we define the normalized importance score ($Imp^*_j$) as 

\begin{align}
Imp'_j &=\left\{\begin{array}{l}
Imp_j \text{ if } Imp_j \geq 0 \\
0, \text{otherwise} 
\end{array}\right. \label{equ:imp_zero} 
\end{align}

\begin{align}
Imp^*_j &=  \frac{Imp'_j}{max_{l=1}^P Imp'_l} 
\end{align}

The new normalized importance score is scaled within the interval $[0,1]$ (see the Appendix for the proof). Equation \ref{equ:imp_zero} sets all negative importance scores from the AIR measure to zero. 
Since we only want to rely on the importance values obtained from the outcome model, we set  $\lambda_0=0$ and $\gamma = 1$. This allows for the heaviest penalization based on outcome-related covariates, namely

\begin{align}
\lambda_{j}^y= Imp^*_j. \label{equ:reg_gain_OARF}
\end{align}


\subsection{Creating the feature space}

Next, we want to use the information obtained in the first step and only use variables that have a high importance on the outcome to estimate the propensity score. Again, we want to use a random forest since the propensity score function can have a similar non-linear structure as the outcome model. 
To guide the feature selection, \cite{deng2012feature} introduce a (guided) regularized random forest (RRF) by proposing to weight the gains of the splits during the recursive procedure. As a result of the random feature selection at each split, after $m$ splits, only a subset $\mathbb{F}$ of features are included in the tree. To limit the feature space the idea is to exclude variables not belonging to $\mathbb{F}$, unless their importance is substantially larger than the maximum of the gain for features already included. \cite{deng2013gene} define the regularized gain as

\begin{align}
\operatorname{Gain}_{R}\left(\mathbf{X}_{j}, t\right)=\left\{\begin{array}{l}
\lambda_{j}^y \Delta_\nu(j, t), X_j \notin \mathbb{F} \text { and } \\
\Delta_\nu(j, t), X_j \in \mathbb{F}
\end{array}\right. \label{equ:Gain_F}
\end{align}

where $\lambda_j \in (0,1]$ is the penalty coefficient that controls the gain for each feature $j$ if this feature was not previously used for a split. Originally, the feature space $\mathbb{F}$ is an empty set at the root note in the first tree. Only if a feature adds enough predictive information it is included. Based on equation \ref{equ:Gain_F}, the smaller the value for $\lambda_j$, the higher the penalty and hence the less likely it is for feature $j$ to be included in the subset. In our setting, we want to include features that may not be that predictive of the treatment but of the outcome. We also want to make sure that the important features are used in the splitting process (at least with a higher probability). Therefore, we already include features in $\mathbb{F}$ that fulfill the following criterion:  

\begin{align}
\kappa_j=\mathds{1}(Imp_j \geq \frac{1}{P}\sum_{j=1}^{P}Imp_j) \\
\mathbb{F} = \{\kappa_1*X_1,\kappa_2*X_2,...,\kappa_P*X_P\}
\end{align}

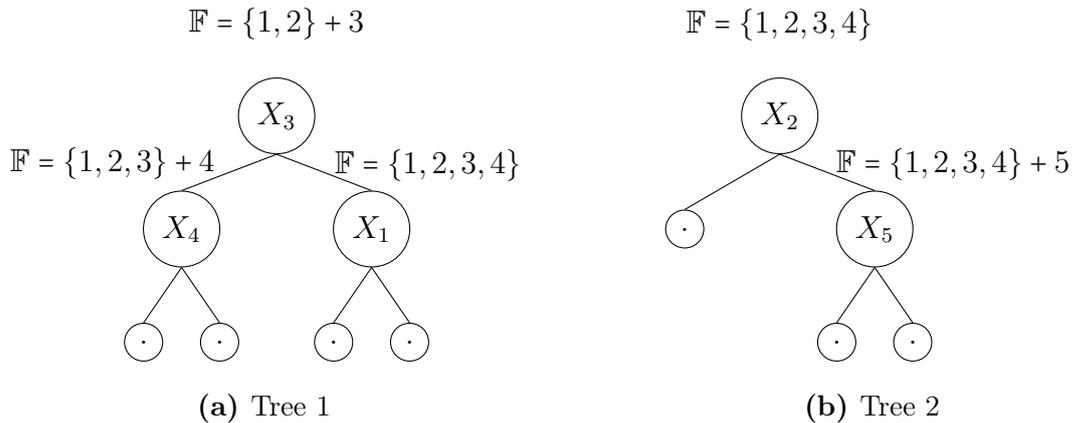
\begin{figure}[ht!]
\begin{subfigure}[b]{0.5\linewidth}
\centering

\begin{tikzpicture}[level distance=1.5cm,
level 1/.style={sibling distance=2.5cm},
level 2/.style={sibling distance=1cm}]
\tikzstyle{every node}=[circle,draw]

\node (Root) [label={ [xshift=0em, yshift=-1.5em] $\mathbb{F}=\{1,2\}+3$}] {$X_3$} 
    child {
    node [label={[xshift=-2.2em, yshift=-2.8em] $\mathbb{F}=\{1,2,3\}+4$}]  {$X_4$} 
    child { node {.}  }
    child { node {.} }
}
child {
    node  [label={[xshift=1.8em, yshift=-2.6em] $\mathbb{F}=\{1,2,3,4\}$}]  {$X_1$} 
    child { node {.} }
    child { node {.} }
};

\end{tikzpicture}
\captionof{figure}{Tree 1}
\end{subfigure}
\begin{subfigure}[b]{0.5\linewidth}
\centering
\begin{tikzpicture}[level distance=1.5cm,
level 1/.style={sibling distance=2.5cm},
level 2/.style={sibling distance=1cm}]
\tikzstyle{every node}=[circle,draw]

\node (Root) [label={ [xshift=0em, yshift=-1.7em] $\mathbb{F}=\{1,2,3,4\}$}] {$X_2$} 
    child {
    node  {.} 
}
child {
    node  [label={[xshift=2.5em, yshift=-3.3em] $\mathbb{F}=\{1,2,3,4\}+5$}]  {$X_5$} 
    child { node {.} }
    child { node {.} }
};

\end{tikzpicture}
\captionof{figure}{Tree 2}
\end{subfigure}
\caption{Initial feature space based on guided regularized random forest (GRRF)}
\label{fig:tree_feature_space}
\end{figure}

If the importance score $X_j$ is at least the mean over all importance scores, we include the variable in the initial feature space and drop values that are zero in $\mathbb{F}$. 
Figure \ref{fig:tree_feature_space} illustrates the guided feature selection process. We start with a non-empty set (here variables 1 and 2 are included). The first split is based on variable $3$. If this variable is not in the feature space, the gain is multiplied by $\lambda_3$. If the penalized gain is higher than the gain from the parent 
the feature is included. As an illustration, we always move from top to bottom and from left to right. The next split is then on $X_4$, again if the penalized gain difference is positive, the feature is included in $\mathbb{F}$. Next, a split on $X_1$ is made, since the variable is already in the feature space the gain is not penalized. Still, it has to be higher than the gain from the parent node to keep the split. Building the first tree, we end up with a feature space containing four variables. The information of the feature space is now used to build further trees. In tree 2, we start with the initial features space as extracted from tree 1. Note that this procedure does not allow to grow trees in parallel since each tree needs the information from the former tree about the feature space to determine if the gain from variables should be penalized or not.

The covariate selection process does not depend on how often a variable can be split and hence if the variable is continuous or, for example, binary. It is sufficient if the gain from one split is above the threshold. Different from calculating the variable importance using the impurity measure we do not average over all split within a tree. This allows the covariates to be of any form, such as binary, categorical or continuous. 

\subsection{Sample splitting}

To avoid overfitting, which can easily happen when using flexible methods such as random forest, we make use of sample splitting and cross-fitting. First, we split our sample into two equal parts, $I_A$ denotes the auxiliary sample and $I_E$ is the estimation sample. We first use the subset $I_A$ to train the propensity score function and $I_E$ to estimate the treatment effect using the predicted propensity score in the IPTW estimator. Let us denote the resulting estimator as $\hat{\theta}(I_A, I_E)$. Now we switch the roles of the auxiliary and estimation sample to obtain a second estimator, called  $\hat{\theta}(I_E, I_A)$. Since both estimators are estimated on only a subset of the observations there might be an efficiency loss. Cross-fitting, which was recently introduced by \cite{chernozhukov2018double}, aims to restore efficiency by simply averaging the two estimates:

\begin{align}
\tilde{\theta} = \frac{1}{2}\left\{\hat{\theta}(I_A,I_E) + \hat{\theta}(I_E,I_A)\right\}
\end{align}

This approach generalizes to $K$ folds where $I_A$ contains $K-1$ folds and $I_E$ the remaining fold. Similar to cross-validation, each fold is used to estimate the ATE by iteratively looping through the folds. The final estimator is the average over the $K$ estimators. 

We use the full sample to get the variable importance from the first step. This is especially helpful if the sample size is small. When using a logit model or the lasso as a benchmark, we also use the full sample and estimate the final treatment effect in one step. 

\section{Simulation study}

To evaluate the performance of our \textbf{OARF} method in more detail, we consider different data generating processes (DGP's) and consider the following methods for comparison and benchmarking: The \textbf{OAL} method by \cite{shortreed2017outcome} and two generalized linear models. The first one uses all covariates in a logit model (\textbf{Lo full}) while the second one only uses target variables ($X_c, X_o$) to estimate the propensity score (\textbf{Lo targ}). We use the same benchmarks for the random forest, denoted by \textbf{RF full} and \textbf{RF targ}. We also use the regularized random forest (\textbf{RRF}) which only sets a penalty based on the first step variable importance but does not make use of an initial feature set, as proposed here by the OARF. We use the following order of variables when we look at the variable selection plots: $X_c, X_o, X_t$. If the amount of the variables are set to two, then the first two are confounders, variable three and four are regressors on the outcome and five and six are regressors on the treatment. We use the $\texttt{ranger}$ package in $\texttt{R}$ for all estimations based on the RF and the $\texttt{RRF}$ package by \cite{deng2013guided} in part for the OARF. The OAL approach is based on the replication file from \cite{shortreed2017outcome}. The tuning parameter $\lambda$ and $\gamma$ for the OAL method are selected using a weighted absolute mean difference (wAMD) which we describe in the Appendix. 

First, we consider two linear settings and generate a binary treatment, $D$, from a Bernoulli distribution with
$\textit{logit}\{P(D=1)\}=\sum_{j=1}^{p} \nu_{j} X_{j}$ and the continuous outcome variable $Y$ as $Y = \theta D + \sum_{j=1}^{p} \beta_{j} X_{j}+\varepsilon,$ where $\varepsilon \sim N(0,1)$ and $\theta = 0.5$. The two settings differ in the strength of the confounding effect. Setting 1 sets $\beta = (0.6, 0.6, 0.6, 0.6,0,0,0,...,0)$ and $\nu = (1,1,0,0,1,1,0,...,0)$, and setting 2 sets $\beta = (0.6, 0.6, 0.6, 0.6,0,0,0,...,0)$ and $\nu = (0.4,0.4,0,0,1,1,0,...,0)$. Setting 2 hence has a weaker confounding relationship than setting 1. These two settings are identical to the one used in \cite{shortreed2017outcome}. Setting 3 aims to have a non-linear relationship between the covariates and the outcome but the same linear structure for the propensity score as in setting 1. Setting 3 is generated as follows: $Y = \theta D + 0.8(X_1 \otimes X_2) + 0.8(X_3 \otimes X_4) +\varepsilon$; $\textit{logit}\{P(D=1)\}=\sum_{j=1}^{p} \nu_{j} X_{j}$, with $\nu = (1,1,0,0,1,1,0,...,0)$. In all three setting we set $N=500$ and  $p = 20$. We find that our OARF performs similar to the benchmark OAL method. The random forest using all covariates and the one that uses the penalization weights from the first step perform worse and are clearly biased. The reason for the bias might be that the full random forest has a higher chance to select variables $X_t$ and $X_s$ than OARF. We do see an improvement in terms of bias when using the RRF, which selects fewer of the above mentioned variables due to the penalization weights. We also find that if we weaken the confounding relationship, all methods have a smaller variance and the RF methods a smaller bias. Boxplots illustrating the IPTW estimator using different methods are shown in Figure \ref{fig:boxplots_lin_OAL}. Selected covariates from the propensity score model are shown in Figure \ref{fig:selected_Var_lin_non_lin}. In both settings, the OARF approach only selects the desired features. This is similar to the OAL method. For comparison, we show that the full RF selects all features and the RRF a higher proportion of all variables while always selecting the desired variables. 

\begin{figure}[ht]
\begin{subfigure}[b]{0.5\linewidth}
\centering
\includegraphics[width=0.8\textwidth]{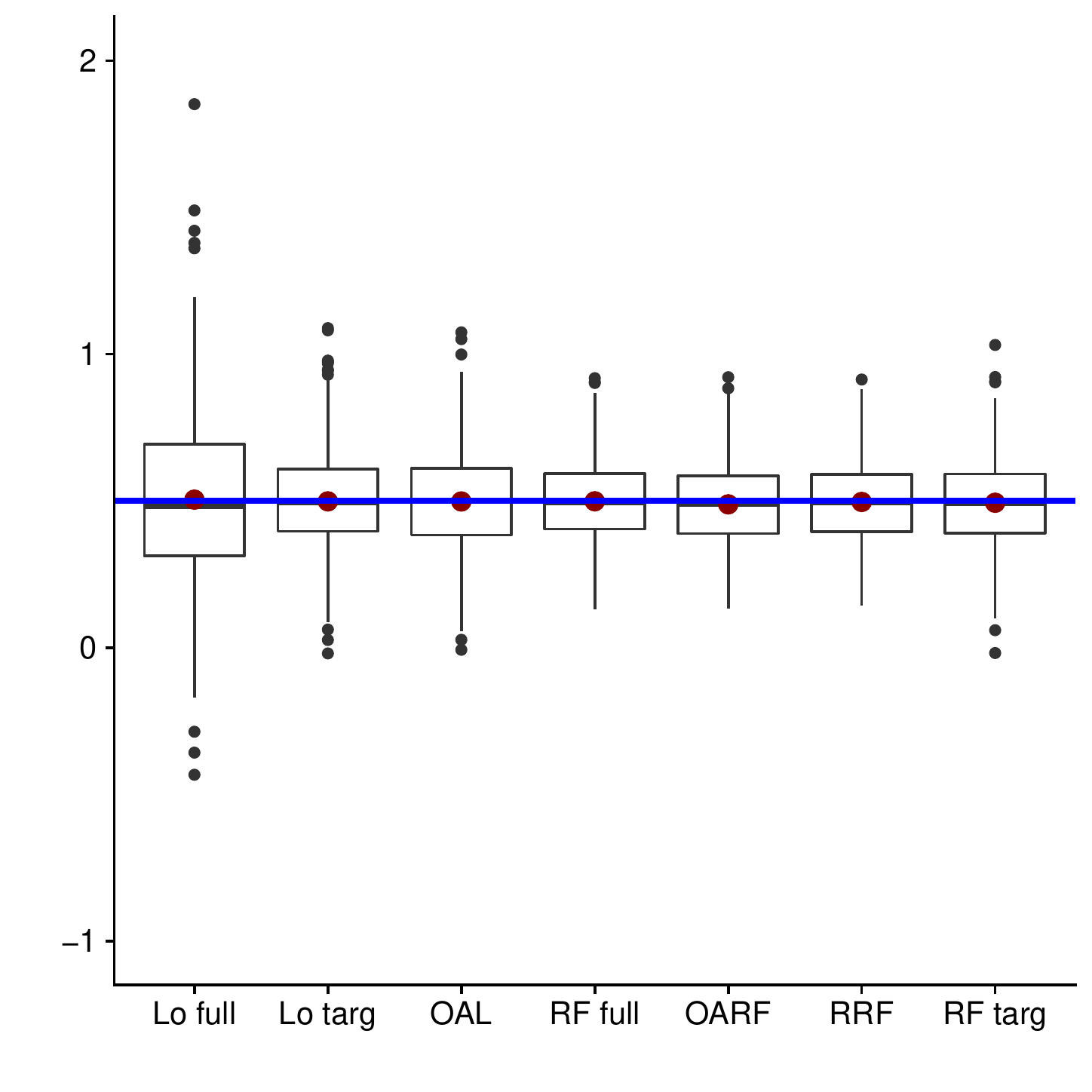}

\captionof{figure}{Treatment effects}
\end{subfigure}
\begin{subfigure}[b]{0.5\linewidth}
\centering
\includegraphics[width=0.8\textwidth]{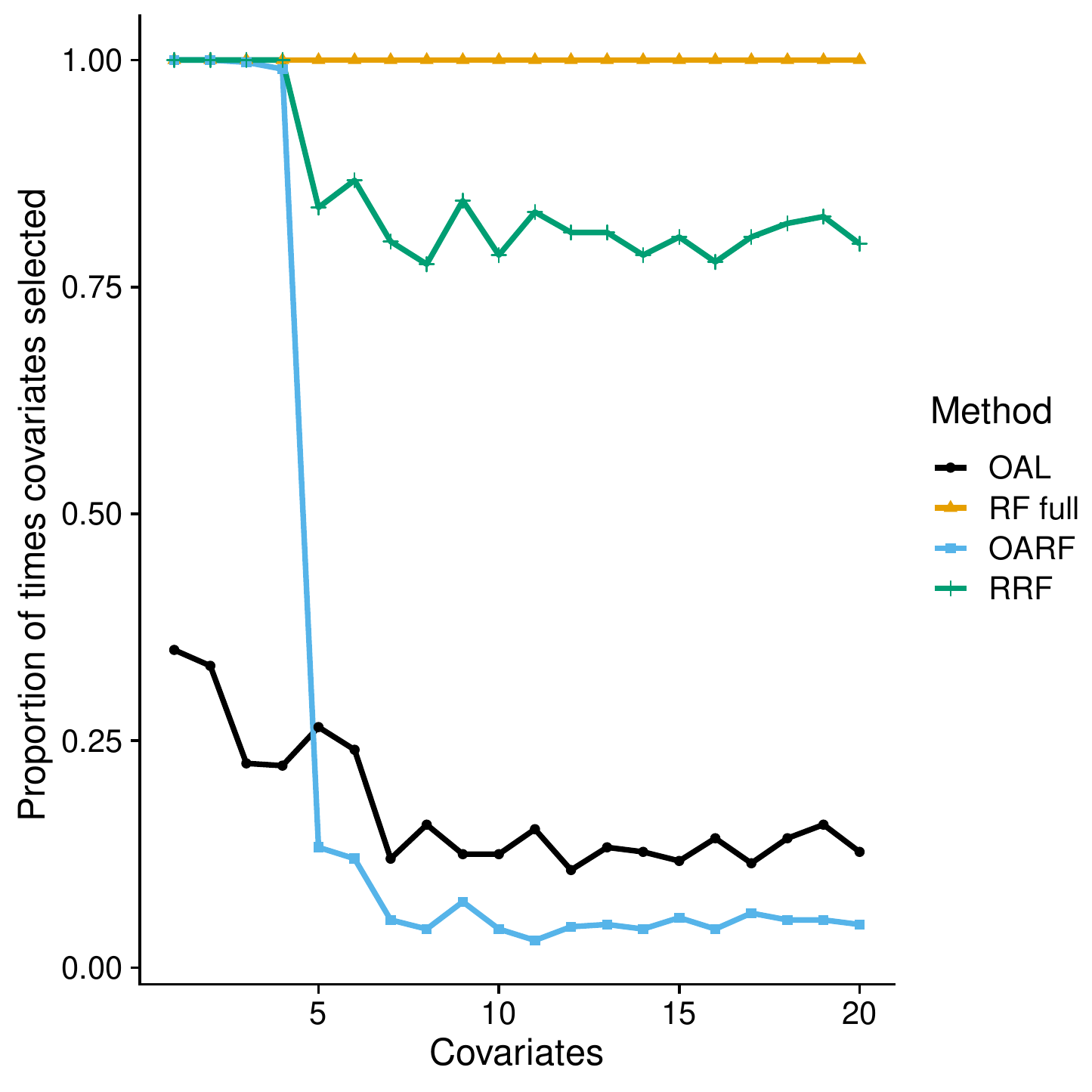}

\captionof{figure}{Selected variables}
\end{subfigure}
\caption{Illustration with non-linear outcome model and linear propensity score model (Setting 3).}
\label{fig:semi_linear}
\end{figure}

In setting 3, illustrated in Figure \ref{fig:semi_linear}, all approaches are unbiased since the propensity score function depends linearly on the covariates. Only the outcome model is non-linear which is why the variance is higher in the linear models compared to the random forest. Next to the treatment effect estimates, the selected variables are illustrated. The OAL model fails to select the correct features as was expected based on the coefficient values from Figure \ref{fig:OLS_vs_RF_outcome}. The RF without regularization uses all variables while the RRF always selects the correct four features but often (in about 80\%) all other variables. Only the OARF selects the correct features and drops the covariates that are not of interest.

\begin{figure}[ht]
\begin{subfigure}[b]{0.5\linewidth}
\centering
\includegraphics[width=0.8\textwidth]{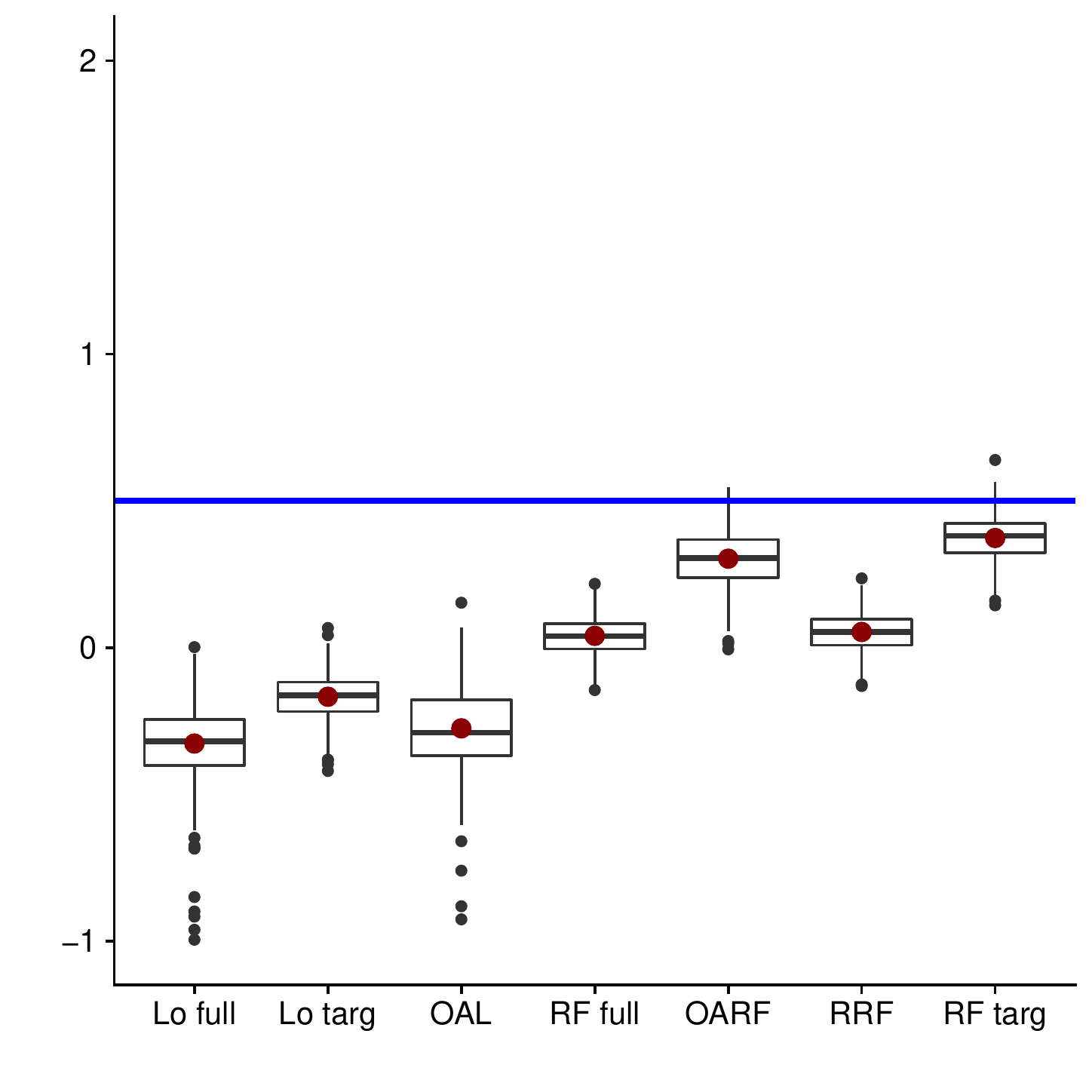}

\captionof{figure}{Setting 4}
\end{subfigure}
\begin{subfigure}[b]{0.5\linewidth}
\centering
\includegraphics[width=0.8\textwidth]{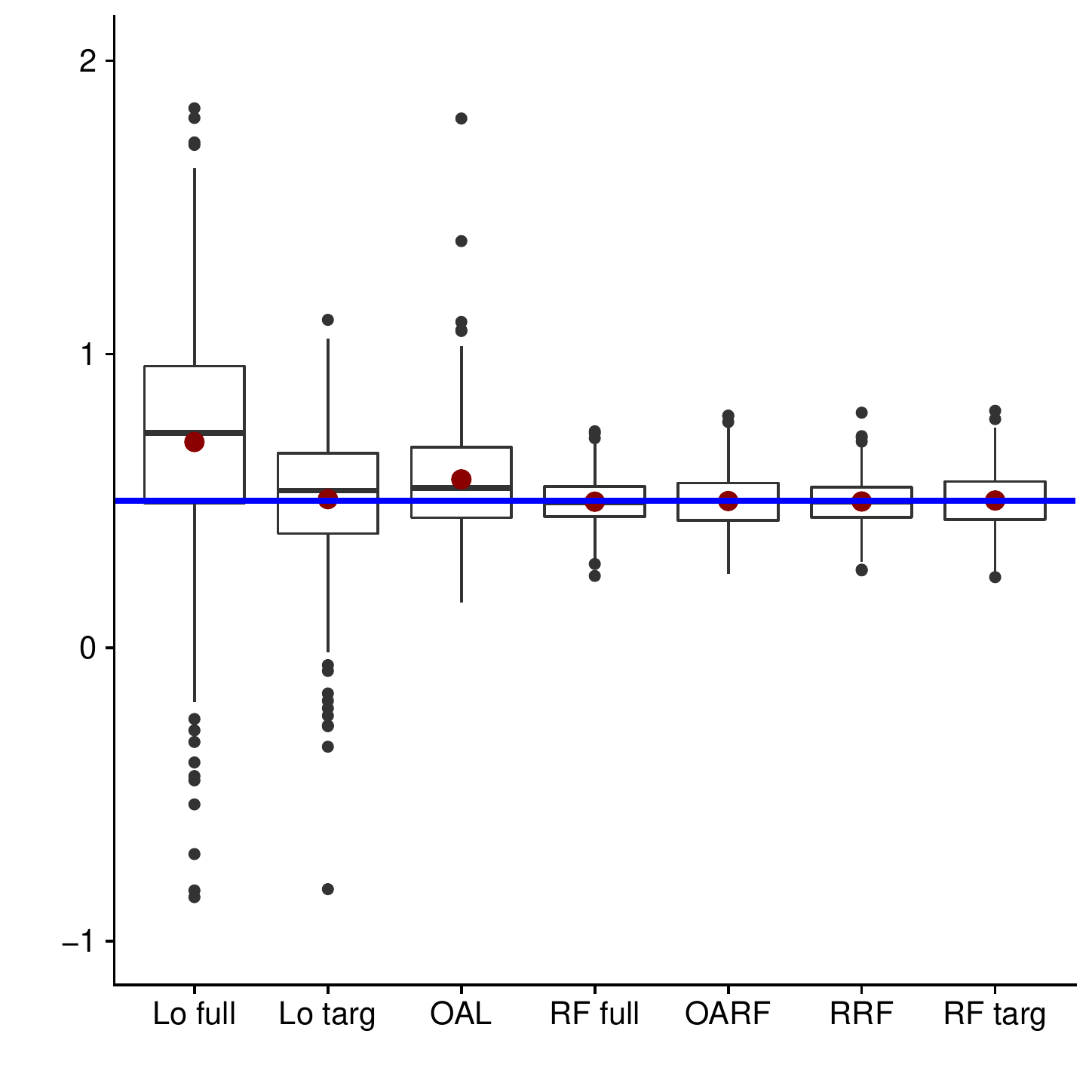}

\captionof{figure}{Setting 5}
\end{subfigure}

\begin{subfigure}[b]{0.5\linewidth}
\centering
\includegraphics[width=0.8\textwidth]{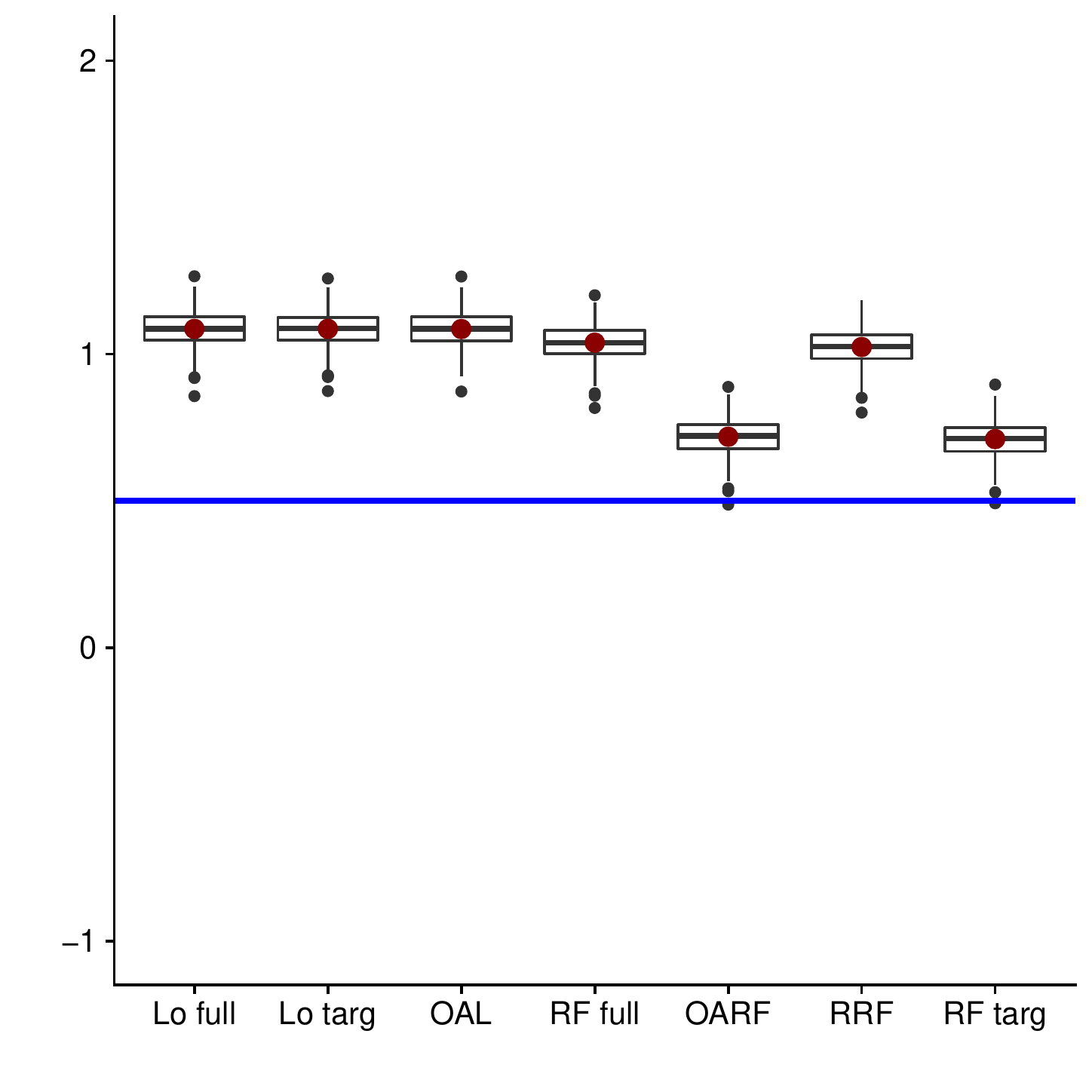}

\captionof{figure}{Setting 6}
\end{subfigure}
\begin{subfigure}[b]{0.5\linewidth}
\centering
\includegraphics[width=0.8\textwidth]{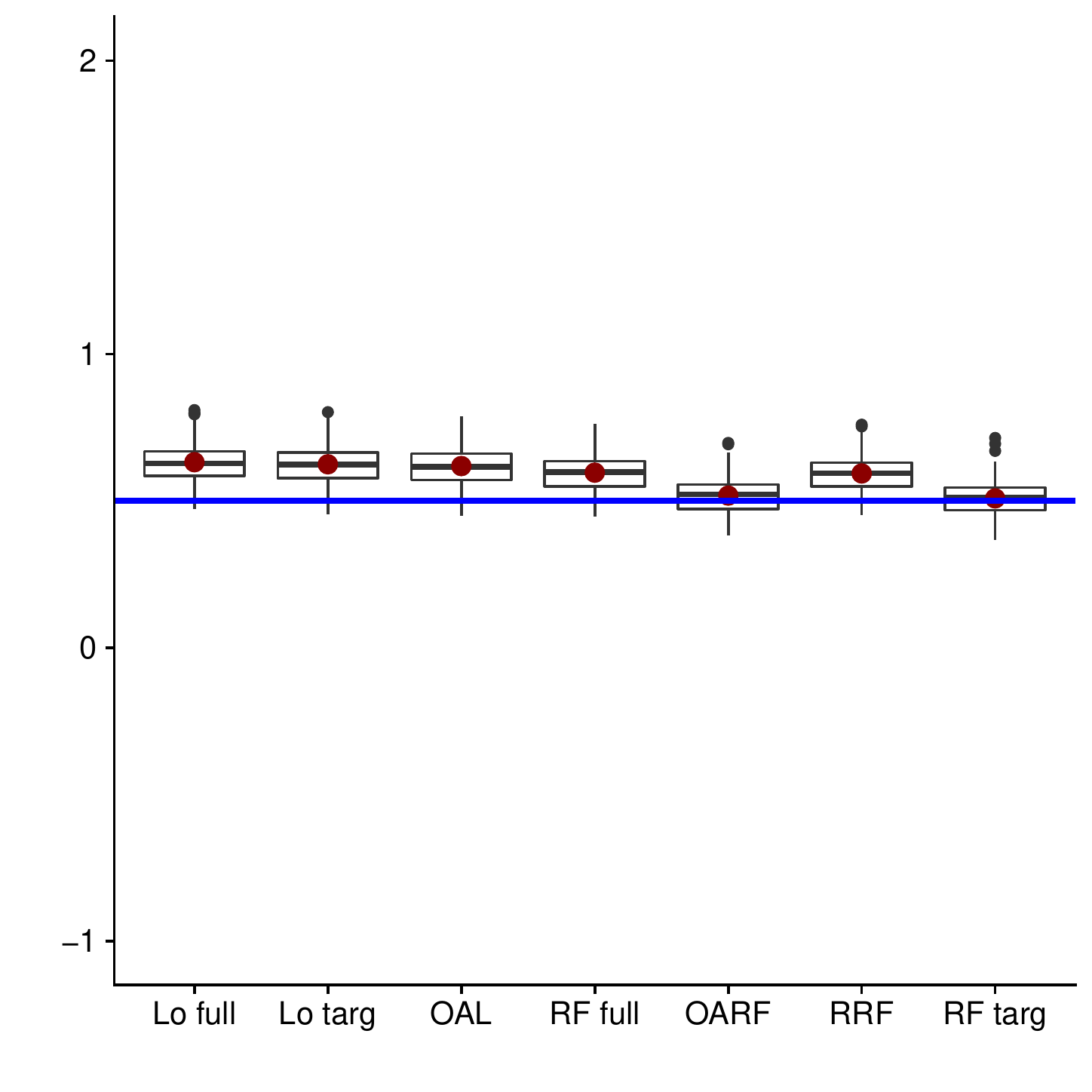}

\captionof{figure}{Setting 7}
\end{subfigure}
\caption{Illustration with non-linear functions for both models.}
\label{fig:non-lin-4-7}
\end{figure}

Next, we generate data processes where both functions are non-linear (settings 4 to 10). These are the settings where we would expect the OARF to perform superior. A complete list of all the DGP's is shown in Table \ref{tab:DGP}. In those settings, we set the sample size to $N = 1000$. Our results are shown for $p =20$ to allow a noticeable visualization. Varying the number of covariates to 50, 100, and 200 does not change the results of the ATE estimates nor the selection of the correct covariates. In setting 4 to 7, we keep the function on the outcome model and only change the propensity score function. Setting 1 to 7 sets the amount of variables for $X_c, X_o$, and $X_t$ to 2 while the amount of spurious variables $X_s = p-(X_c + X_o + X_t)$. Setting 8 to 10 uses $X_c = 6, X_o = X_t =2$ covariates and again the remaining set for $X_s$. The favourable covariates to select are $X_1$ to $X_4$ for settings 1 to 7 and $X_1$ to $X_8$ for setting 8 to 10. Figure \ref{fig:non-lin-4-7} shows ATE estimates for settings where all functions are non-linear and potentially complicated while Figure \ref{fig:boxplots_moreXc} shows results for similar functions but with more depending covariates. The proportion of selected covariates over all simulations are illustrated in Figure \ref{fig:selected_Var_lin_non_lin} and Figure \ref{fig:selected_Var_moreXc}. Overall, we find that the OARF performs best and is close to the RF that only uses $X_c$ and $X_o$ (RF targ). Setting 4 and 5 show that all methods are biased while the OARF is closed to the true ATE. In setting 4, all methods are downward biased, which leads to a negative estimate using the OAL. Only the RF approaches estimate the correct sign of the treatment effect. In setting 6, the linear methods are slightly upward biased while the RF approaches show no bias. Setting 7 shows a similar effect of bias where only the OARF estimates unbiased treatment effects. Using more covariates allows making the functions even more complicated since the flexibility can be increased. Setting 8 and 10 again show some bias for the linear methods as well as a higher variance compared to the RF approaches. Setting 9 is comparable to setting 4 in the sense that all methods are downward biased. In this setting, even the OARF is biased and shows a higher variance. We notice that in some settings there is not so much difference between the OARF and the full RF. The advantage of the OARF remains since it achieves the same accuracy using fewer variables as the full RF. In all settings that have at least one non-linear function, only the OARF selects the correct features with high precision. The full RF selects all covariates and the RRF uses unnecessary covariates in about 80\% of the cases.

Introducing a correlation between the covariates increases the bias and variance. We investigate the effect in a linear setting with moderate and heavy correlation. The results are shown in Figure \ref{fig:boxplots_correlations}. If the correlation is moderate ($\rho =0.2$) the OARF is still unbiased while all other methods show a slight bias and an increase in variance. It is reassuring that the random forest can find the correct importance score even with moderate correlation among the variables. If we introduce a heavy correlation ($\rho =0.5$) also the random forest approaches are biased while the variance in the linear models increases heavily. The OARF is still closest to the target method. 
The variable selection is still correct as illustrated in Figure \ref{fig:selected_Var_correlation}.  Last, we show results when the outcome function is more complicated and non-linear. Boxplots that illustrate the variance over 500 repetitions are shown in Figure \ref{fig:boxplots_complex_Y} while the corresponding variable coverage rates are shown in Figure \ref{fig:selected_Var_complex_Y}. 
Using a more complex outcome function increases the bias for the linear methods. Even the RRF shows a high bias in setting 13 and 14. The OARF is closest to the oracle RF. Setting 15 does show an equally small bias among all methods, again with the OARF and the oracle RF closest to the true treatment effect. The coverage rates regarding the selected variables show a similar picture as for all other settings. The vanilla RF does select all variables, the RRF selects the four correct variables in almost 100\% of the cases but also all other variables in more than 75\% of the cases. The OAL is not able to select the correct variables (e.g. variable 4 is selected in only 12.5\% of the cases. For setting 13 and 14, the OARF selects the desired variables in 100\% of the cases and drops all other variables with the same accuracy. Only in setting 15, one variable ($X_2$) is only selected around 37\% of the time while all other variables are selected as desired. 

In Table \ref{tab:CR_CI} we show coverage rates of 95\% confidence intervals and the width of the interval (in parentheses). Confidence intervals for IPW were constructed using a percentile‐based nonparametric bootstrap. For the OAL method, we use a smoothed non-parametric bootstrap approach that takes the model selection procedure into account. This procedure is described in \cite{efron2014estimation}. The confidence intervals for all RF approaches are based on non-parametric bootstrapping. We use 500 bootstrap samples for each method. We then take the $0.025$  $(\alpha/2)$ and $0.975$ $(1-\alpha/2)$ quantile from the empirical distribution as the lower and upper bound for the confidence interval. We apply three non-parametric versions, the \textbf{RF full} (without regularization), the \textbf{RRF} (which uses regularization but no initial feature set), and our proposed \textbf{OARF}. We also apply the \textbf{IPW} method using a linear model to estimate the propensity score and the \textbf{OAL} method using the outcome-adaptive lasso to estimate the propensity score.  We notice that the width of the confidence interval for the OARF is smaller than for the OAL method (in some settings only half as wide). The vanilla RF and the RRF do not achieve a coverage rate of 95\% for any settings, while the OARF achieve the rate in 8 out of the 15 settings. The results show that some data generating processes might be too complex and hence are biased for the IPTW estimator. This is why increasing the sample size does not lead to a higher coverage rate. In all other settings, we see an increase in the coverage rate and a decrease in the width of the confidence intervals when increasing the sample size from N=500 to N=2000. 
\begin{sidewaystable}
\centering
\caption{Coverage rates and width for 95\% confidence intervals}
\label{tab:CR_CI}
\resizebox{1\linewidth}{!}{%
\begin{tabular}{lrrrrrrrrrrrrrrr}
\hline \hline
\multicolumn{1}{c}{\begin{tabular}[c]{@{}c@{}}Setting / \\ Estimator\end{tabular}} & \multicolumn{1}{c}{1} & \multicolumn{1}{c}{2} & \multicolumn{1}{c}{3} & \multicolumn{1}{c}{4} & \multicolumn{1}{c}{5} & \multicolumn{1}{c}{6} & \multicolumn{1}{c}{7} & \multicolumn{1}{c}{8} & \multicolumn{1}{c}{9} & \multicolumn{1}{c}{10} & \multicolumn{1}{c}{11} & \multicolumn{1}{c}{12} & \multicolumn{1}{c}{13} & \multicolumn{1}{c}{14} & \multicolumn{1}{c}{15} \\ \hline
\multicolumn{16}{c}{\textbf{\textbf{N = 500}}} \\
IPW     & 0.81 (0.98) & 0.96 (0.58) & 0.94 (1.07) & 0.05 (0.85) & 0.00 (0.53)& 0.84 (1.90) & 0.84 (0.55)  & 0.95 (2.59) &0.10 (0.91)             & 0.91 (1.28) & 0.65 (1.09) & 0.94 (0.55) & 0.10 (1.43) & 0.07 (1.16) & 0.52 (0.39) \\
OAL     & 0.97 (0.72) & 0.99 (0.59) & 1.00 (1.01) & 0.18 (1.00) & 0.00 (0.74) & 1.00 (1.33) & 0.97 (0.76)  & 0.99 (1.38) & 0.14 (1.11)            & 0.99 (1.23) & 0.98 (0.71) & 0.99 (0.58) & 0.02 (1.32) & 0.04 (1.31) & 0.89 (0.67) \\
RF full & 0.00 (0.46) & 0.04 (0.45) & 0.93 (0.49) & 0.04 (0.49) & 0.00 (0.47) & 0.94 (0.57) & 0.82( 0.48)  & 0.93 (0.62) & 0.04 (0.65)            & 0.89 (0.62) & 0.00 (0.45) & 0.00 (0.45) & 0.00 (0.82) & 0.00 (0.44) & 0.38 (0.35) \\
RRF     & 0.00 (0.46) & 0.04 (0.45  & 0.94 (0.49) & 0.05 (0.49) & 0.00 (0.47) & 0.94 (0.58) & 0.83 ( 0.48) & 0.92 (0.63) &0.04 (0.65)             & 0.89 (0.62) & 0.00 (0.46) & 0.00 (0.45) & 0.00 (0.83) & 0.00 (0.44) & 0.38 (0.35) \\
OARF    & 0.83 (0.67) & 0.92 (0.58) & 0.96 (0.51) & 0.13 (0.51) & 0.03 (0.48) & 0.97 (0.66) & 0.87 ( 0.48) & 0.95 (0.62) &0.07 (0.70)             & 0.92 (0.66) & 0.82 (0.68) & 0.93 (0.55) & 0.23 (1.02) & 0.63 (0.75) & 0.71 (0.36) \\
\multicolumn{16}{c}{\textbf{N = 2000}}    \\
IPW     & 0.54 (0.49) & 0.96 (0.25) & 0.89 (0.51) & 0.05 (0.43) &0.00 (0.26)      & 0.68 (1.20) & 0.48 (0.27)  & 0.95 (1.28) & 0.00 (0.43) & 0.94 (0.66) & 0.5 (0.48)  & 0.92 (0.26) & 0.12 (0.63) & 0.04 (0.90)           &0.19 (0.28)             \\
OAL     & 1.00 (0.59) & 1.00 (0.47) & 1.00 (0.91) & 0.20 (1.15) & 0.00 (0.58)     & 1.00 (1.72) & 1.00 (0.60)  & 1.00 (0.83) & 0.00 (0.97) & 1.00 (1.19) & 1.00 (0.59) & 1.00 (0.47) & 0.01 (0.95) & 0.06 (1.29)            &0.95 (0.62)             \\
RF full & 0.20 (0.23) & 0.00 (0.22) & 0.93 (0.24) & 0.00 (0.24) & 0.00 (0.23)    & 0.92 (0.29) & 0.55 (0.24)  & 0.93 (0.31) & 0.00 (0.32) & 0.88 (0.31) & 0.00 (0.23) & 0.00 (0.22) & 0.00 (0.45) &0.00 (0.32)             &0.12 (0.24)             \\
RRF     & 0.14 (0.23) & 0.00 (0.22) & 0.93 (0.24) & 0.00 (0.24) &0.00 (0.23)      & 0.92 (0.29) & 0.55 (0.24)  & 0.93 (0.31) & 0.00 (0.32) & 0.88 (0.31) & 0.00 (0.23) & 0.00 (0.22) & 0.00 (0.45) & 0.00 (0.32)            &0.12 (0.24)             \\
OARF    & 0.96 (0.38) & 0.95 (0.30) & 0.98 (0.33) & 0.14 (0.31) & 0.00 (0.25)     & 0.99 (0.42) & 0.94 (0.25)  & 0.97 (0.22) & 0.00 (0.33) & 0.95 (0.33) & 0.85 (0.38) & 0.95 (0.29) & 0.25 (0.62) & 0.59 (0.55)            & 0.78 (0.27)            \\
\hline \hline
\end{tabular}
}

\end{sidewaystable}

\begin{figure}[ht]
\begin{subfigure}[b]{0.5\linewidth}
\centering
\includegraphics[width=0.8\textwidth]{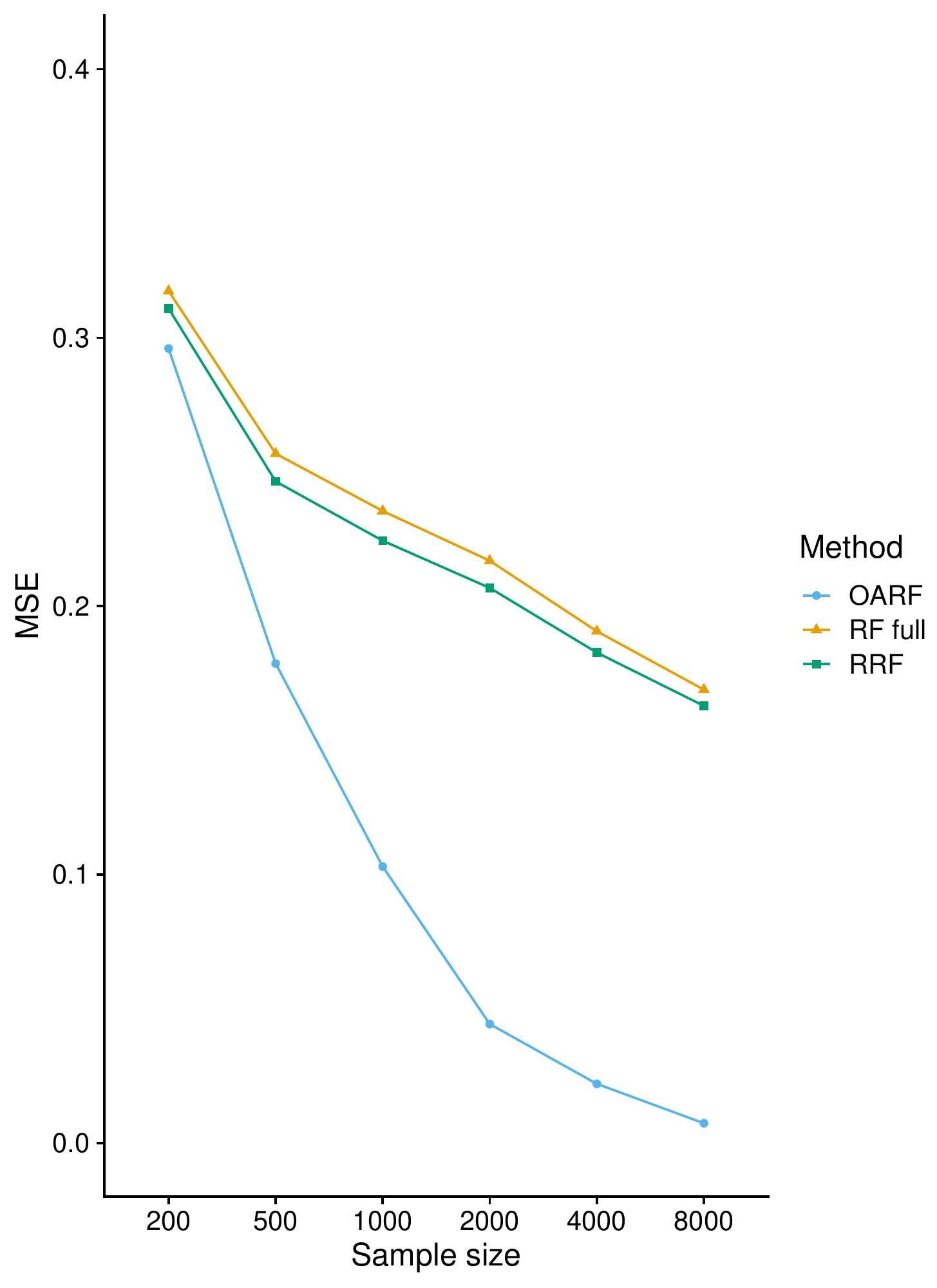}

\captionof{figure}{Setting 4}
\end{subfigure}
\begin{subfigure}[b]{0.5\linewidth}
\centering
\includegraphics[width=0.8\textwidth]{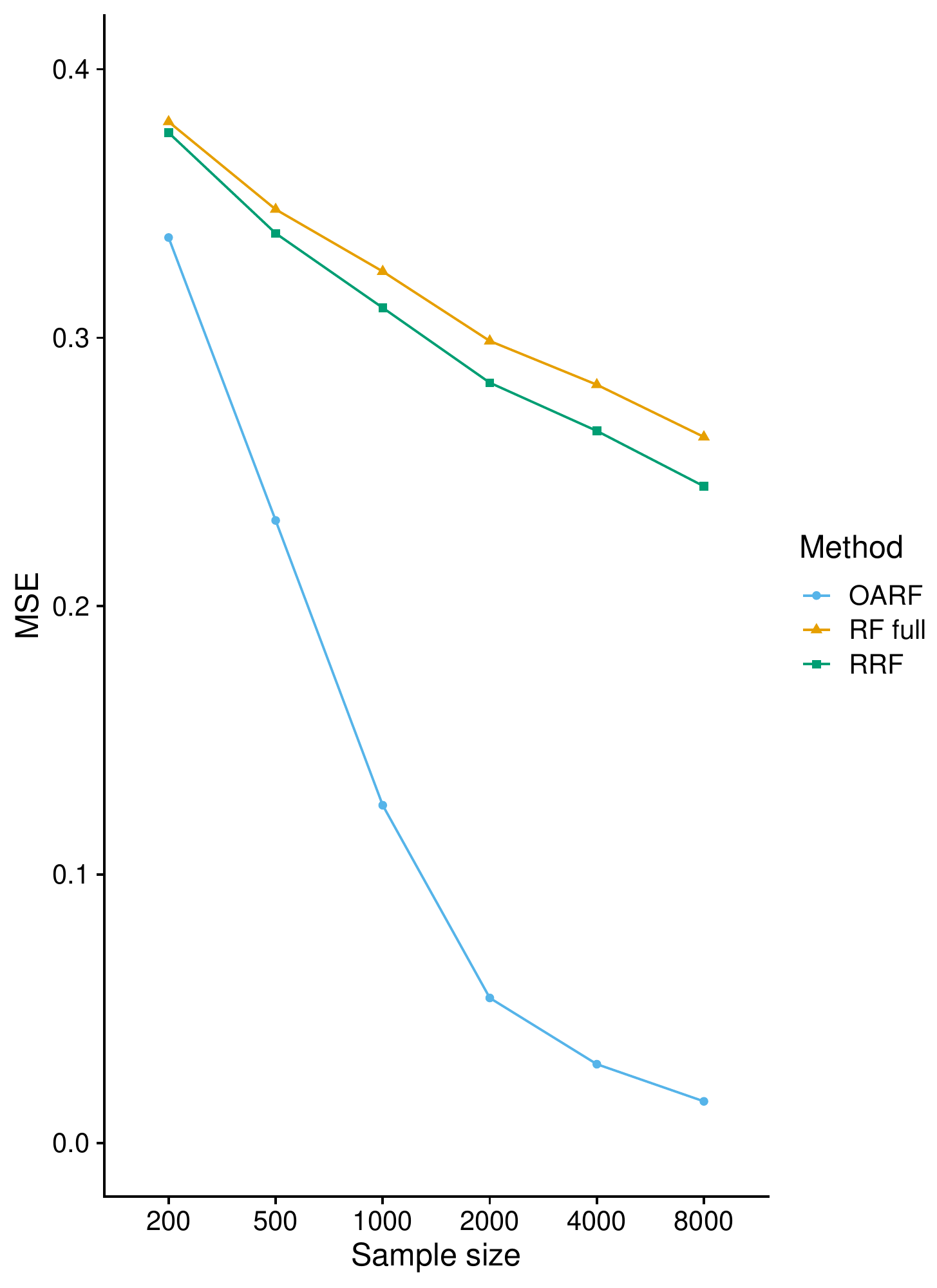}

\captionof{figure}{Setting 5}
\end{subfigure}
\caption{MSE given different sample sizes.}
\label{fig:MSE_sam_size}
\end{figure}

The OARF approach does not only decrease the variance when increasing the sample size but also the bias. Figure \ref{fig:MSE_sam_size} shows the mean squared error over 400 Monte Carlo replications for six different samples sizes (from 200 to 8000 observations). Here we use setting 4 and 5 as the data generating process since those settings have a low coverage rate of the confidence intervals. We compare the three random forest approaches, the full version, the regularized, and the OARF. The results show that the OARF achieves a significant decrease in MSE when increasing the sample size (e.g. for setting 5: from MSE = 0.35 for 200 observations, to MSE  = 0.03 for 8000). The other two algorithms always have a higher MSE and the decrease is only slightly. The main reason for the high decrease in MSE when using the OARF is both, a decrease in variance and bias. To visualize this result we show boxplots in Figure \ref{fig:boxplot_sam_size_S4} and \ref{fig:boxplot_sam_size_S5} in the Appendix. They show the ATE estimation using the three random forest algorithms under varying sample sizes for each of the 400 Monte Carlo iterations. While the variance decreases in all algorithms when increasing the sample size, only the OARF has a noticeable decrease in bias.

In the Appendix, we discuss the possibility of tuning certain parameters in the random forest while keeping the objective to get balanced covariates rather than maximize the prediction accuracy. We also discuss a generalization of the OARF to other methods like the double machine learning, introduced by \cite{chernozhukov2018double}. We find that using the OARF to estimate the propensity score decreases the bias compared to a full RF. In settings where even the OARF is biased, we additionally use the OARF to estimate the conditional mean of $Y$. This approach further decreases the bias.

\section{Empirical Examples}
In this section, we revisit two empirical examples. Both datasets are observational and contain a rich set of characteristics that may lead to selection into treatment. 
We use the same 5 methods as in the simulation study above. We report the ATE and a 95\% confidence interval (CI).  The datasets we use are freely accessible: RHC (\url{https://hbiostat.org/data/}), birth weight (\url{http://www.stata-press.com/data/r13/cattaneo2.dta}).

\subsection{Re-analysis of SUPPORT data on Right Heart Catheterization}
Right heart catheterization (RHC) is a diagnostic procedure used for critically ill patients. \cite{connors1996effectiveness}used a propensity score matching approach to study the effectiveness of right heart catheterization in an observational setting. The authors found that after controlling for selection bias by using a rich set of covariates, RHC appeared to lead to higher mortality than not performing RHC. This conclusion is in contrast to popular belief among practitioners that RHC was beneficial. 

The SUPPORT study collected data on hospitalized adult patients at five medical centers in the US. Based on information from a panel of experts, a rich set of characteristics, believed to be related to the decision on whether to perform the right heart catheterization or not, was collected.
The study consists of data on 5735 individuals, RHC was performed on 2184 (the treatment group) while the remaining 3551 individuals did not get RHC (the control group). Treatment is equal to 1 if right heart catheterization was applied within 24 hours of admission, and 0 otherwise. The outcome of interest is an indicator for survival at 30 days. In total, we observe 72 covariates (including many dummy variables). After excluding variables with more than 50\% missing values, we have 68 covariates left. 

\cite{connors1996effectiveness} matched treated and untreated patients based on the propensity score, with each unit, matched at most once.  \cite{hirano2001estimation} use different matching estimators (one of them is the exact matching) and their regression adjusted estimates range from  $(-0.062, -0.053)$. \cite{crump2009dealing} apply different samples based on the propensity score overlap and estimate an ATE from $-0.059$ to $-0.060$. \cite{ramsahai2011extending} use propensity score and generic matching to investigate the effect of RHC on mortality within 180 days (note that the outcome variable is now an indicator for mortality, not survival). Their results are consistent with the previous findings, namely an estimated ATE using propensity score matching of $0.063$ and $0.046$ when using generic matching. A more recent study by \cite{li2018balancing} considers different weighting strategies for covariate balancing. These strategies are based on the propensity score to estimate different target parameters. Their results also confirm previous estimates. For example, they estimate an ATE using overlap weighting of $-0.065$ and $-0.067$ when using optimal matching. All methods mentioned above use a logit model to estimate the propensity score. Given the rich set of covariates that consists of characteristics like age, race, income, and medical characteristics, there might be interaction effects and non-linear dependence. It would also be useful to know which covariates are true confounders and are selected when we do not assume any parametric form for both, the outcome model and the propensity score model. 

As shown in Table \ref{tab:RHC_ATE}, all methods estimate a negative treatment effect, suggesting that performing RHC does decrease survival within the first 30 days. The results from the IPW and OAL are consistent with results from previous findings where the propensity score is estimated via a logit model. The RF results are smaller in magnitude with the lowest ATE of $-0.041$ by the OARF. The confidence intervals for the RF methods are tighter compared to the IPW or OAL method, while the latter method even includes zero. Among the 68 covariates, we find that the full RF selects the same variables as the RRF, in total 42 covariates with a proportion of $90-100\%$. The OAL method selects only 3 variables and 6 variables around 50\% of the time (proportion between $50-60\%$). The OARF selects 9 variables in almost all iterations. All variables selected by OAL are also selected by OARF and additional variables are age, Duke Activity Status Index (DASI), APACHE score, Glasgow Coma Score (scoma1), white blood cell count Day 1 (wblc1), and bilirubin Day 1 (bili1). Variables that are not selected by OAL and OARF but with the RF and RRF are, among others, sex, race, education, and income. Especially characteristic variables that might be uncorrelated with a person's well being are not selected by the OARF but are still selected using the RRF using the same weights as the OARF. The complete list of covariates and their inclusion proportion is listed in Table \ref{tab:RHC_cov_full}. Compared to OAL, the OARF includes age, DASI, white blood cells, and three other covariates.

\begin{table}[ht]
\centering
\caption{Estimates for average treatment effects in RHC study}
\label{tab:RHC_ATE}
\resizebox{0.5\textwidth}{!}{%
\begin{tabular}{lrc}
\hline \hline
Method  & Estimate & 95\% CI 		  \\ \hline
IPW     & -0.055 & (-0.087, -0.026)             \\
OAL     & -0.057 & (-0.158, +0.031)            \\
RF full & -0.045 & (-0.074, -0.026)             \\
RRF     & -0.045 & (-0.074, -0.027)            \\
OARF    & -0.041 & (-0.071, -0.022)              \\
       \hline \hline
\end{tabular}
}
\end{table}

\begin{table}[ht]
\centering
\caption{Covariate selection (at least 90\%)}
\label{tab:RHC_Cov_Sel}
\resizebox{0.7\textwidth}{!}{%
\begin{tabular}{lcc}
\hline \hline
Method  & \# Covariates & excluded covariates		  \\ \hline     
OAL     & 3 &   e.g. sex, race, education, income           \\
RF full & 42 &    e.g. trauma, rental, hema         \\
RRF     & 42 &     e.g. trauma, rental, hema           \\
OARF    & 9 & e.g. sex, race, education, income            \\
       \hline \hline
\end{tabular}
}
\end{table}

\subsection{Effect of smoking on birth weight}

In this example, we reinvestigate the effect of maternal smoking status during pregnancy, the treatment variable, on babies’ birth weight, the outcome variable. We use a publicly available dataset that consists of 4642 singleton births in the USA. Additionally, we observe a rich set of characteristics like age,  marital status, race, education, number of prenatal care visits, months since last birth, an indicator of firstborn infant, and indicator of alcohol consumption during pregnancy. All covariates are for the mother, except education, which we also observe for the father. The full dataset, containing more observations, was first used by \cite{almond2005costs} who found a strong negative effect of maternal smoking during pregnancy on the weights of babies (about 200 -- 250 gram lighter for a baby with a mother smoking). In their study, the authors use a logit model to estimate the propensity score. We focus on estimating the propensity score without assuming any parametric form and again base the variable selection on features that are associated with the outcome. Table \ref{tab:CAT_ATE} shows the ATE estimates along with 95\% CI`s. We find similar negative effects as \cite{almond2005costs}. The OARF estimates a decreased birth weight of -224 grams. Compared to the other RF models and the OAL method, the OARF has the tightest confidence intervals. Only the classic IPW estimator has a slightly tighter upper bound.

The OAL model and the full RF include 17 out of 19 variables while the RRF model includes 16 covariates. The OARF only includes 7 covariates and excludes, for example, an indicator for marital status, the education of mother and father, and if there were prenatal visits. The excluded variables are shown in Table \ref{tab:BW_CovSel} and the complete list with inclusion probability is shown in Table \ref{tab:BW_cov_full}. The OARF is the only method that excludes the variable alcohol, meaning that it is not a confounding variable nor is it predictive on the outcome. The results, from a recent study by \cite{lundsberg2015low}, suggest low-to-moderate alcohol exposure during early and late gestation is not associated with increased risk of low birth weight. Such findings are quite interesting since they allow a better understanding and interpretation of true confounding variables and of such that are not informative. 

\begin{table}[ht]
\centering
\caption{Estimates for average treatment effects in birth weight study}
\label{tab:CAT_ATE}
\resizebox{0.4\textwidth}{!}{%
\begin{tabular}{lrc}
\hline \hline
Method  & Estimate & 95\% CI 		  \\ \hline
IPW     & -236 & (-286,-187)             \\
OAL     & -236 & (-357,-115)            \\
RF full & -221 & (-287,-162)             \\
RRF     & -205 & (-347,-83)            \\
OARF    & -224 & (-286,-165)              \\
       \hline \hline
\end{tabular}
}
\end{table}

\begin{table}[ht]
\centering
\caption{Covariate selection (at least 90\%)}
\label{tab:BW_CovSel}
\resizebox{0.9\textwidth}{!}{%
\begin{tabular}{lcc}
\hline \hline
Method  & \# Covariates & excluded covariates                                                                                                            \\ \hline
OAL     & 17            & mother age, birth month                                                                                                        \\
RF full & 17            & mother hispanic, foreign,                                                                                                      \\
RRF     & 16            & mother hispanic, father hispanic, foreign,                                                                                     \\
OARF    & 7             & \begin{tabular}[t]{@{}c@{}}all the above and married, alcohol, \\ m. and f. education, first baby, prenatal visit\end{tabular} \\ 
       \hline \hline
\end{tabular}
}
\end{table}

\section{Discussion}

We propose a non-parametric variable selection procedure for the estimation of treatment effects from observational studies. Building on outcome-adaptive penalization, we use a random forest to define variables that are predictive of the outcome. This allows the outcome model to deviate from a linear dependence on the covariates. We use a modified variable importance score to generate coefficients from the outcome model. We use the importance score to penalize variables that are spurious or that only predict the treatment but not the outcome. We show how to use penalty weights for each covariate to regularize the random forest that estimates the propensity score. Additional to the penalty, we show the importance of an initial feature space and how to include it in the RF. A Monte Carlo simulation shows that our proposed method, the OARF, has a smaller variance and produces unbiased estimates. In cases where all estimators are biased, the OARF produces the smallest bias and variance. The second goal of our proposed approach is to select only variables that have a relationship with the outcome (including all confounding variables) while excluding variables that predict the treatment and unimportant variables. Based on the simulation, we find that only the OARF selects the correct covariates and disregards all others. This holds for linear and non-linear settings and even if there is a strong correlation between the variables. 

We apply the OARF and all other benchmark methods in two empirical examples. The ATE is comparable between all methods while the OARF shows tighter confidence intervals compared to the OAL and other RF methods. Regarding the variable selection, we find that the OARF selects and drops different variables compared to all other methods. This allows for a detailed evaluation of which variable might be responsible for selection bias and which variables are just spurious or only affect the propensity score.

\clearpage
\FloatBarrier 
\newpage

\phantomsection \addcontentsline{toc}{section}{References}
\bibliographystyle{plainnat}
\bibliography{references}


\clearpage

\appendix

\section{Tables}

\begin{table}[ht]
\centering
{\renewcommand{\arraystretch}{1.5}%
\resizebox{0.8\textwidth}{!}{%
    \begin{threeparttable}
\caption{Data generating processes.}
\label{tab:DGP}
\begin{tabular}{lll}
\hline \hline
DGP & Propensity score model & Outcome model \\
\hline
1 & $\sum_{j=1}^{p} \nu_{j} X_{j}; \quad \nu = (1,1,0,0,1,1,0,...,0)$ & $\theta A + \sum_{j=1}^{p} \beta_{j} X_{j}+\varepsilon$ \\
2 & $\sum_{j=1}^{p} \nu_{j} X_{j}; \quad \nu = (0.4,0.4,0,0,1,1,0,...,0)$ & $\theta A + \sum_{j=1}^{p} \beta_{j} X_{j}+\varepsilon$ \\
3 & $\sum_{j=1}^{p} \nu_{j} X_{j}; \quad \nu = (1,1,0,0,1,1,0,...,0)$ & $\theta A +0.8X_1\otimes X_2 + 0.8X_3\otimes X_4 +\varepsilon$ \\
 4 & $X_1(1-X_2) + X_5(1-X_6)$ & $\theta A +0.8X_1\otimes X_2 + 0.8X_3\otimes X_4 +\varepsilon$ \\
 5 &$0.8X_1 \otimes X_2 + 0.8X_5 \otimes X_6$   & $\theta A +0.8X_1\otimes X_2 + 0.8X_3\otimes X_4 +\varepsilon$ \\
 6 &$2\cos(X_2) + \sum_{j=1}^{p} \nu_{j} X_{j}; \quad \nu = (1,1,0,0,1,1,0,...,0)$ & $\theta A +0.8X_1\otimes X_2 + 0.8X_3\otimes X_4 +\varepsilon$ \\
 7 & $2\mathds{1}_{\{X_1>0\}}\mathds{1}_{\{X_2>1\}} + 2\mathds{1}_{\{X_5>0\}}\mathds{1}_{\{X_6>1\}} + X_1 \otimes X_6$ & $\theta A +0.8X_1\otimes X_2 + 0.8X_3\otimes X_4 +\varepsilon$ \\
 8 &\begin{tabular}[t]{@{}l@{}}  $2X_2 \otimes (1-X_6) + 2X_1 \mathds{1}_{\{X_9>1\}} +\sum_{j=1}^{p} \nu_{j} X_{j};$ \\
  $ \nu = (1,1,1,1,1,1,0,0,1,1,0,...,0)$ \end{tabular} & \begin{tabular}[t]{@{}l@{}} $\theta A + 0.8X_1 \otimes X_2 + 0.8X_3 \otimes X_4 +$ \\  
$0.8X_5 \otimes X_6 + 0.8X_7 \otimes X_8 +\varepsilon$ \end{tabular} \\
 9 &$0.5X_1^2 + 0.5X_2 - X_3 \otimes X_4 + 0.5X_5 + 0.5X_6 + 0.5X_9^2 + 0.5X_{10}$  &\begin{tabular}[t]{@{}l@{}} $\theta A + X_1 \otimes X_2 + X_3 \otimes X_4 +$ \\ $0.5X_2 + 0.5X_6 + X_7 \otimes X_8 +\varepsilon$ \end{tabular} \\
 10 &$-e^{(X_1)} + 0.4X_2 + e^{(X_3)} + 0.4X_4 + 0.5X_5^2 + X_6 \otimes X_9 + 0.4X_{10}$  &  \begin{tabular}[t]{@{}l@{}} $\theta A + 0.8X_1 \otimes X_2 + 0.8X_3 \otimes X_4 +$ \\  
$0.8X_5 \otimes X_6 + 0.8X_7 \otimes X_8 +\varepsilon$ \end{tabular}\\
11 & \multicolumn{2}{c}{As setting 1 but with a correlation between the variables of around 0.2}  \\
12 & \multicolumn{2}{c}{As setting 2 but with a correlation between the variables of around 0.2} \\
13 & $X_1(1-X_2) + X_5(1-X_6)$  &$X_1(1-X_2) + X_3(1-X_4)$ \\
14 &$2\cos(X_2) + \sum_{j=1}^{p} \nu_{j} X_{j}; \quad \nu = (1,1,0,0,1,1,0,...,0)$ & \begin{tabular}[t]{@{}l@{}} $2\cos(X_2) + \sum_{j=1}^{p} \beta_{j} X_{j};$ \\ $\beta = (0.6,0.6,0.6,0.6,0,0,0,...,0)$ \end{tabular} \\
15 &$2\mathds{1}_{\{X_1>0\}}\mathds{1}_{\{X_2>1\}} + 2\mathds{1}_{\{X_5>0\}}\mathds{1}_{\{X_6>1\}} + X_1 \otimes X_6$ & $2\mathds{1}_{\{X_1>0\}}\mathds{1}_{\{X_2>1\}} + 2\mathds{1}_{\{X_3>0\}}\mathds{1}_{\{X_4>1\}} + X_1 \otimes X_4$\\
\hline \hline
\end{tabular}
\begin{tablenotes}
      \small
      \item \textit{Notes: Only setting 1 and 2 have a linear DGP. Setting 1 to 7 and 11 to 15 set $X_c = X_o = X_t =2$ while setting 8 to 10 set $X_c = 6, X_o = X_t = 2$. } 
    \end{tablenotes}
\end{threeparttable}
}
}
\end{table}

\begin{table}[ht]
\centering
\caption{RHC study: Selected covariates by categories}
\label{tab:RHC_cov_full}
\resizebox{0.8\textwidth}{!}{%
\begin{tabular}{lrrrr||lrrrr}
\hline \hline
              & \multicolumn{4}{c}{\% Selected} &            & \multicolumn{4}{c}{\% Selected} \\
Covariates    & RF full  	 & RRF     & OARF     & OAL 	 & Covariates & RF full  & RRF   & OARF  & OAL  \\ \hline
age           & 1.0 & 1.0 & 1.0 & 0.0    & DASI    & 1.0 & 1.0 & 1.0 & 0.0  \\
sex           & 1.0 & 1.0 & 0.0 & 0.0    & APACHE score        & 1.0 & 0.8 & 1.0 & 1.0  \\
raceblack     & 0.6 & 0.4 & 0.0 & 0.0  & ca\_yes       & 0.8 & 0.7 & 0.0 & 0.2\\
raceother     & 0.0 & 0.0 & 0.0 & 0.0  & ca\_meta      & 0.0 & 0.0 & 0.0 & 0.2\\
edu           & 1.0 & 1.0 & 0.0 & 0.0    & surv2md1    & 1.0 & 1.0 & 1.0 & 1.0  \\
income1       & 0.9 & 0.8 & 0.0 & 0.0   & aps1         & 1.0 & 1.0 & 1.0 & 0.0 \\
income2       & 0.9 & 0.9 & 0.0 & 0.0  & scoma1        & 1.0 & 1.0 & 1.0 & 0.0\\
income3       & 0.3 & 0.2 & 0.0 & 0.0  & wtkilo1       & 1.0 & 1.0 & 0.0 & 0.0\\
ins\_care     & 0.9 & 0.9 & 0.0 & 0.0   & temp1        & 1.0 & 1.0 & 0.0 & 0.0 \\
ins\_pcare    & 1.0 & 0.8 & 0.0 & 0.0    & meanbp1     & 1.0 & 1.0 & 0.4 & 0.0  \\
ins\_caid     & 0.7 & 0.4 & 0.0 & 0.0    & resp1       & 1.0 & 1.0 & 0.0 & 0.0  \\
ins\_no       & 0.0 & 0.0 & 0.0 & 0.0    & hrt1        & 1.0 & 1.0 & 0.0 & 0.0  \\
ins\_carecaid & 0.1 & 0.0 & 0.0 & 0.0    & pafi1       & 1.0 & 1.0 & 0.3 & 0.0  \\
cat1\_copd    & 1.0 & 0.9 & 0.0 & 0.0    & paco21      & 1.0 & 1.0 & 0.7 & 0.0  \\
cat1\_mosfsep & 1.0 & 1.0 & 0.0 & 0.0    & ph1         & 1.0 & 1.0 & 0.0 & 0.8  \\
cat1\_mosfmal & 0.0 & 0.0 & 0.8 & 0.5    & wblc1       & 1.0 & 1.0 & 1.0 & 0.0  \\
cat1\_chf     & 0.9 & 0.9 & 0.0 & 0.0    & hema1       & 1.0 & 1.0 & 0.0 & 0.0  \\
cat1\_coma    & 0.2 & 0.1 & 1.0 & 1.0    & sod1        & 1.0 & 1.0 & 0.0 & 0.0  \\
cat1\_cirr    & 0.1 & 0.1 & 0.0 & 0.0    & pot1        & 1.0 & 1.0 & 0.0 & 0.0  \\
cat1\_lung    & 0.0 & 0.0 & 0.0 & 0.0    & crea1       & 1.0 & 1.0 & 0.6 & 0.0  \\
cat2\_mosfsep & 1.0 & 1.0 & 0.0 & 0.0    & bili1       & 1.0 & 1.0 & 1.0 & 0.0  \\
cat2\_coma    & 0.0 & 0.0 & 0.0 & 0.0    & alb1        & 1.0 & 1.0 & 0.0 & 0.0  \\
cat2\_mosfmal & 0.0 & 0.0 & 0.2 & 0.2    & cardiohx    & 1.0 & 0.9 & 0.0 & 0.0  \\
cat2\_lung    & 0.0 & 0.0 & 0.0 & 0.0    & chfhx       & 1.0 & 0.9 & 0.0 & 0.1  \\
cat2\_cirr    & 0.0 & 0.0 & 0.0 & 0.0    & dementhx    & 0.4 & 0.4 & 0.0 & 0.2  \\
resp          & 1.0 & 1.0 & 0.0 & 0.0    & psychhx     & 0.1 & 0.1 & 0.0 & 0.5  \\
card          & 1.0 & 1.0 & 0.0 & 0.0    & chrpulhx    & 1.0 & 0.8 & 0.0 & 0.0  \\
neuro         & 1.0 & 1.0 & 0.0 & 0.0    & renalhx     & 0.0 & 0.0 & 0.0 & 0.6  \\
gastr         & 0.9 & 0.9 & 0.0 & 0.0    & liverhx     & 0.1 & 0.1 & 0.0 & 0.4  \\
renal         & 0.1 & 0.0 & 0.0 & 0.0    & gibledhx    & 0.0 & 0.0 & 0.0 & 0.5  \\
meta          & 0.0 & 0.0 & 0.0 & 0.0    & malighx     & 1.0 & 0.9 & 0.0 & 0.7  \\
hema          & 0.1 & 0.0 & 0.0 & 0.0    & immunhx     & 1.0 & 0.9 & 0.0 & 0.0  \\
seps          & 1.0 & 1.0 & 0.0 & 0.0    & transhx     & 1.0 & 1.0 & 0.0 & 0.7  \\
trauma        & 0.0 & 0.0 & 0.0 & 0.0    & amihx       & 0.0 & 0.0 & 0.0 & 0.6  \\
\hline \hline
\end{tabular}
}
\end{table}

\begin{table}[ht]
\centering
\caption{Birth weight study: Selected covariates by categories}
\label{tab:BW_cov_full}
\begin{tabular}{lrrrr}
  \hline
              & \multicolumn{4}{c}{\% Selected}  \\ 
  Covariates    & RF full  	 & RRF     & OARF     & OAL \\ \hline
married & 1.0 & 1.0 & 0.8 & 1.0 \\ 
mhisp & 0.4 & 0.3 & 0.0 & 0.9 \\ 
fhisp & 0.9 & 0.8 & 0.0 & 0.9 \\ 
foreign & 0.8 & 0.7 & 0.1 & 0.9 \\ 
alcohol & 1.0 & 1.0 & 0.0 & 0.9 \\ 
deadkids & 1.0 & 1.0 & 0.6 & 0.9 \\ 
mage & 1.0 & 1.0 & 1.0 & 0.8 \\ 
medu & 1.0 & 1.0 & 0.6 & 0.9 \\ 
fage & 1.0 & 1.0 & 1.0 & 0.9 \\ 
 fedu & 1.0 & 1.0 & 0.6 & 0.9 \\ 
 nprenatal & 1.0 & 1.0 & 1.0 & 1.0 \\ 
 monthslb & 1.0 & 1.0 & 1.0 & 0.9 \\ 
 order & 1.0 & 1.0 & 0.2 & 1.0 \\ 
 mrace & 1.0 & 1.0 & 1.0 & 1.0 \\ 
 frace & 1.0 & 1.0 & 1.0 & 1.0 \\ 
 prenatal & 1.0 & 1.0 & 0.5 & 1.0 \\ 
 birthmonth & 1.0 & 1.0 & 1.0 & 0.8 \\ 
 fbaby & 1.0 & 1.0 & 0.1 & 1.0 \\ 
 prenatal1 & 1.0 & 1.0 & 0.1 & 1.0 \\ 
   \hline
\end{tabular}
\end{table}

\clearpage
\section{Proofs}

\textbf{Proof of unconfoundedness based on the propensity score.}

We  show that $Pr(D_i = 1|Y_i^0,Y_i^1,e(X_i)) = Pr(D_i = 1|e(X_i)) = e(X_i)$,
implying independence of $(Y_i^0, Y_i^1)$ and $D_i$ conditional on $e(X_i)$. First, note that

\begin{align}
\mathsf{E}_{D \mid e(X)}\left[D_{i} \mid e\left(X_{i}\right)\right] &=\mathsf{E}_{X \mid e(X)}\left\{\mathsf{E}_{D \mid X, e(X)}\left[D_{i} \mid X_{i}, e\left(X_{i}\right)\right] \mid e\left(X_{i}\right)\right\} \nonumber \\
&=\mathsf{E}\left[e\left(X_{i}\right) \mid e\left(X_{i}\right)\right]=e\left(X_{i}\right)
\end{align}

For simplification, we show the proof for $Y^0$ and note that the same logic follows for $Y^1$ or both.
Using the law of iterated expectation and noticing that $e(X_i)$ is a function of $X_i$, it follows that, 
\begin{align}
\mathsf{E}\left[D_{i} \mid Y_{i}^{0}, e\left(X_{i}\right)\right] &=\mathsf{E}_{X \mid Y_{i}^{0}, e(X)}\left\{\mathsf{E}_{D \mid Y^{0}, X, e(X)}\left[D_{i} \mid Y_{i}^{0}, X_{i}, e\left(X_{i}\right)\right] \mid Y_{i}^{0}, e\left(X_{i}\right)\right\} \nonumber \\
&=\mathsf{E}_{X \mid Y_{i}^{0}, e(X)}\left\{\mathsf{E}_{D \mid Y^{0}, X}\left[D_{i} \mid Y_{i}^{0}, X_{i}\right] \mid Y_{i}^{0}, e\left(X_{i}\right)\right\} \label{equ: uncon_2}
\end{align}

Using the assumption of conditional independence of $D_i$ and $Y^0_i$ given $X_i$ allows us to neglect the conditioning in equation \ref{equ: uncon_2}: 
\begin{align}
\mathsf{E}_{X \mid Y_{i}^{0}, e(X)}\left\{\mathsf{E}_{D \mid Y^{0}, X}\left[D_{i} \mid Y_{i}^{0}, X_{i}\right] \mid Y_{i}^{0}, e\left(X_{i}\right)\right\} &= \nonumber \\
\mathsf{E}_{X \mid Y_{i}^{0}, e(X)}\left\{\mathsf{E}_{D \mid X}\left[D_{i} \mid X_{i}\right] \mid Y_{i}^{0}, e\left(X_{i}\right)\right\} &= \nonumber \\
\mathsf{E}_{X \mid Y_{i}^{0}, e(X)}\left[e\left(x_{i}\right) \mid Y_{i}^{0}, e\left(X_{i}\right)\right] &=e\left(X_{i}\right)
\end{align}

Combining equation 3 and 2 shows that, $\mathsf{E}\left[D_{i} \mid Y_{i}^{0}, e\left(X_{i}\right)\right] = e(X_i)$. Based on this result and using equation 1 shows that given the propensity score, $D_i$ is independent from $Y_i^0$.

\textbf{Propensity score model for linear example:}

Dependence is linear: $a(x) = 1X_1 + 1X_2 + 1X_5 + 1X_6$.
Calculate the probability distribution for the vector $a$ from the 	logit distribution function:  
\begin{align*}
e_{0}(x) = \frac{exp\{a(x)\}}{(1+exp\{a(x)\})}
\end{align*}
Apply a random number generator from a Binomial function $B\{N,e_0(x)\}$ with probability for success = $e_{0}(x)$. This creates a vector $D \in \{0;1\}$ such that
\begin{align*}
D \overset{ind.}{\sim} \text{Bernoulli}\{e_{0}(x)\}.
\end{align*}

\textbf{Min-Max normalization to the interval $[0,1]$: }

\begin{align*}
x' = a + \frac{(x - \text{min}(x))(b-a)}{\text{max}(x)-\text{min}(x)}
\end{align*}

Since the $min(VarImp) = 0$, $a=0$ and $b=1$, the expression simplifies to 

\begin{align*}
x' =  \frac{x}{\text{max}(x)}
\end{align*}

If $min(VarImp) \neq$ 0:

\begin{align*}
x' = \frac{(x - \text{min}(x))}{\text{max}(x)-\text{min}(x)}
\end{align*}

\clearpage
\section{Figures}

\begin{figure}[ht]
\begin{subfigure}[b]{0.5\linewidth}
\centering
\includegraphics[width=0.8\textwidth]{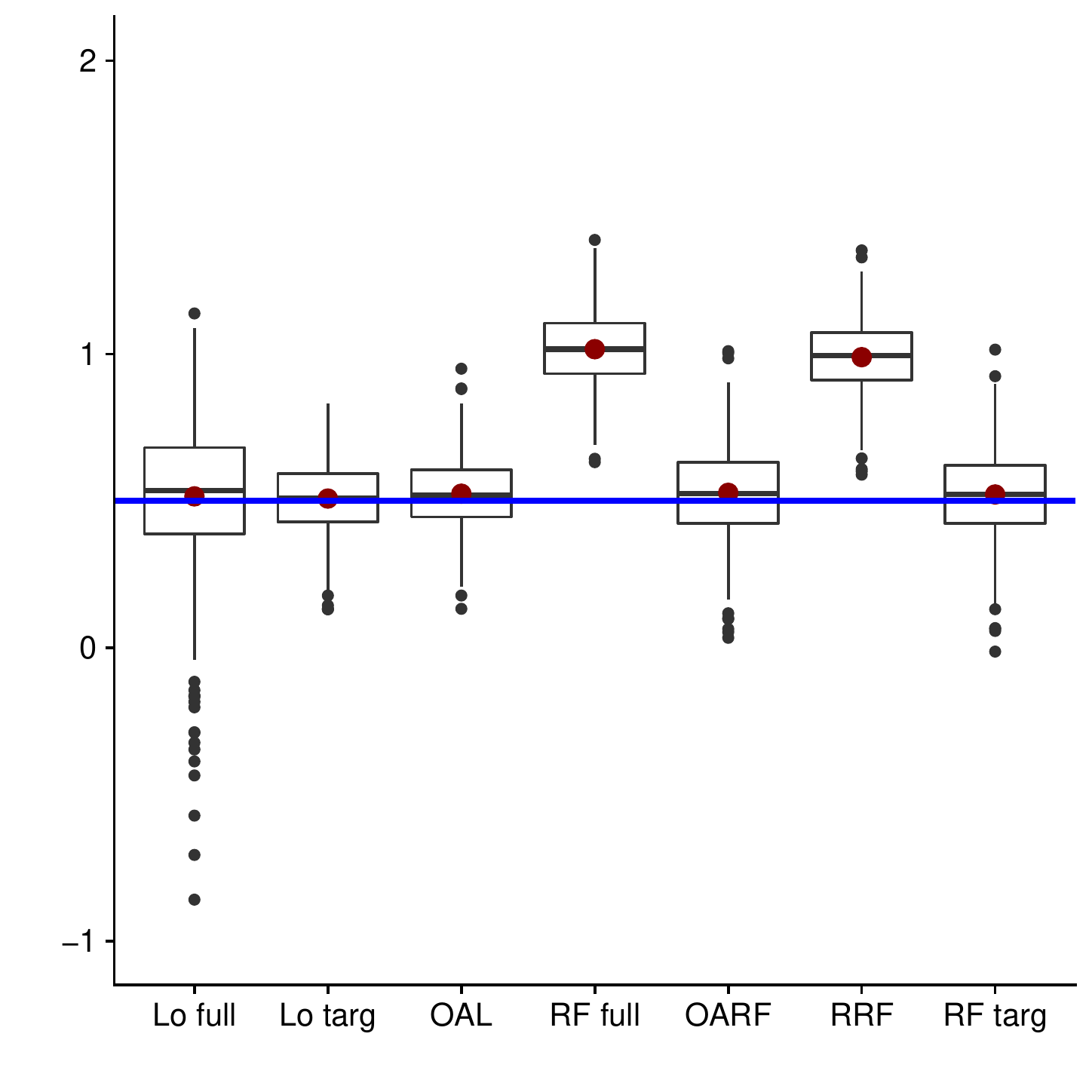}

\captionof{figure}{Linear Setting 1}
\end{subfigure}
\begin{subfigure}[b]{0.5\linewidth}
\centering
\includegraphics[width=0.8\textwidth]{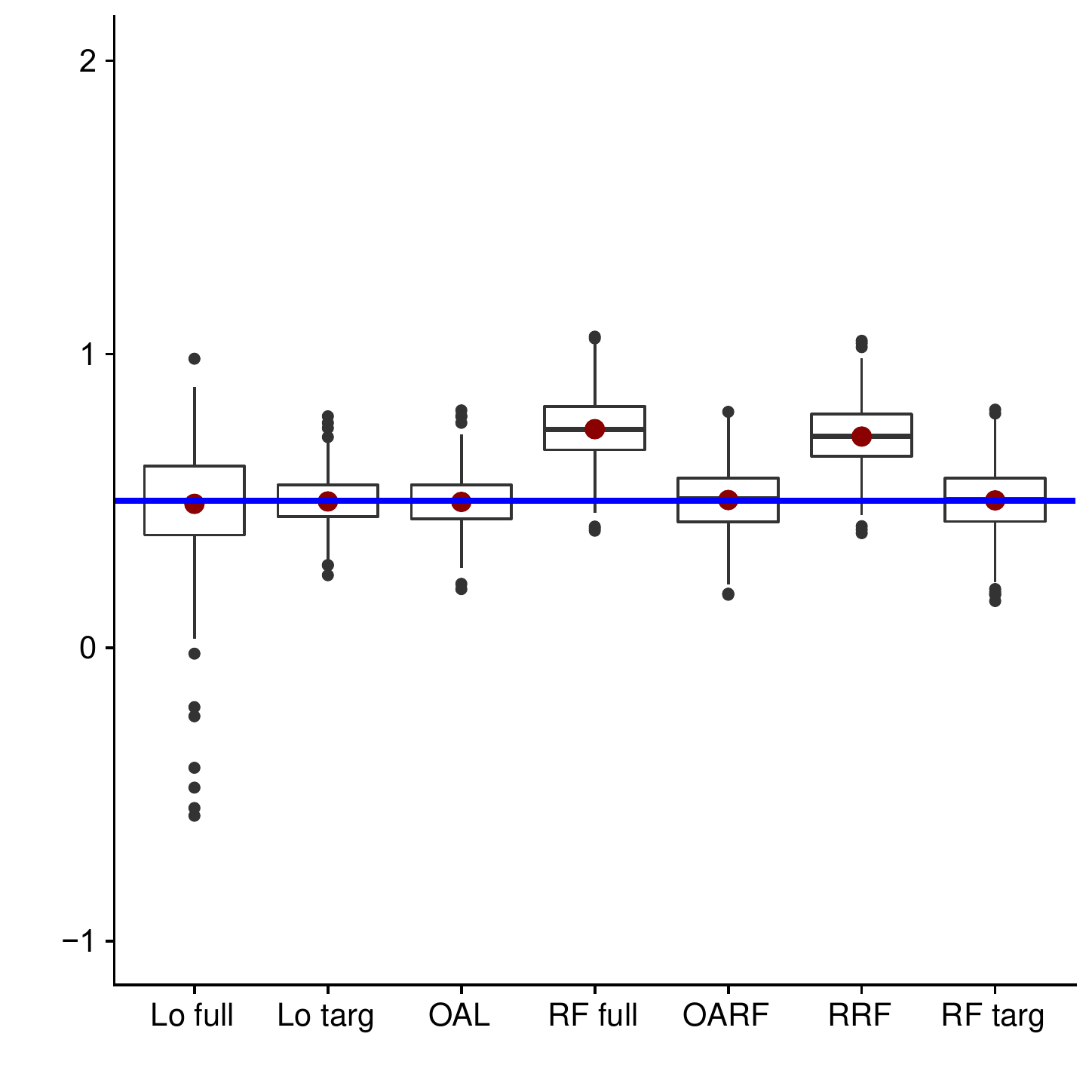}

\captionof{figure}{Linear Setting 2}
\end{subfigure}
\caption{Illustrations in linear settings. ATE via IPTW.}
\label{fig:boxplots_lin_OAL}
\end{figure}

\begin{figure}[ht]
\begin{subfigure}[b]{0.33\linewidth}
\centering
\includegraphics[width=1\textwidth]{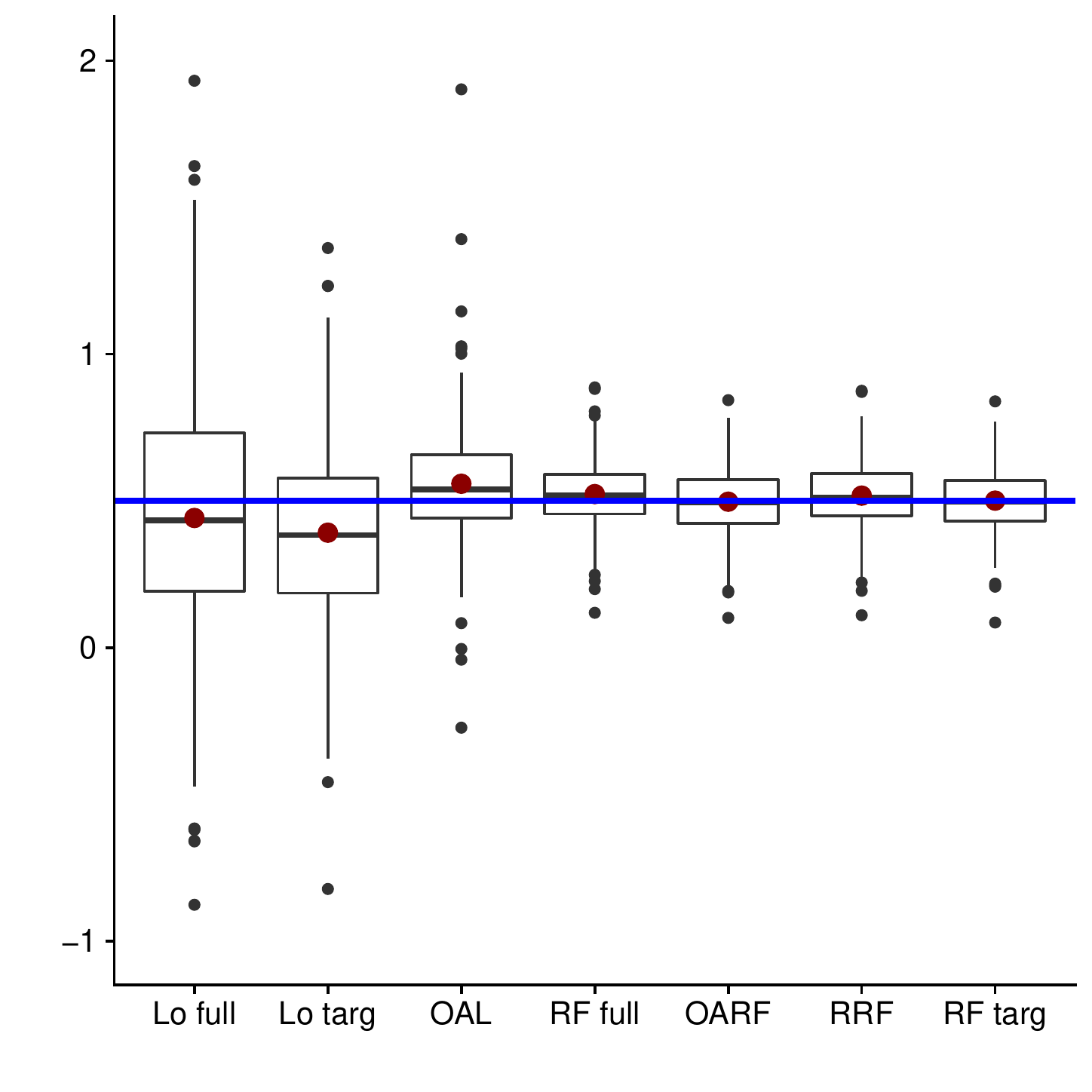}
\captionof{figure}{Setting 8}
\end{subfigure}%
\begin{subfigure}[b]{0.33\linewidth}
\centering
\includegraphics[width=1\textwidth]{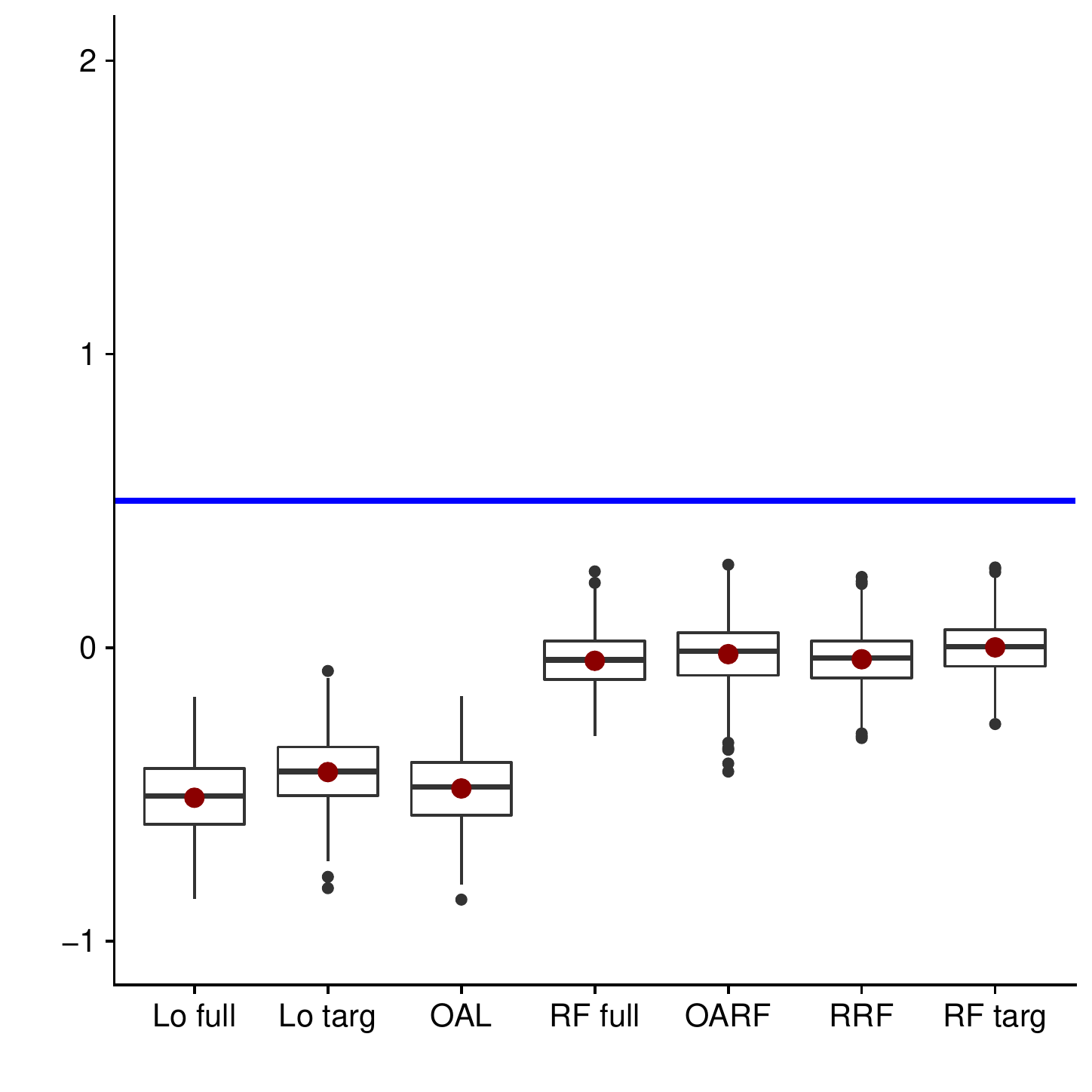}
\captionof{figure}{Setting 9}
\end{subfigure}%
\begin{subfigure}[b]{0.33\linewidth}
\centering
\includegraphics[width=1\textwidth]{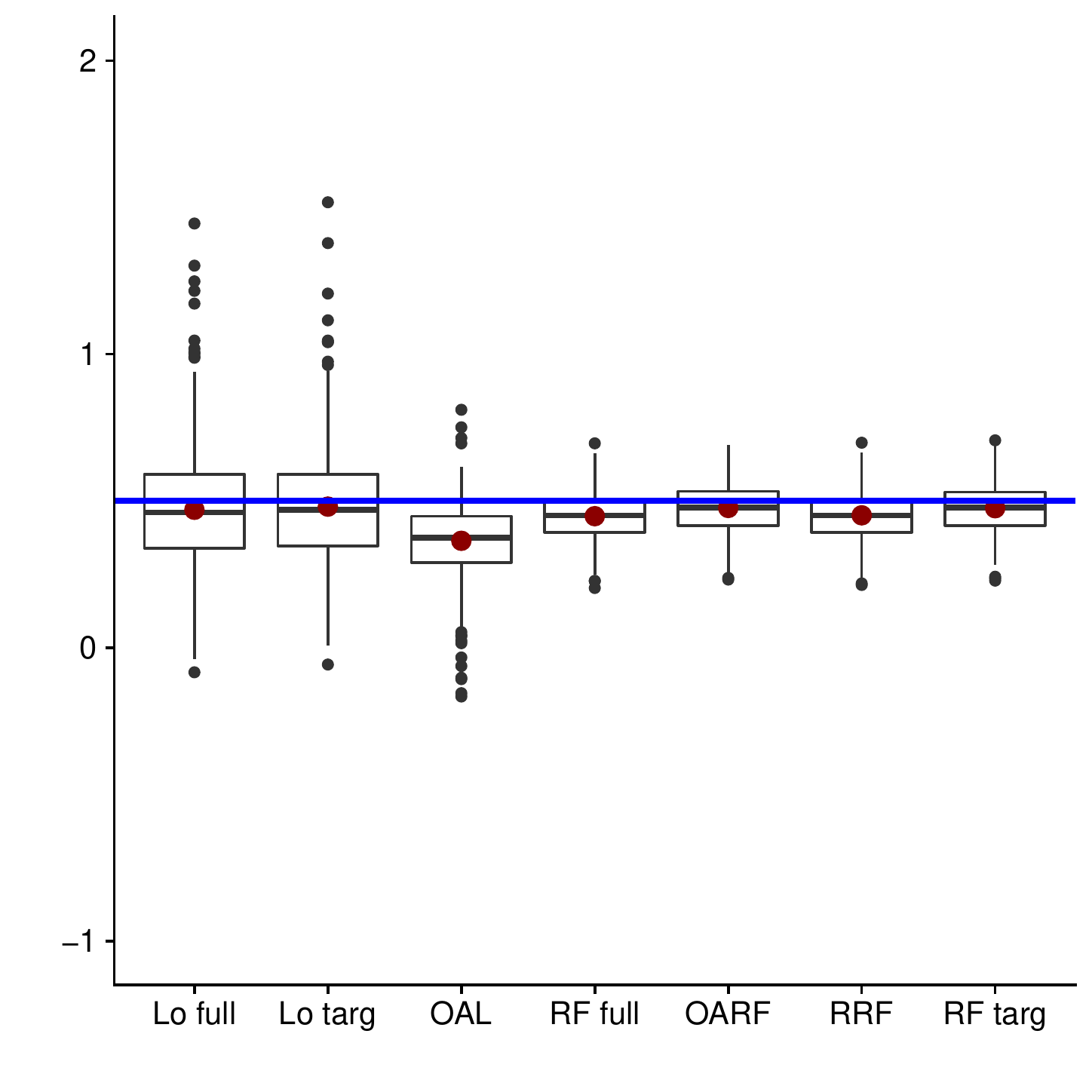}
\captionof{figure}{Setting 10}
\end{subfigure}
\caption{Illustrations with more dependent covariates.}
\label{fig:boxplots_moreXc}
\end{figure}

\begin{figure}[ht]
\begin{subfigure}[b]{0.5\linewidth}
\centering
\includegraphics[width=0.8\textwidth]{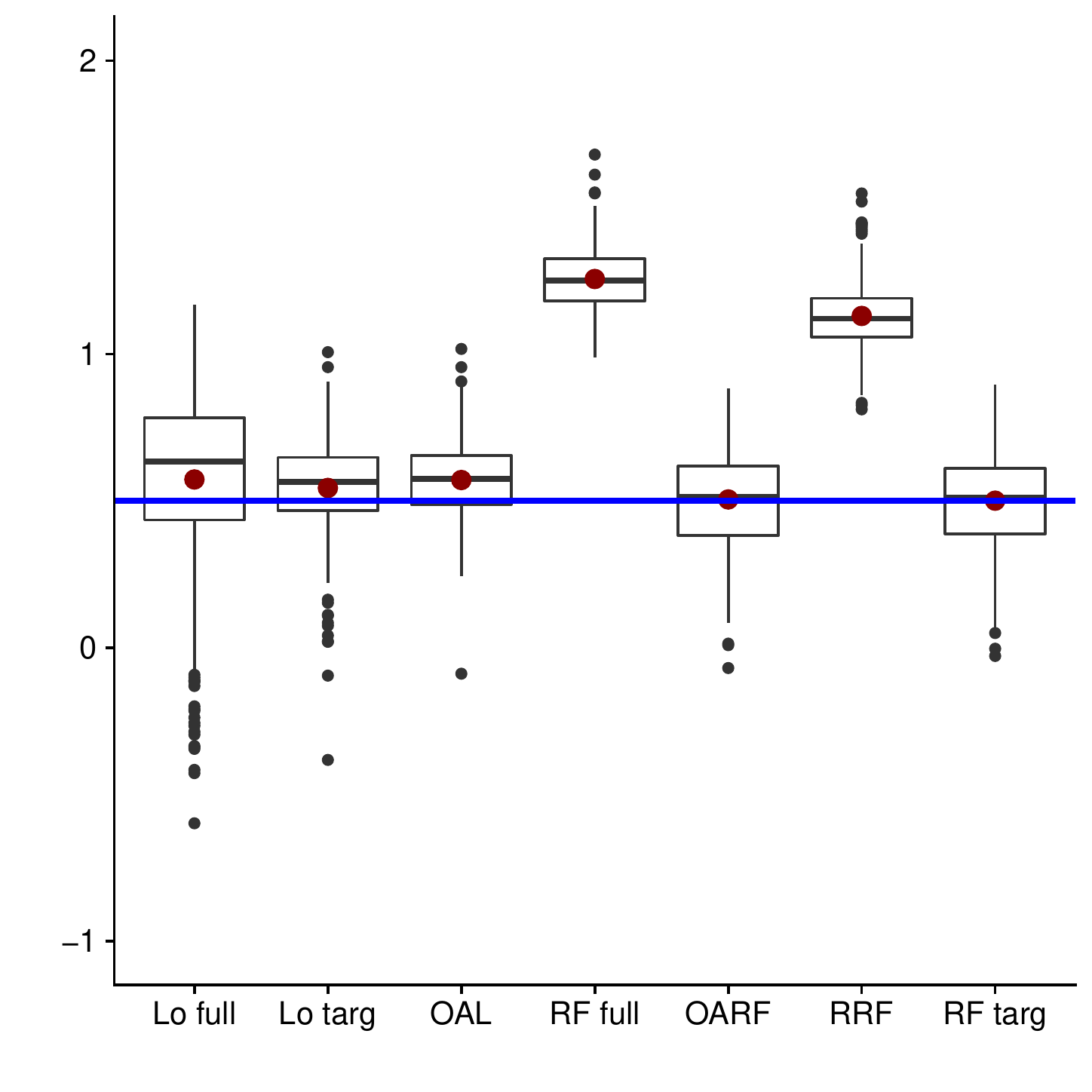}
\captionof{figure}{Setting 1 with correlation $\rho = 0.2$}
\end{subfigure}%
\begin{subfigure}[b]{0.5\linewidth}
\centering
\includegraphics[width=0.8\textwidth]{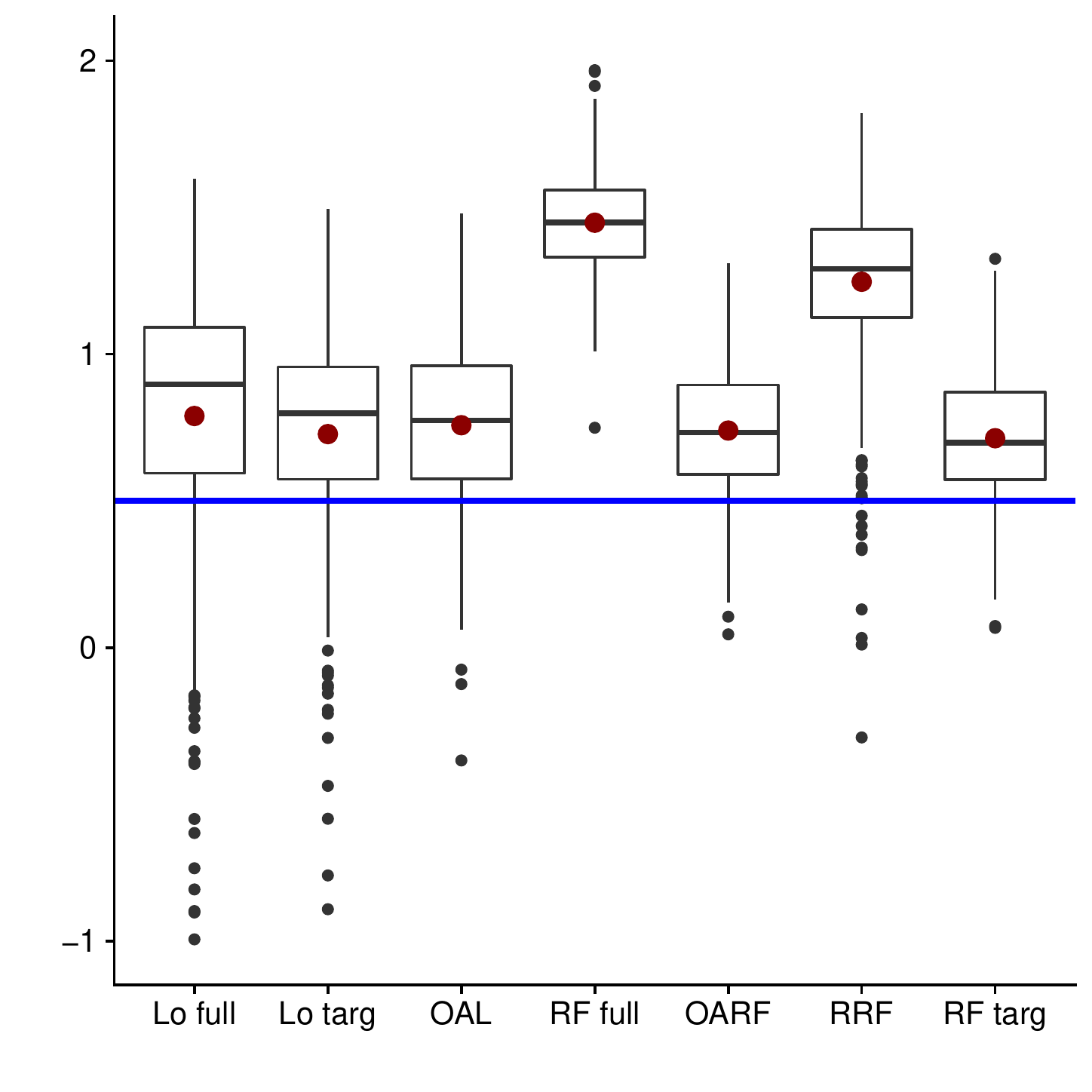}
\captionof{figure}{Setting 1 with correlation $\rho = 0.5$}
\end{subfigure}
\caption{Illustrations with positive correlation between the covariates.}
\label{fig:boxplots_correlations}
\end{figure}

\begin{figure}[ht]
\begin{subfigure}[b]{0.33\linewidth}
\centering
\includegraphics[width=1\textwidth]{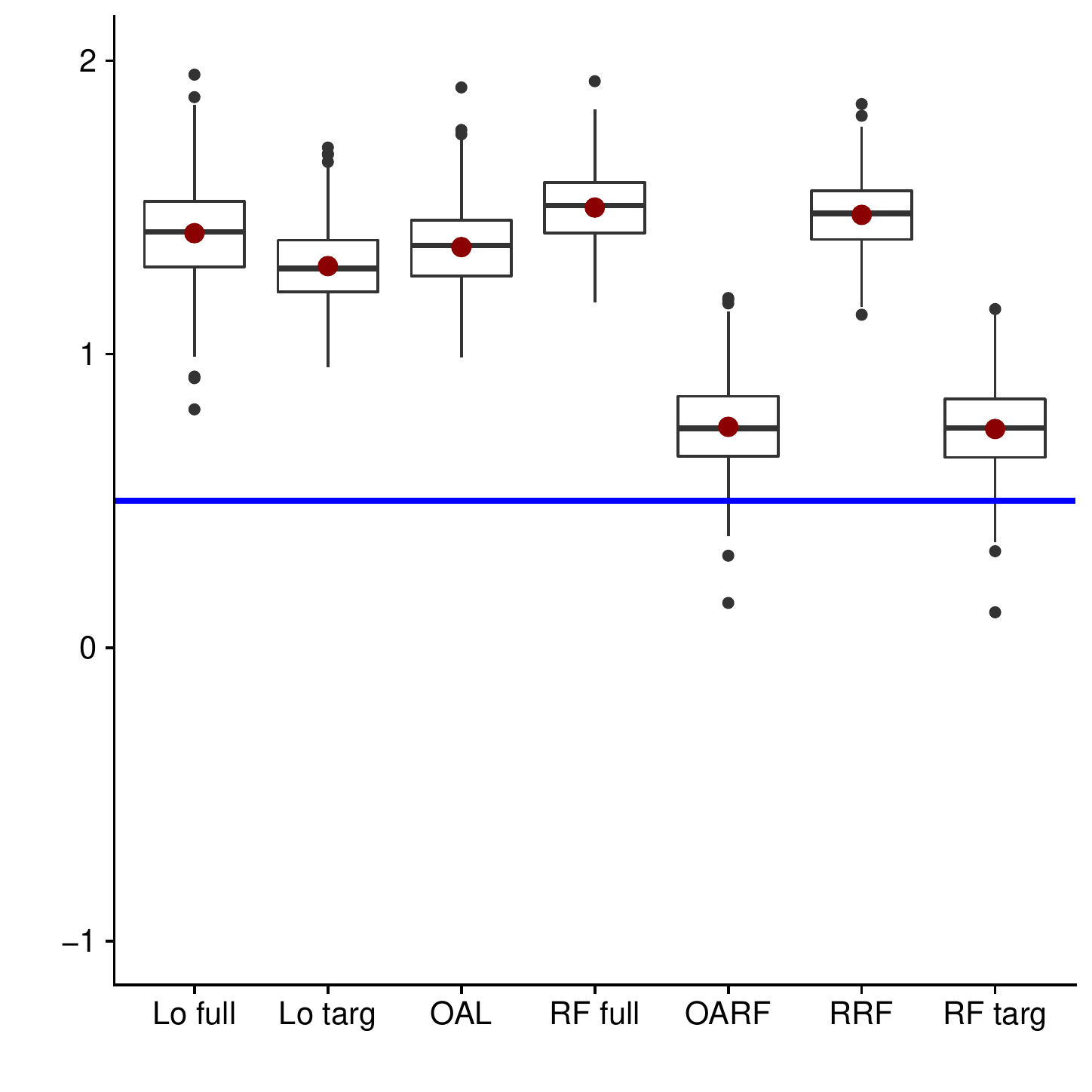}
\captionof{figure}{Setting 13}
\end{subfigure}%
\begin{subfigure}[b]{0.33\linewidth}
\centering
\includegraphics[width=1\textwidth]{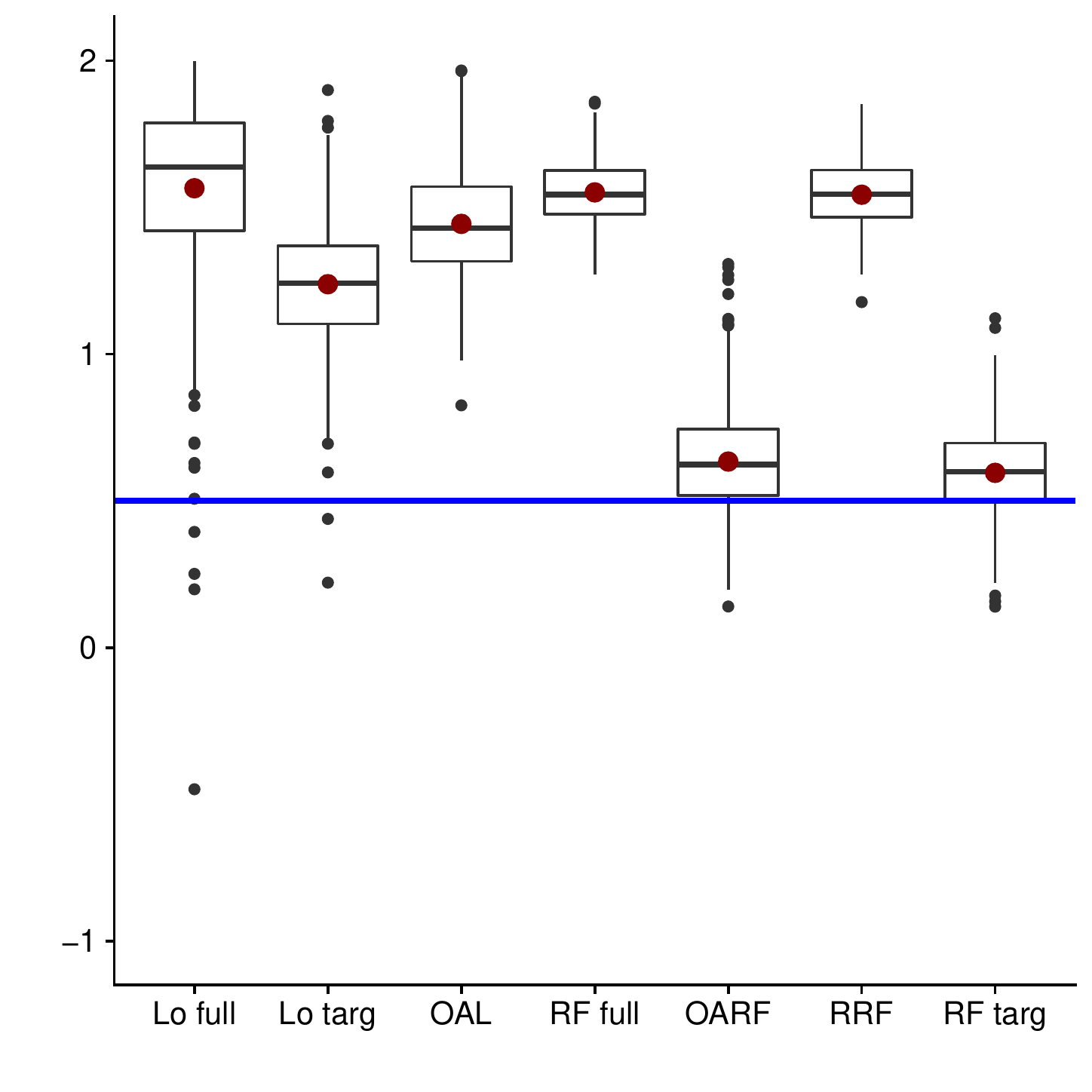}
\captionof{figure}{Setting 14}
\end{subfigure}%
\begin{subfigure}[b]{0.33\linewidth}
\centering
\includegraphics[width=1\textwidth]{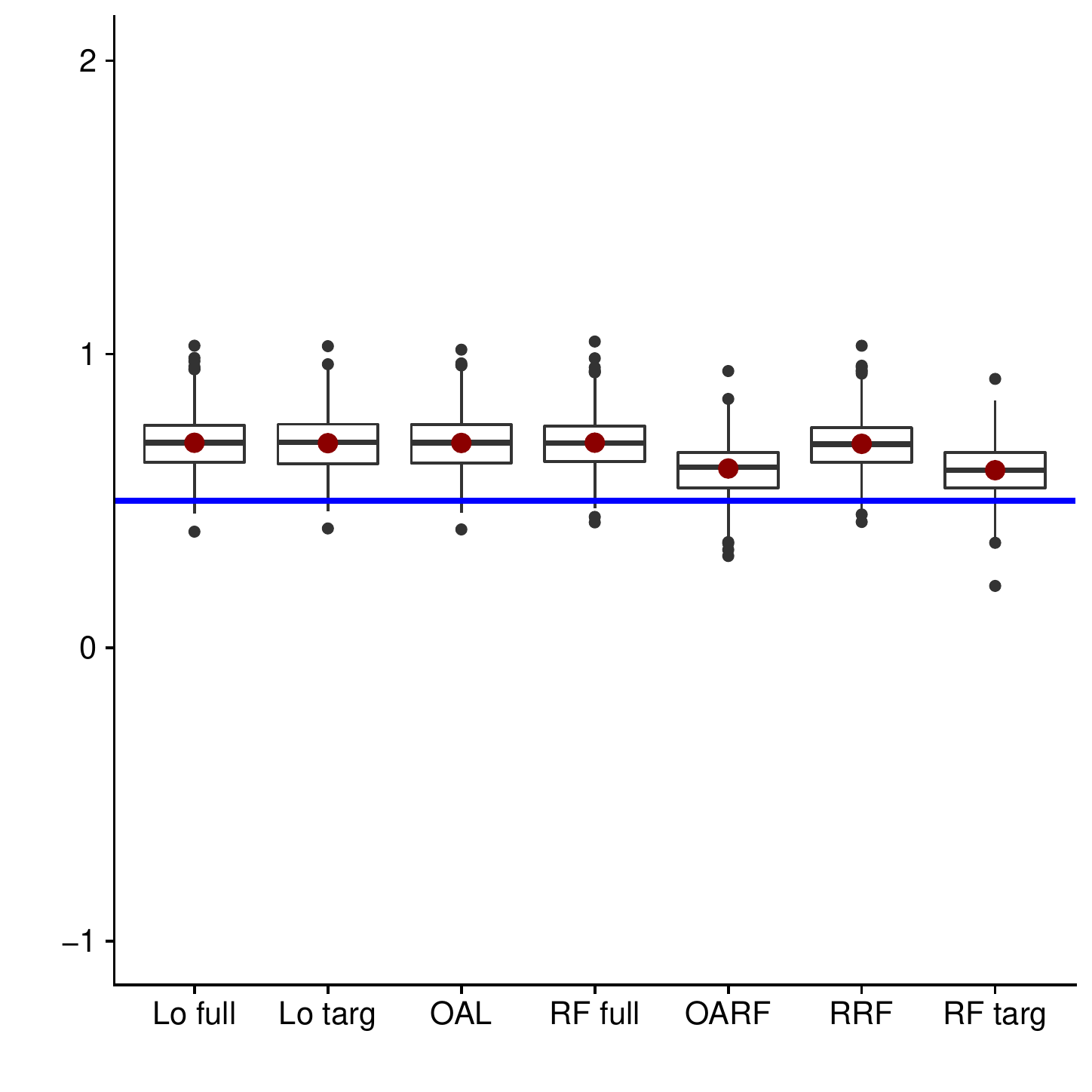}
\captionof{figure}{Setting 15}
\end{subfigure}
\caption{Illustrations with complex outcome function.}
\label{fig:boxplots_complex_Y}
\end{figure}

\begin{figure}[ht]
\begin{subfigure}[b]{0.33\linewidth}
\centering
\includegraphics[width=1\textwidth]{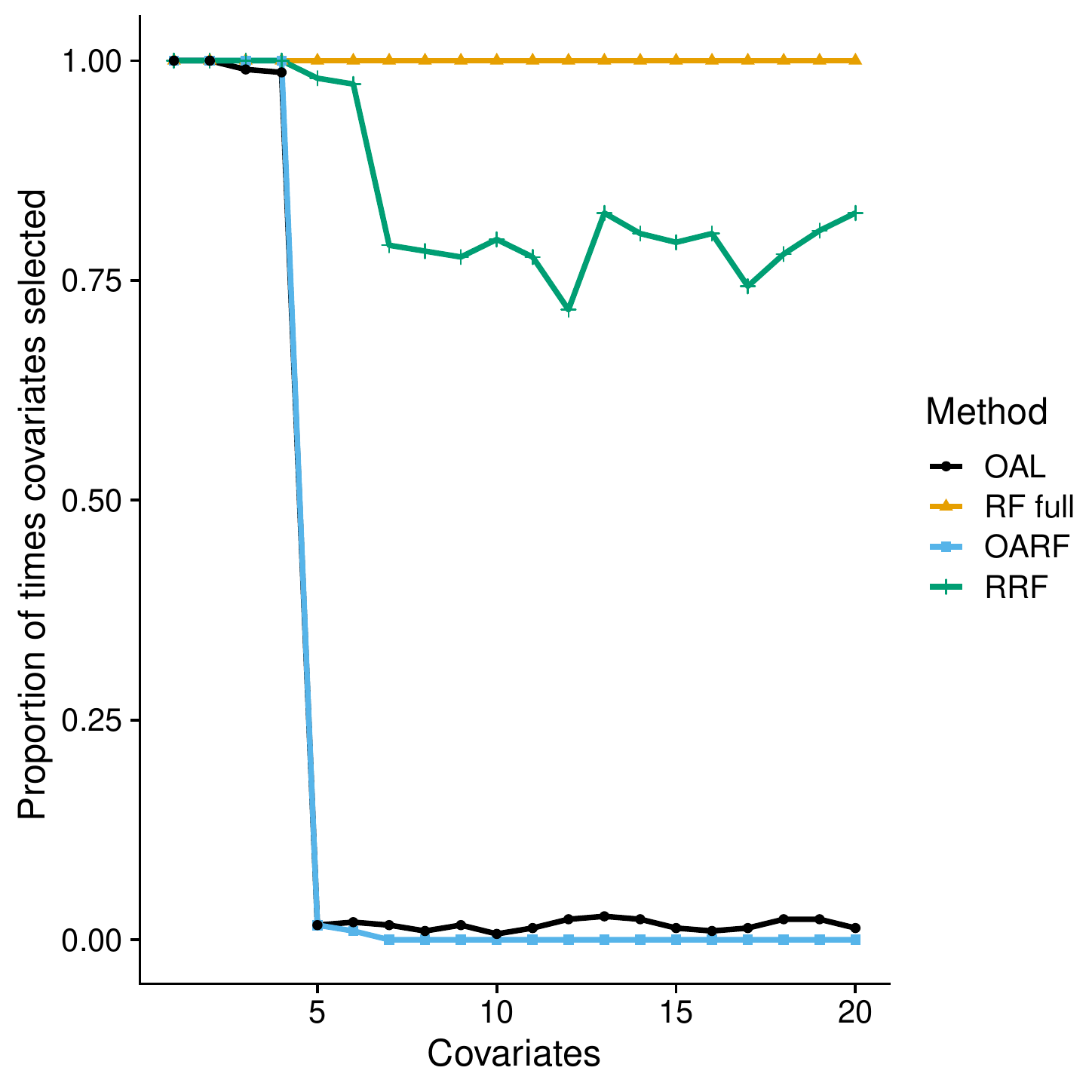}
\captionof{figure}{Setting 1}
\end{subfigure}%
\begin{subfigure}[b]{0.33\linewidth}
\centering
\includegraphics[width=1\textwidth]{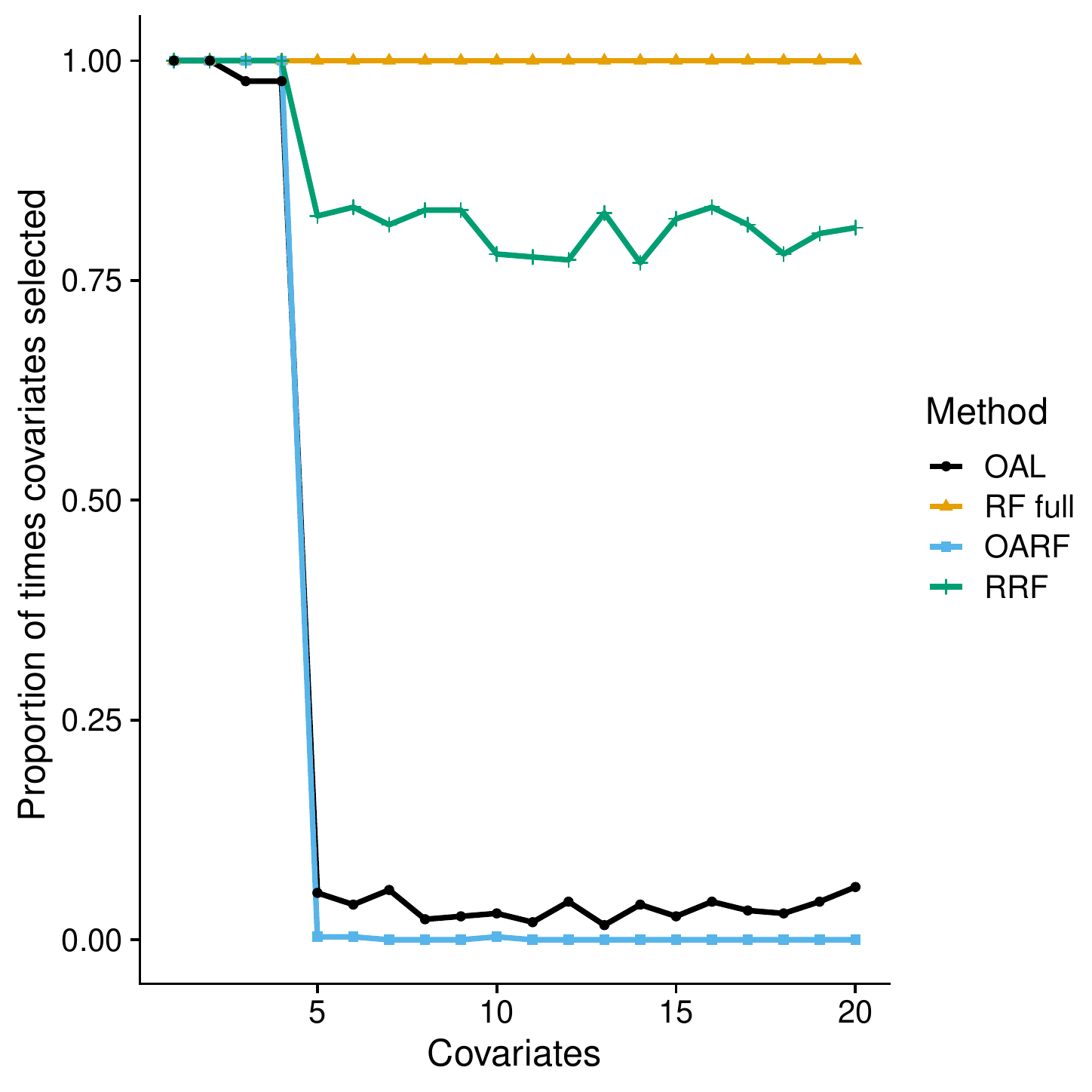}
\captionof{figure}{Setting 2}
\end{subfigure}%
\begin{subfigure}[b]{0.33\linewidth}
\centering
\includegraphics[width=1\textwidth]{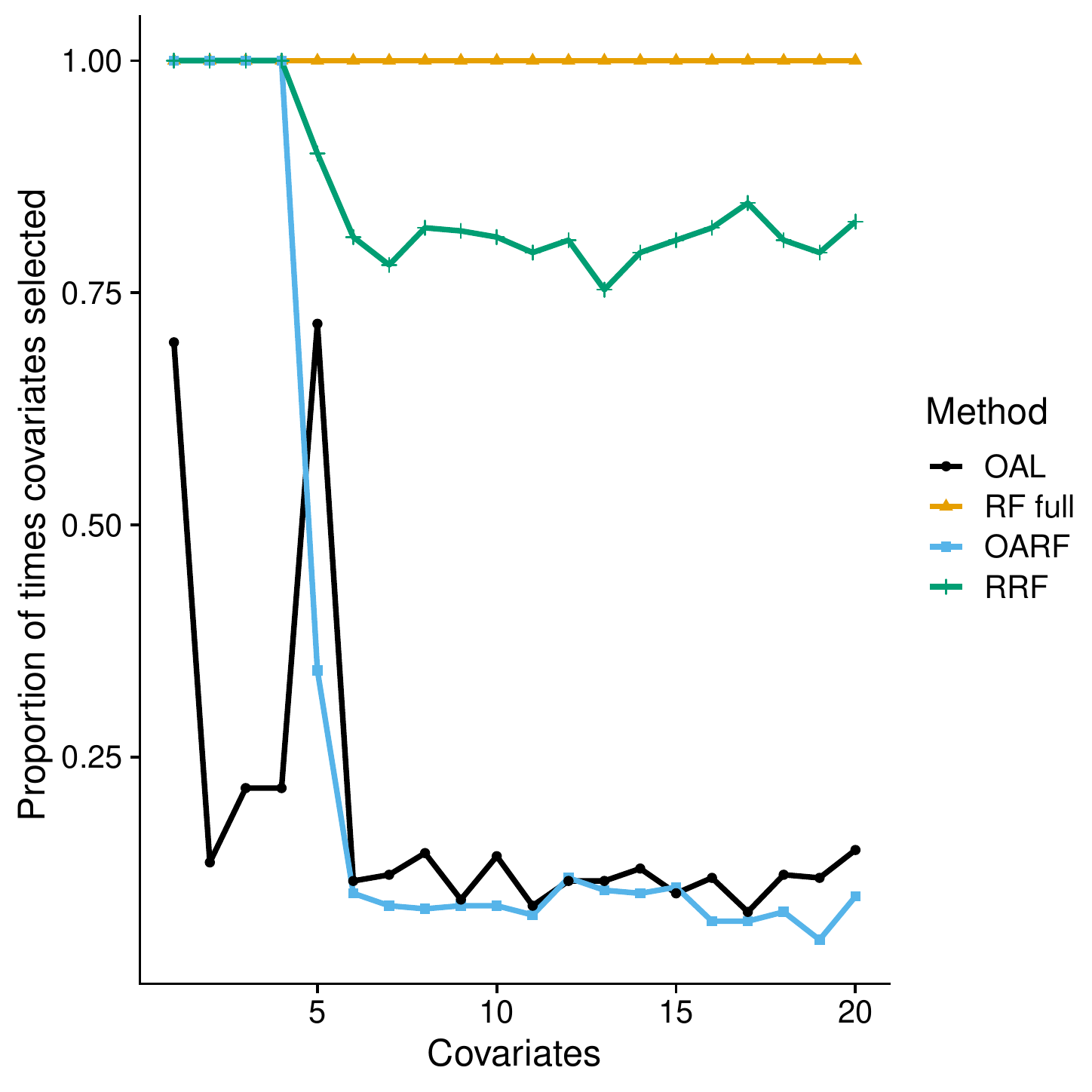}
\captionof{figure}{Setting 4}
\end{subfigure}

\begin{subfigure}[b]{0.33\linewidth}
\centering
\includegraphics[width=1\textwidth]{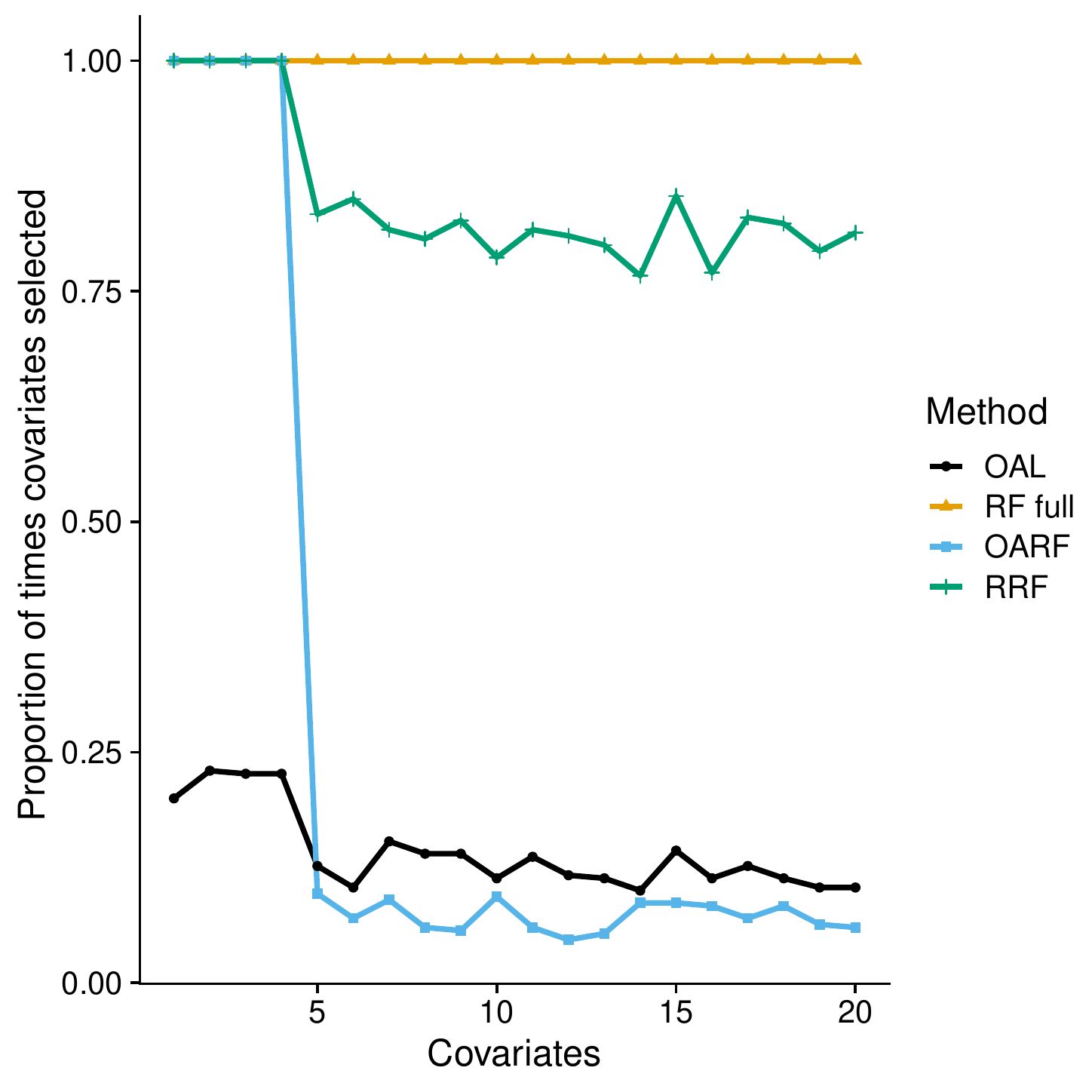}
\captionof{figure}{Setting 5}
\end{subfigure}
\begin{subfigure}[b]{0.33\linewidth}
\centering
\includegraphics[width=1\textwidth]{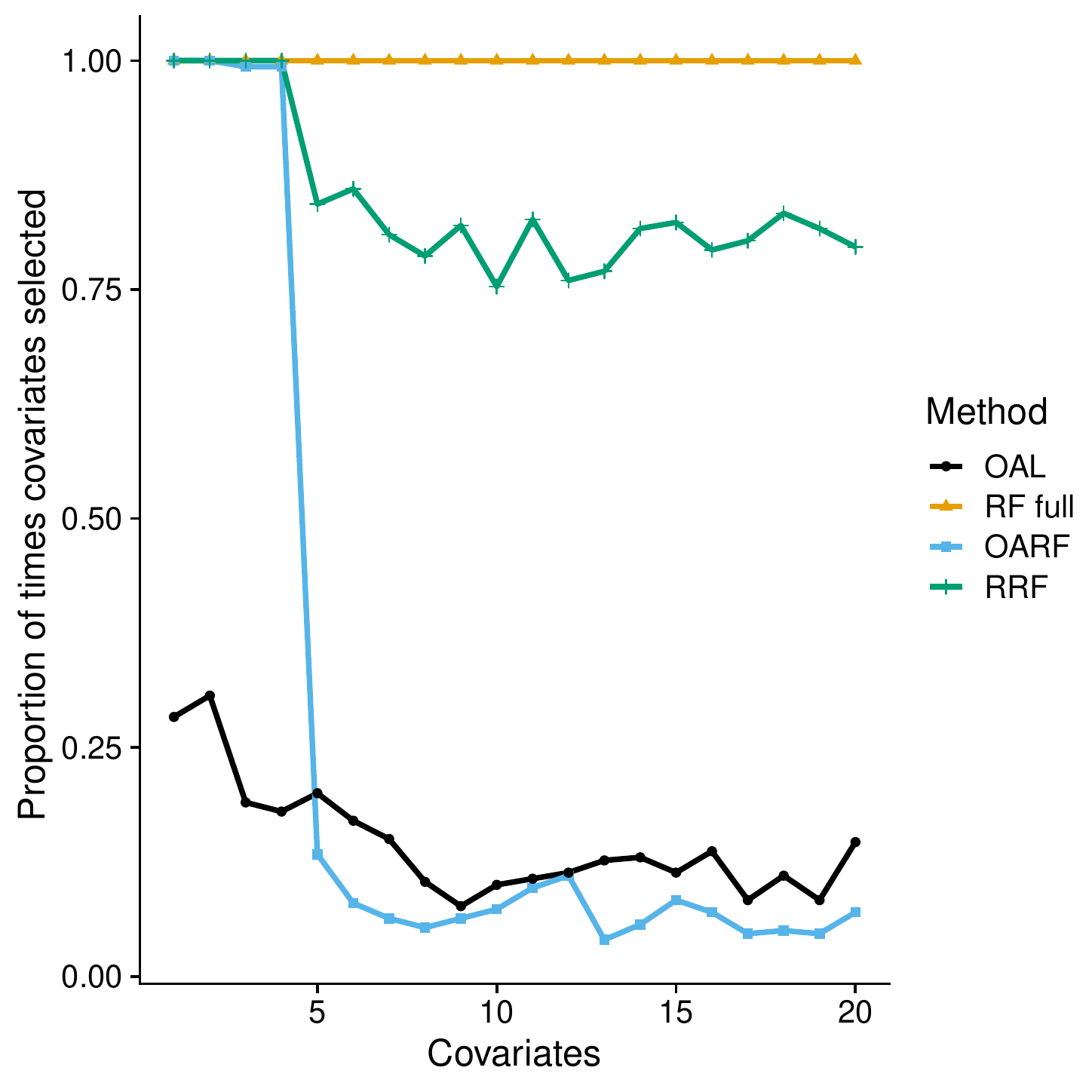}
\captionof{figure}{Setting 6}
\end{subfigure}%
\begin{subfigure}[b]{0.33\linewidth}
\centering
\includegraphics[width=1\textwidth]{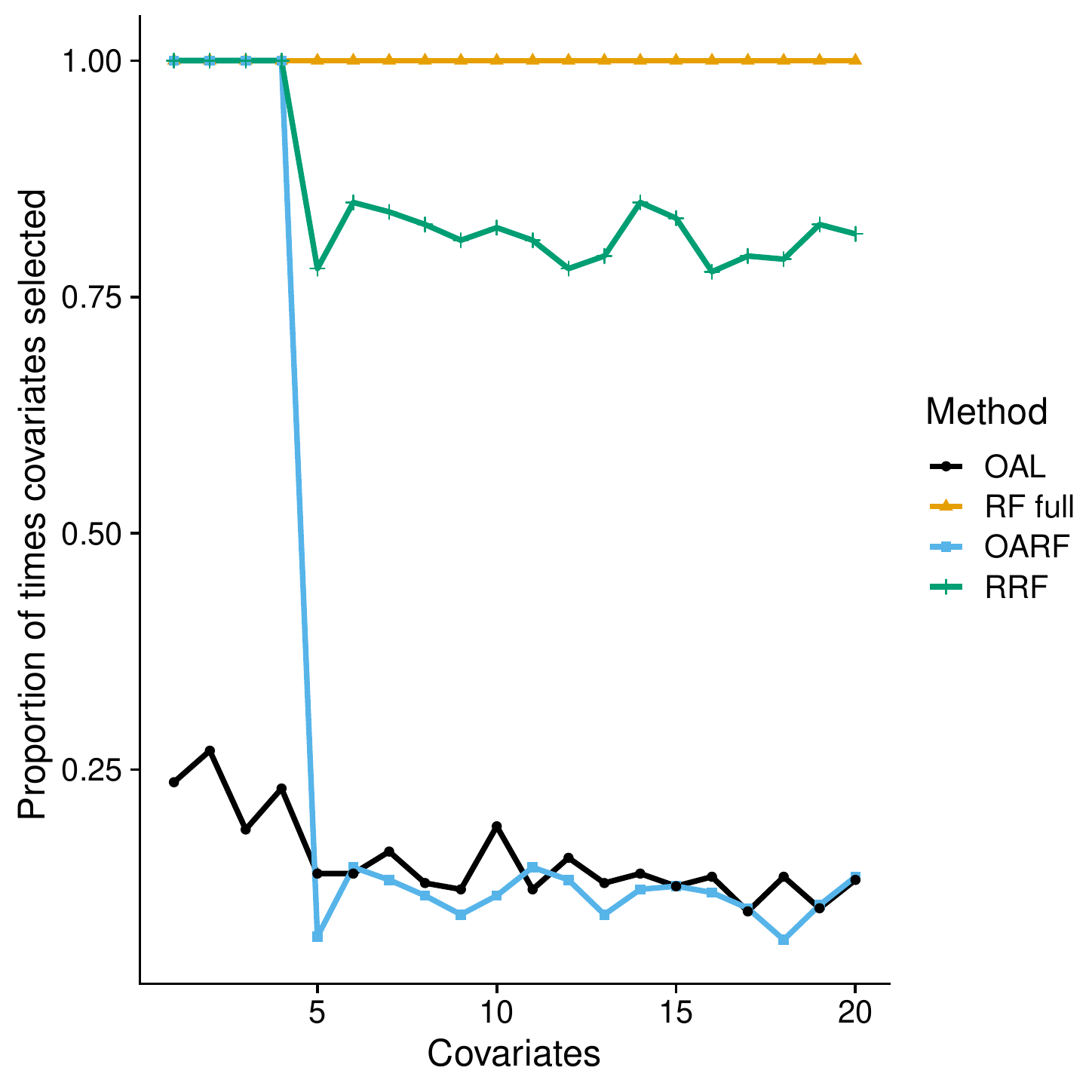}
\captionof{figure}{Setting 7}
\end{subfigure}

\caption{Illustrations of selected variables over 500 simulations.}
\label{fig:selected_Var_lin_non_lin}
\end{figure}

\begin{figure}[ht]
\begin{subfigure}[b]{0.33\linewidth}
\centering
\includegraphics[width=1\textwidth]{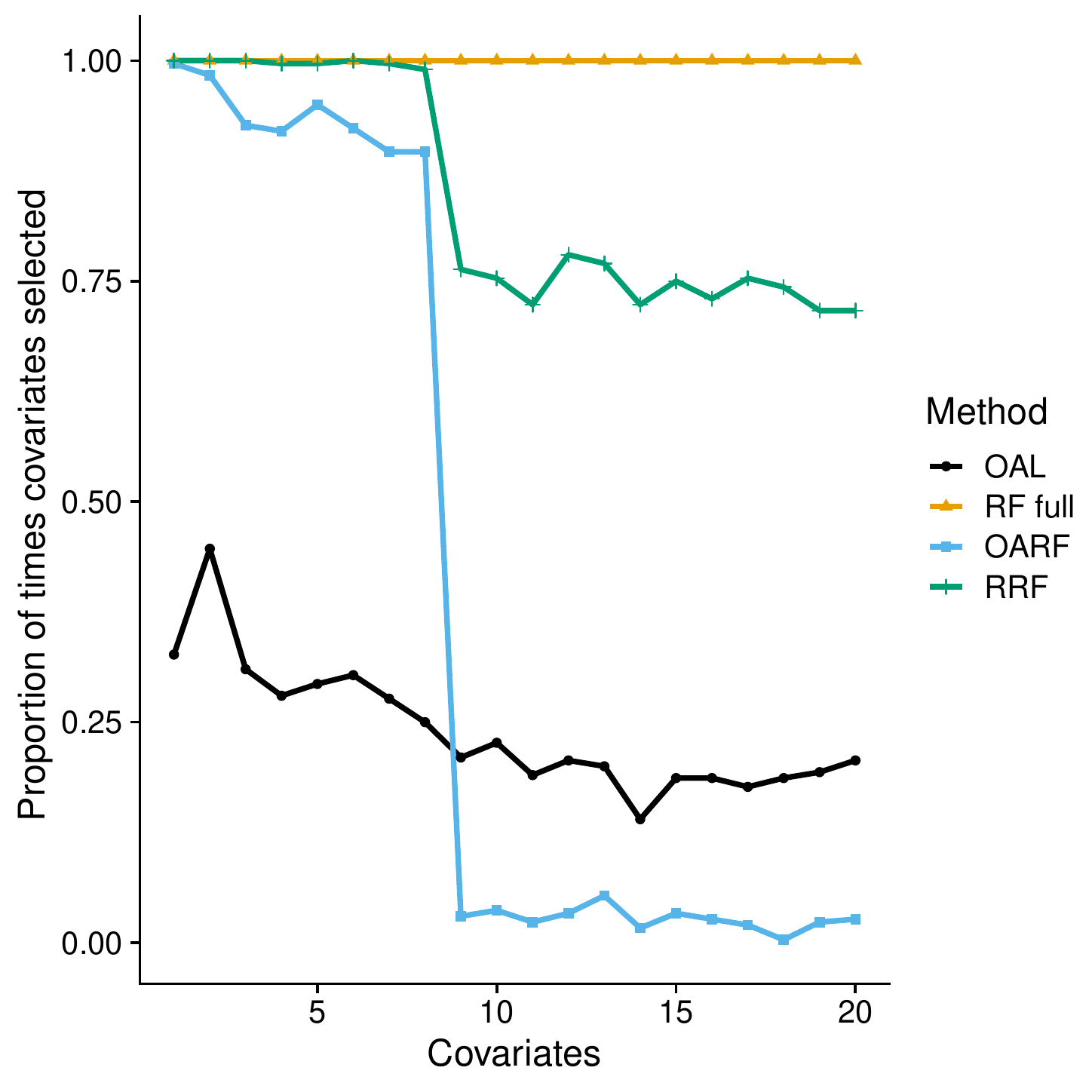}
\captionof{figure}{Setting 8}
\end{subfigure}%
\begin{subfigure}[b]{0.33\linewidth}
\centering
\includegraphics[width=1\textwidth]{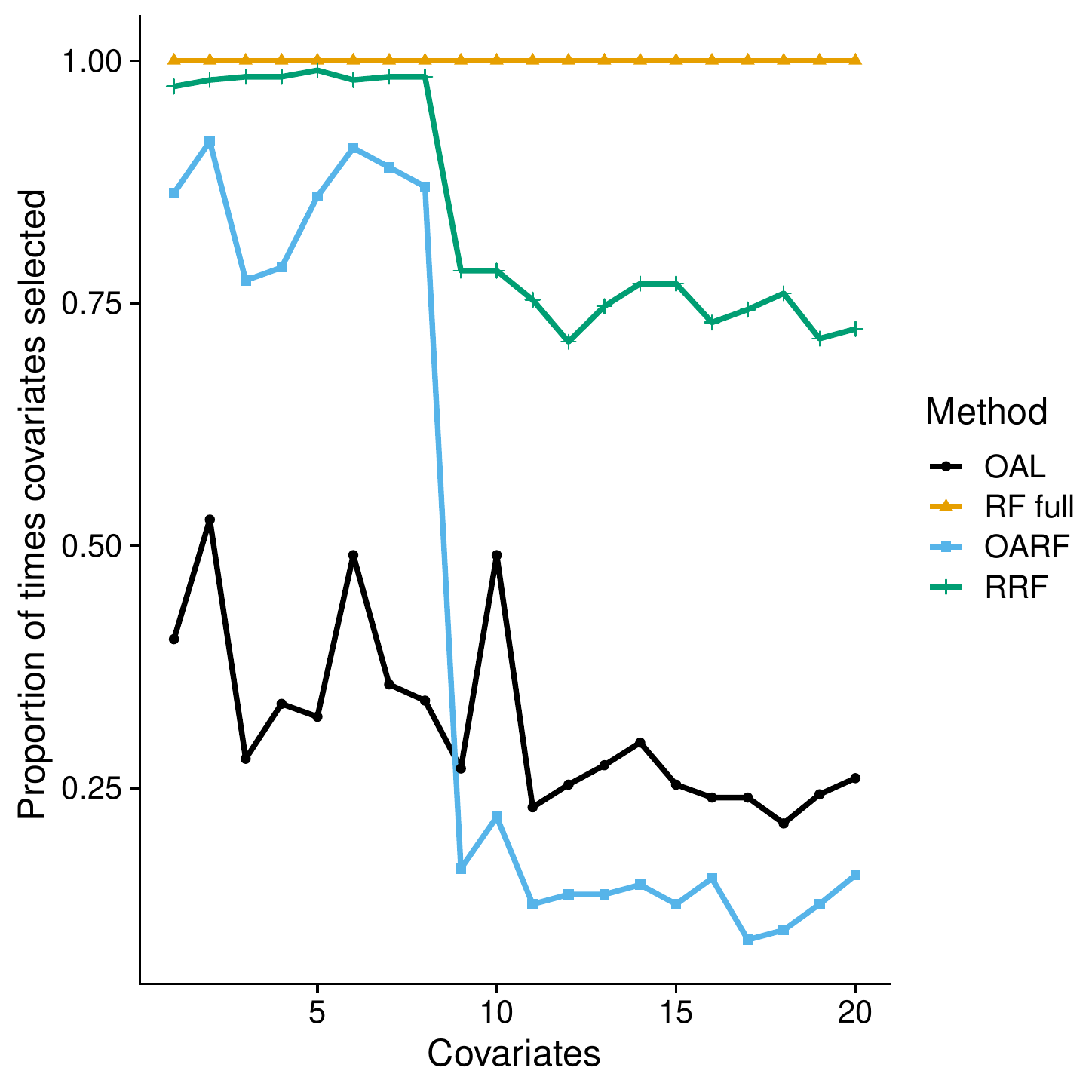}
\captionof{figure}{Setting 9}
\end{subfigure}%
\begin{subfigure}[b]{0.33\linewidth}
\centering
\includegraphics[width=1\textwidth]{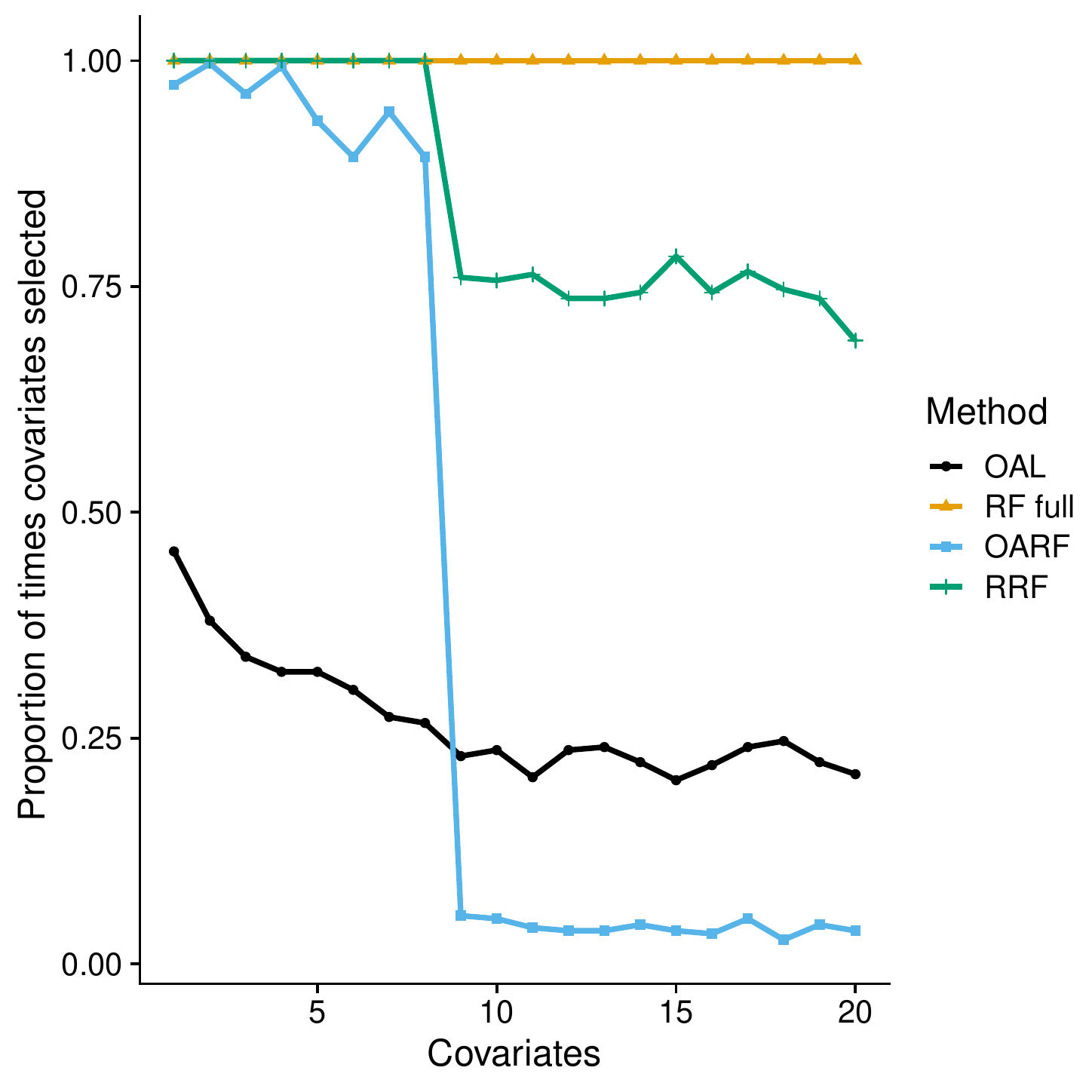}
\captionof{figure}{Setting 10}
\end{subfigure}
\caption{Illustrations of selected variables over 500 simulations. Favourable covariates are $X_1$ to $X_8$. }
\label{fig:selected_Var_moreXc}
\end{figure}

\begin{figure}[ht]
\begin{subfigure}[b]{0.5\linewidth}
\centering
\includegraphics[width=0.8\textwidth]{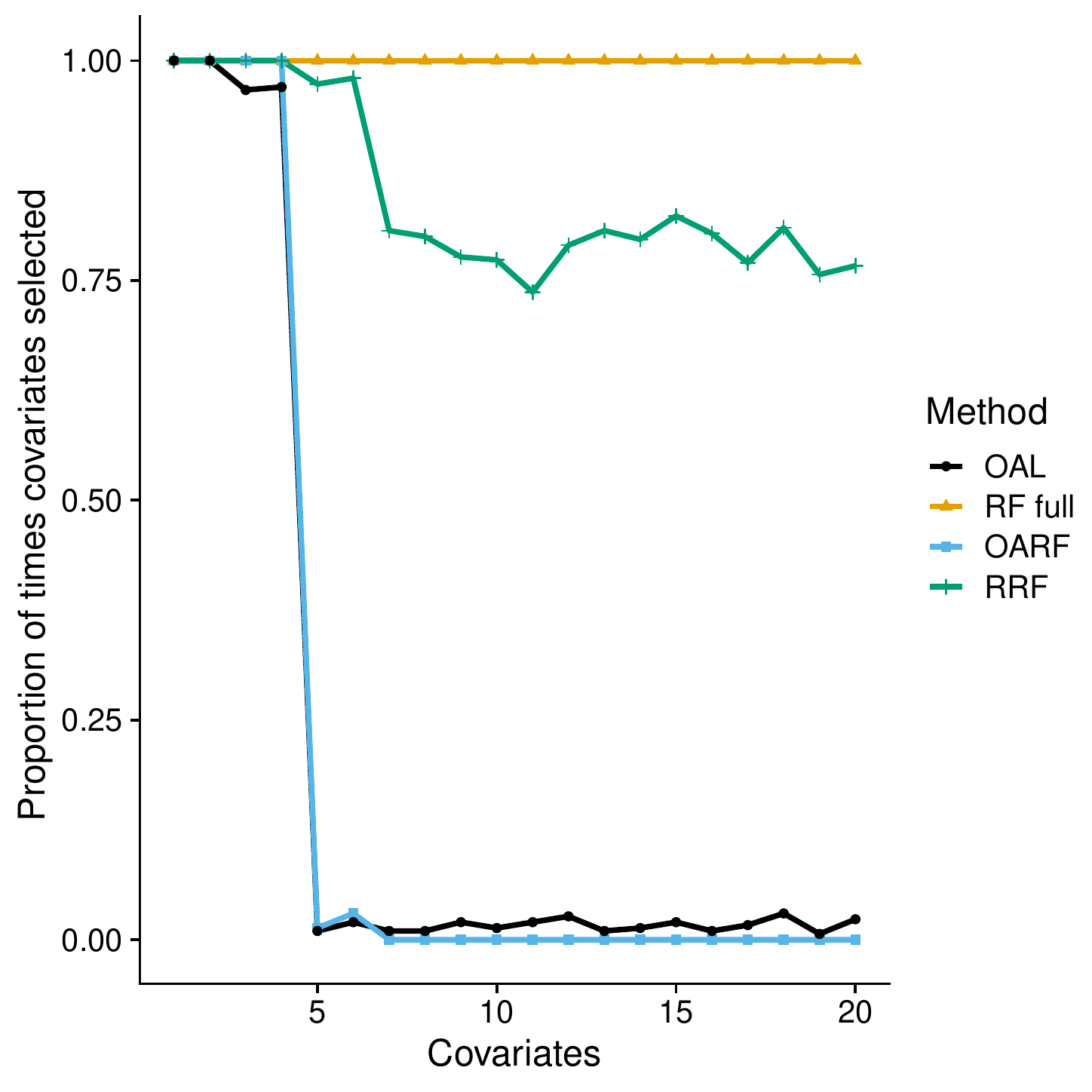}
\captionof{figure}{Setting 1 with correlation $\rho = 0.2$}
\end{subfigure}%
\begin{subfigure}[b]{0.5\linewidth}
\centering
\includegraphics[width=0.8\textwidth]{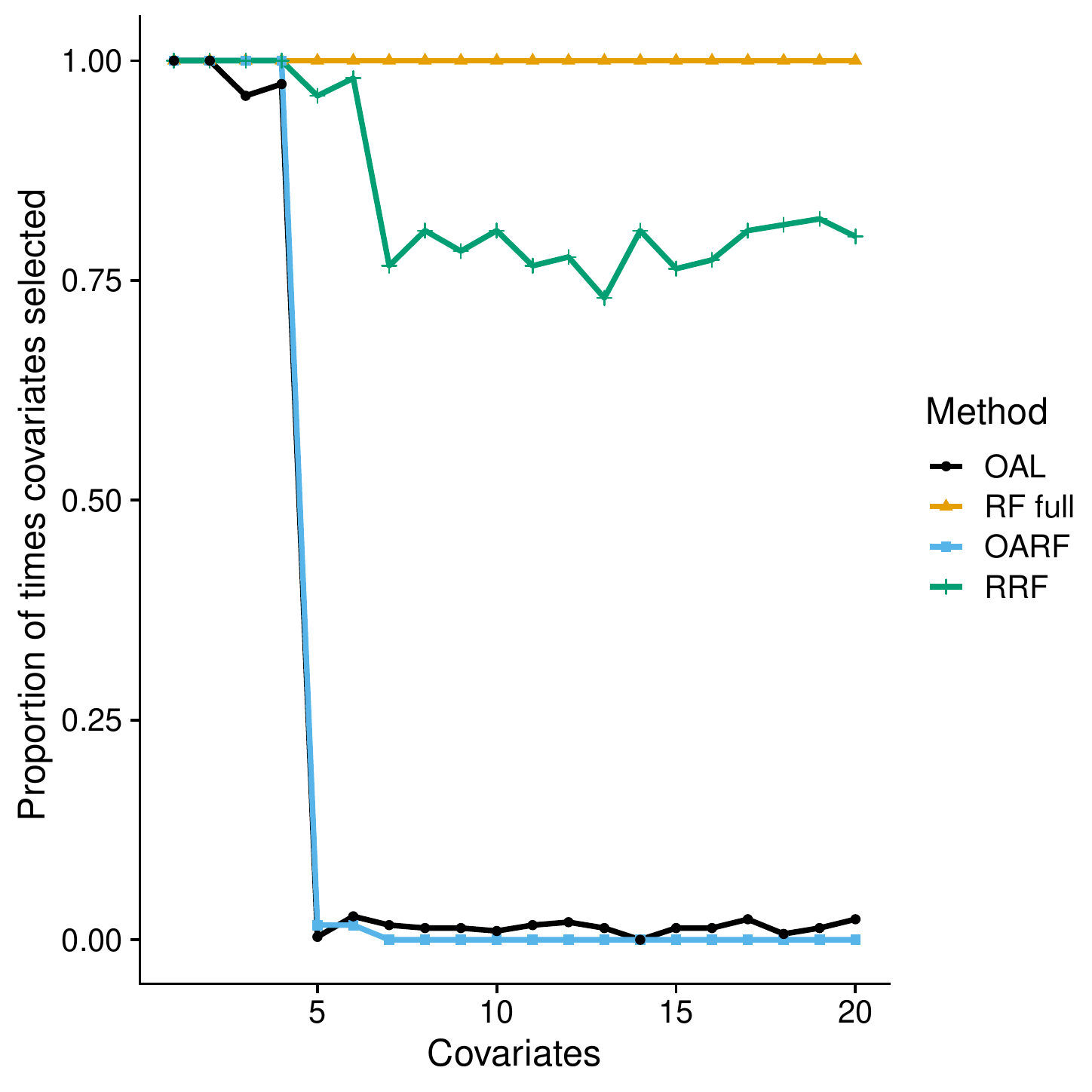}
\captionof{figure}{Setting 1 with correlation $\rho = 0.5$}
\end{subfigure}
\caption{Illustrations of selected variable coverage over 500 simulations and under positive correlation. Favourable covariates are $X_1$ to $X_4$.}
\label{fig:selected_Var_correlation}
\end{figure}

\begin{figure}[ht]
\begin{subfigure}[b]{0.33\linewidth}
\centering
\includegraphics[width=1\textwidth]{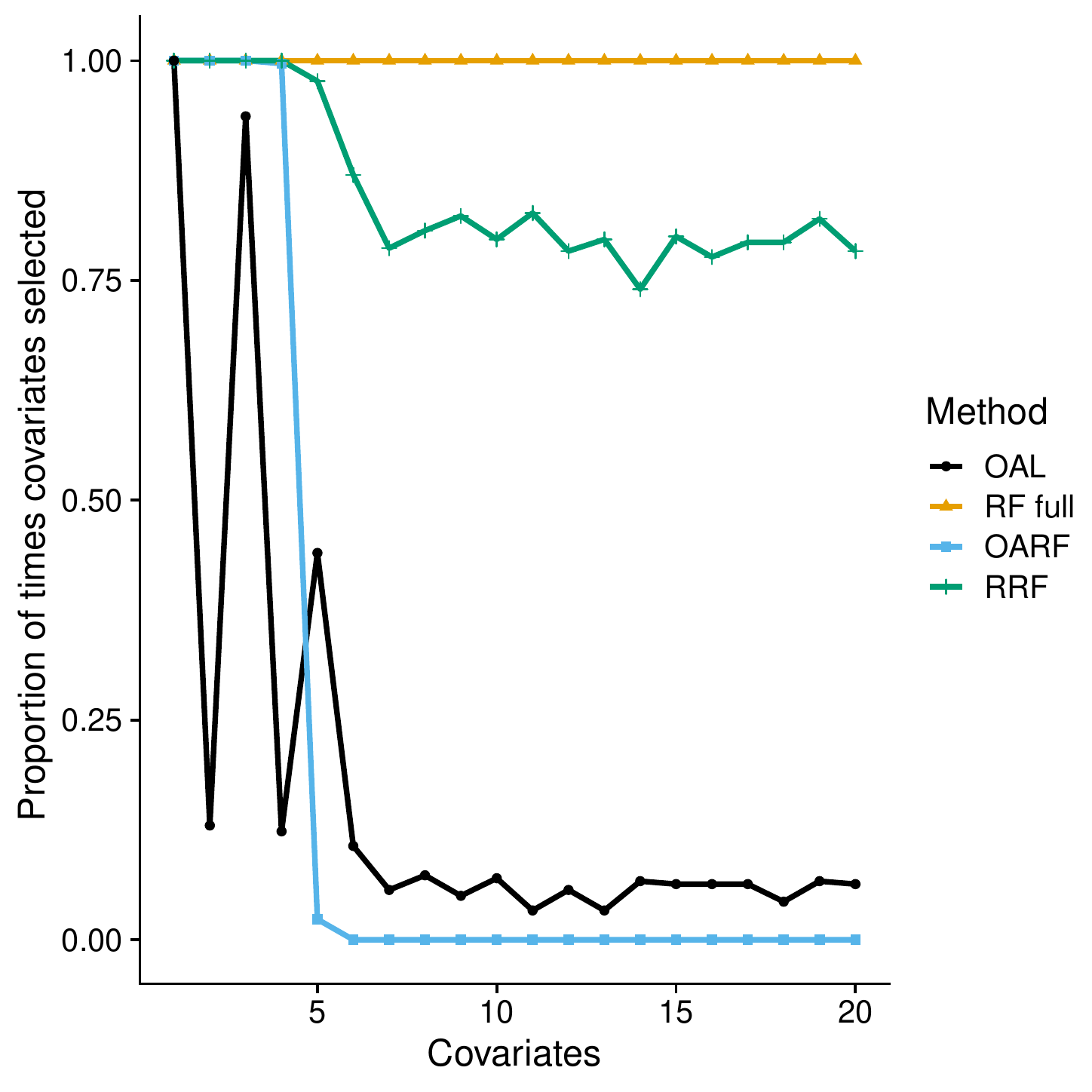}
\captionof{figure}{Setting 13}
\end{subfigure}%
\begin{subfigure}[b]{0.33\linewidth}
\centering
\includegraphics[width=1\textwidth]{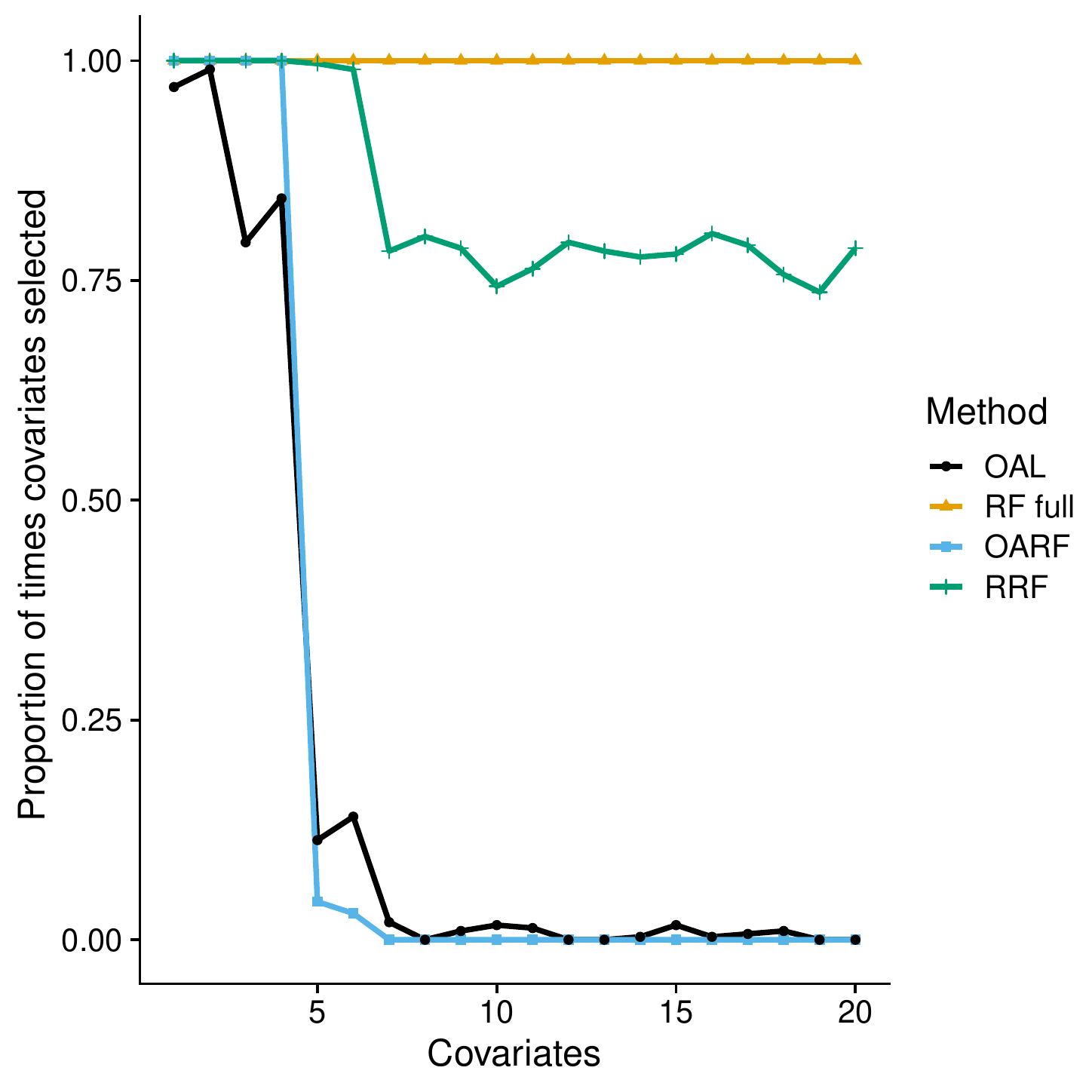}
\captionof{figure}{Setting 14}
\end{subfigure}%
\begin{subfigure}[b]{0.33\linewidth}
\centering
\includegraphics[width=1\textwidth]{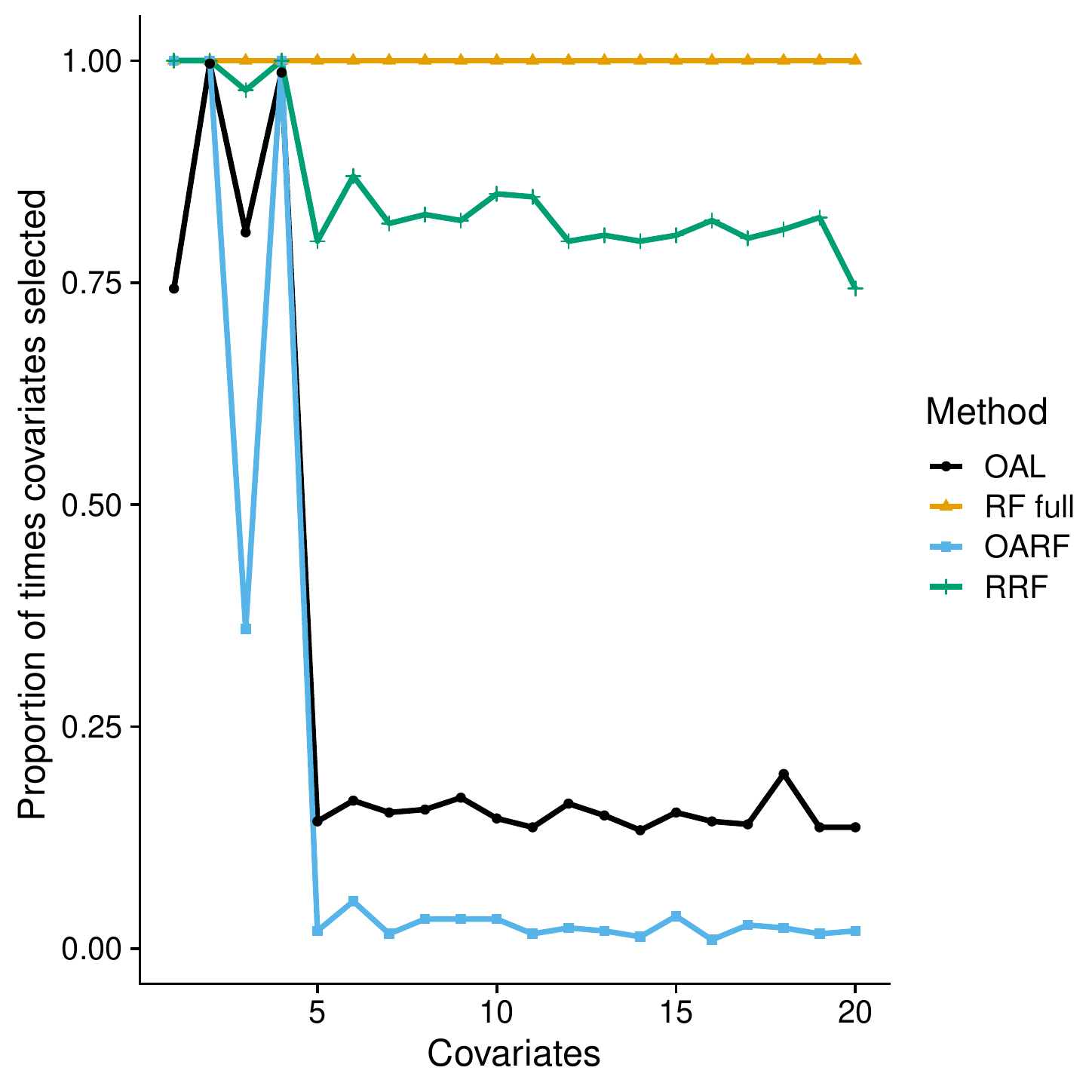}
\captionof{figure}{Setting 15}
\end{subfigure}
\caption{Illustrations of selected variable coverage over 500 simulations. Favourable covariates are $X_1$ to $X_4$. }
\label{fig:selected_Var_complex_Y}
\end{figure}

\begin{figure}[ht]
\begin{center}
\includegraphics[width=0.8\textwidth]{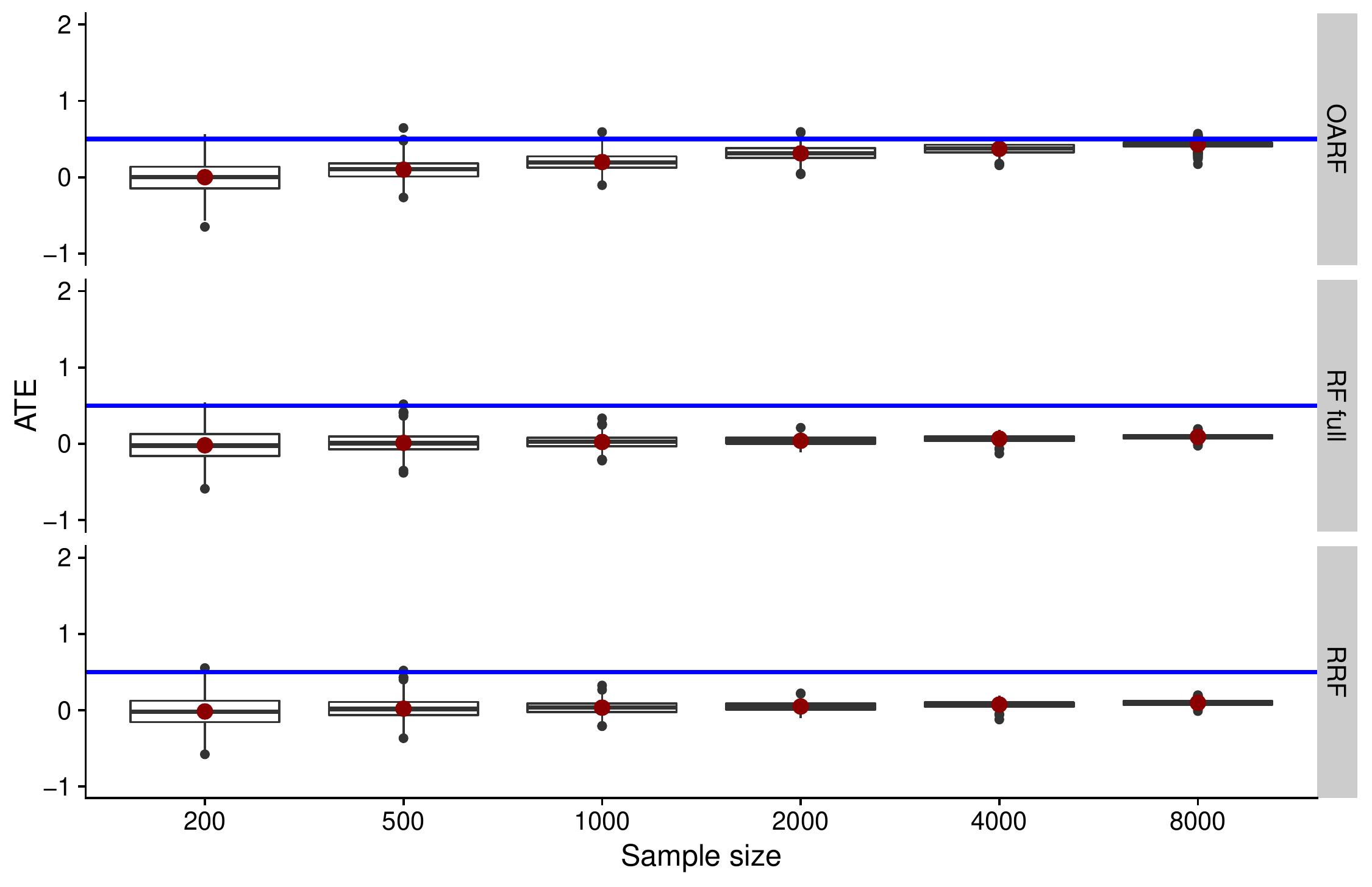}
\caption{Setting 4: Boxplots of ATE for varying sample size.}
\label{fig:boxplot_sam_size_S4}
\end{center}
\end{figure}

\begin{figure}[ht]
\begin{center}
\includegraphics[width=0.8\textwidth]{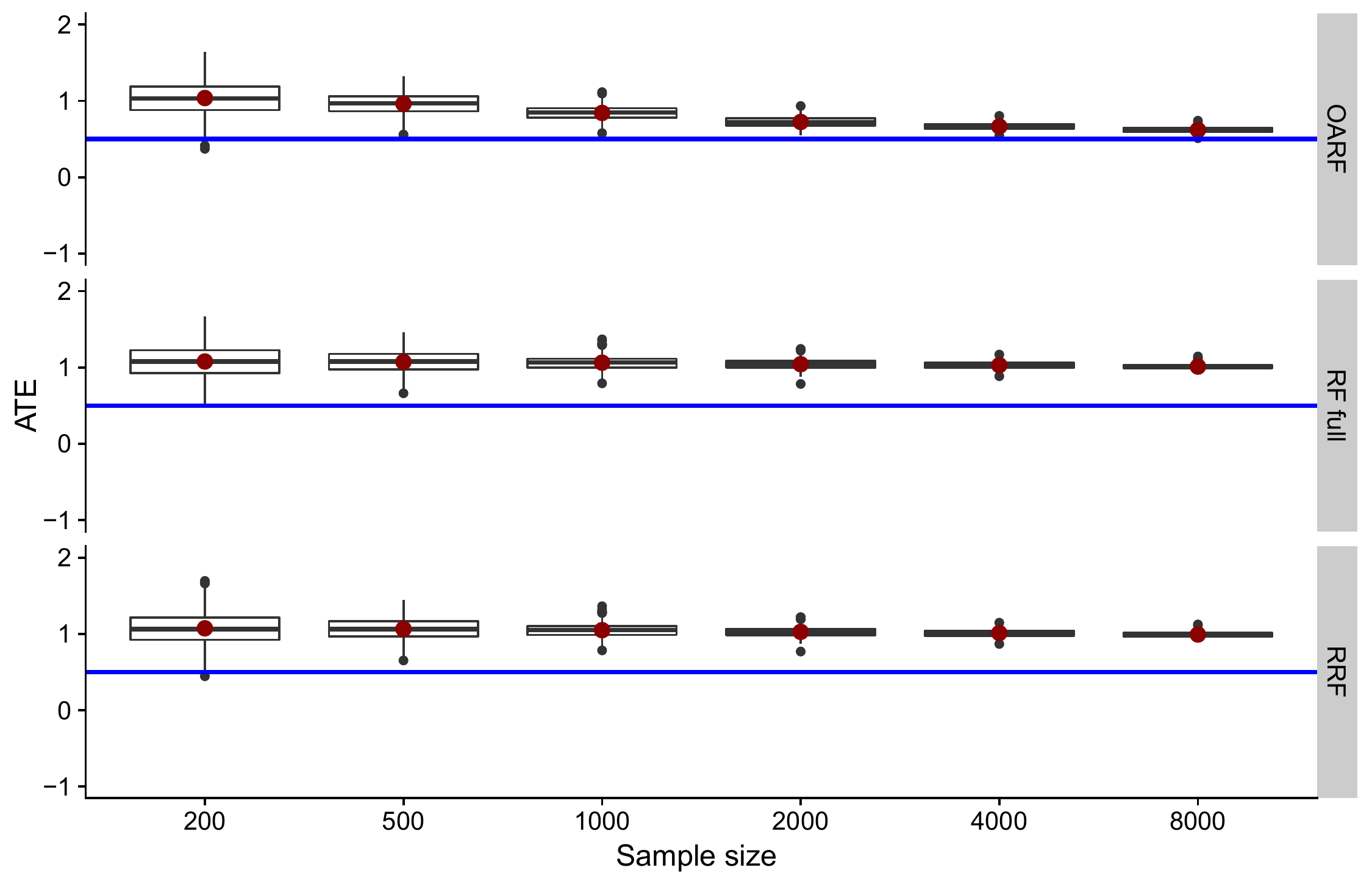}
\caption{Setting 5: Boxplots of ATE for varying sample size.}
\label{fig:boxplot_sam_size_S5}
\end{center}
\end{figure}

\clearpage
\section{Selection of tuning parameters}

The outcome-adaptive lasso (OAL) approach has to important tuning parameters. The parameter $\lambda_n$ is the regularization parameter that needs to be optimised while the parameter $\gamma$ is set to fulfil  $\lambda_n n^{\gamma/2 - 1} = n^2$, with $N =n$. In the IPTW estimator, the propensity score is used to balance the covariate distribution between the treatment and the control group. \cite{shortreed2017outcome} propose to select $\lambda_n$ by minimizing a weighted absolute mean difference (wAMD) using the covariates and the propensity score for the treatment and control group:

\begin{align}
\operatorname{wAMD}\left(\lambda_{n}\right)=\sum_{j=1}^{d}\left|\widetilde{\beta}_{j}\right|\left|\frac{\sum_{i=1}^{n} \widehat{\tau}_{i}^{\lambda_{n}} X_{i j} D_{i}}{\sum_{i=1}^{n} \widehat{\tau}_{i}^{\lambda_{n}} D_{i}}-\frac{\sum_{i=1}^{n} \widehat{\tau}_{i}^{\lambda_{n}} X_{i j}\left(1-D_{i}\right)}{\sum_{i=1}^{n} \widehat{\tau}_{i}^{\lambda_{n}}\left(1-D_{i}\right)}\right| \label{equ:wAMD}
\end{align}

\begin{align}
\widehat{\tau}_{i}^{\lambda_{n}}=\frac{D_{i}}{\hat{e}_{i}^{\lambda_{n}}\left\{\mathbf{X}_{i}, \hat{\alpha}(O A L)\right\}}+\frac{1-D_{i}}{1-\hat{e}_{i}^{\lambda_{n}}\left\{\mathbf{X}_{i}, \widehat{\alpha}(O A L)\right\}} \label{equ:OAL_weights}
\end{align}

Equation \ref{equ:OAL_weights} represents the IPT-weights obtained from the propensity score model using the OAL method for variable selection. 
The $\lambda_n$ value that minimizes the wAMD is used to estimate the ATE using the propensity score estimates given the specific $\lambda_n$ and $\gamma$. 

In equation \ref{equ:wAMD} the beta coefficients are used to weight the covariate balancing based on the strengths of the coefficients. Since we do not require exact coefficients from a linear model for the weighting, we could use the wAMD to tune the OARF. Instead of the coefficients, the $\left|\widetilde{\beta}_{j}\right|$ could contain the variable importance scores (they don't even need to be standardized). Since we mainly want to find a good penalization procedure for the propensity score function, possible tuning parameters could include the threshold for the initial feature space, the normalization of the importance score, and whether to apply different penalty weights based on the depth of the tree. For example, now we use the penalty $Imp^*_j$ for each node that contains the variable $j$ and is not in the initial feature set $\mathbb{F}$. Another possibility would be to use $(Imp^*_j)^{\xi}$ where $\xi$ states the depth of the node. Considering the depth of a node would make sense if we believe that splits near the end of a tree are less important and hence are heavier penalized. In the simulation settings that we consider, we find that the default values lead to a significant decrease in bias and variance. Even if the $\texttt{ranger}$ implementation of the RF is quite fast, the tuning of parameters is computationally expensive (in comparison to the tuning of the lasso). These are the main reasons why we do not consider parameter tuning at this stage but provide a possible approach to do so if necessary.

\begin{figure}[ht]
\begin{center}
\includegraphics[width=0.8\textwidth]{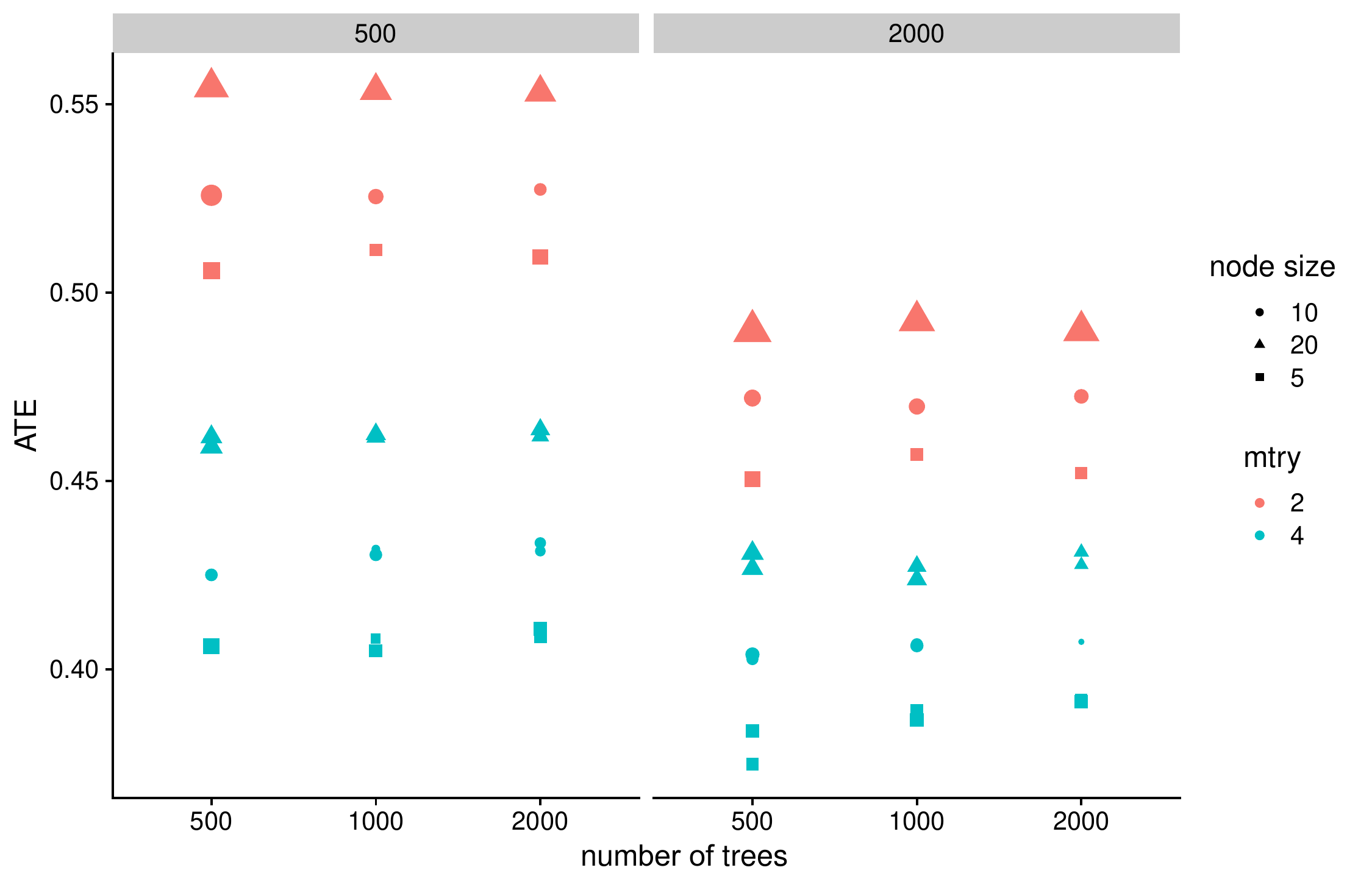}
\caption{ATE for different tuning parameters based on wAMD.}
\label{fig:tuning_wAMD}
\end{center}
\end{figure}

The RF tuning parameters aim to get high prediction accuracy. We do not aim for a high classification rate using the propensity score rather that it balances the covariate distribution between the treated and control group observations. Still, we can use the wAMD and tune, for example, the number of random variables to choose at each split (mtry), the number of trees, the node size, or imbalance methods, as long as we maximize the weighted absolute mean difference. Figure \ref{fig:tuning_wAMD} shows the ATE applying different tuning parameters on the OARF. The ATE is a median ATE over 200 Monte Carlo iterations. We also show the relative amount of tuning combinations that were chosen when minimizing the wAMD. We have 18 combinations of tuning parameters to choose from. The more often each combination was chosen based on minimizing the wAMD, the bigger the symbol. We find that, with 500 observations, the most often chosen combination has a node size of 20 and selected 2 variables at each split. The same holds when increasing the sample size to 2000 observations. With the latter amount of observations, the best tuning parameter combination is closest to the true ATE of 0.5. The number of trees seems not to be important since the results are mainly constant.

\section{Generalization of OARF}

The balancing of covariates through the propensity score generalizes to other methods besides the IPTW estimator. We illustrate the ATE estimation using the double machine learning (DML) approach proposed by \cite{chernozhukov2018double}. The estimation is based on the residual-on-residual approach to cancel out the effect from confounding covariates. If more variables are at choice from which only a few are true confounders it might be more beneficial to select variables for the propensity score estimation. Since this approach needs to estimate two functions (the conditional mean of the outcome and the propensity score function) we do not consider this approach as a direct comparison in the main part of the paper.  

The treatment effect is estimated as follows: First, we estimate the conditional mean of the outcome by regressing $X$ on $Y$. This results in the function $\hat{\ell}(X)$. Second, we estimate two propensity score models. One that uses all covariates and the second based on the OARF. We then estimate the residuals $\hat{U} = Y - \hat{\ell}(X)$ and $\hat{V}_m = Y - \hat{e}_m(X)$. Note that only $\hat{V}$ depends on the method $m$ but $\hat{U}$ is only estimated once. 
The treatment effect is then estimated by:

\begin{align}
\check{\theta}=\left(\frac{1}{n} \sum_{i =1} \hat{V}_{i} \hat{V}_{i}\right)^{-1} \frac{1}{n} \sum_{i = 1} \hat{V}_{i}\left(\hat{U}\right)
\end{align}

\begin{figure}[ht]
\begin{subfigure}[b]{0.33\linewidth}
\centering
\includegraphics[width=1\textwidth]{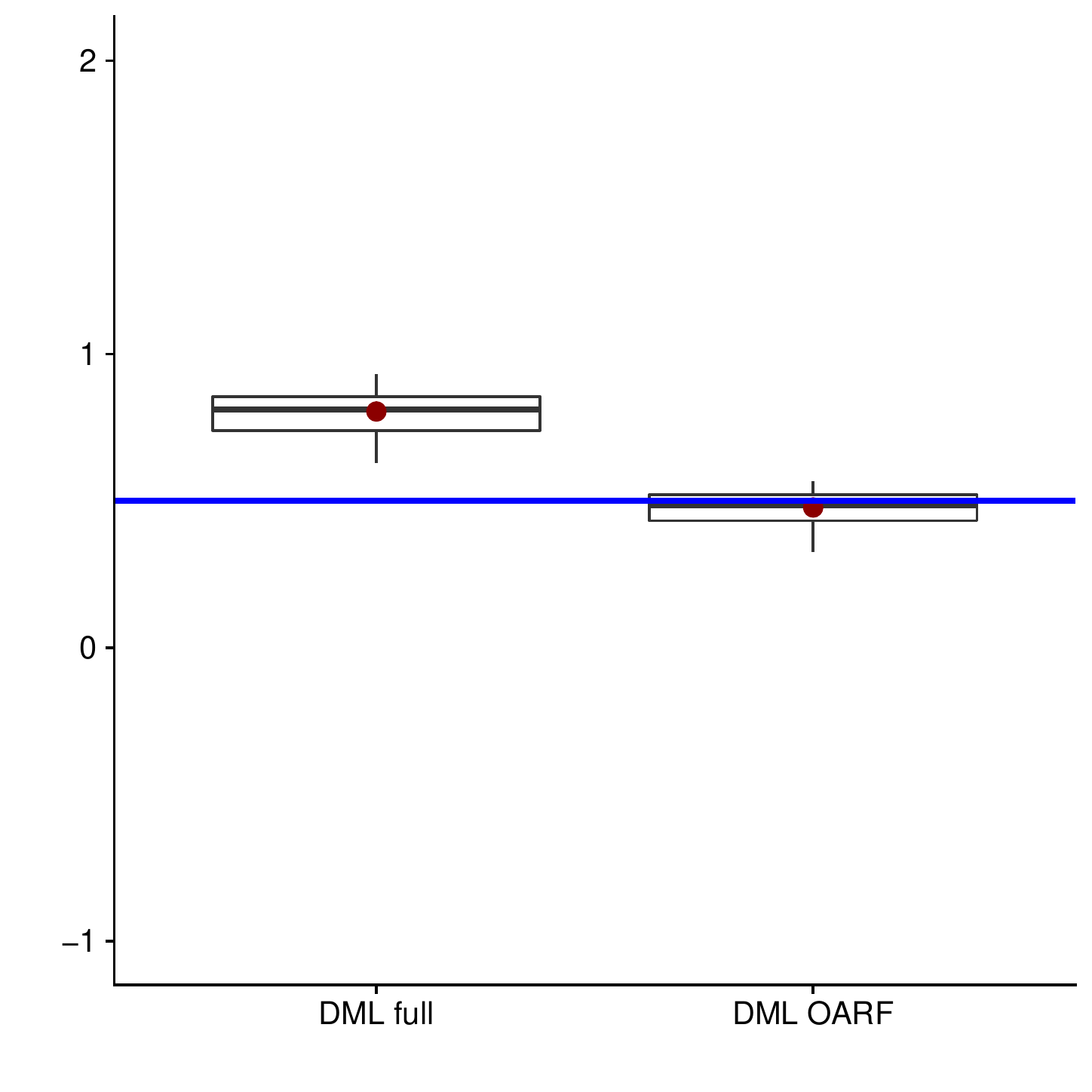}
\captionof{figure}{Setting 1}
\end{subfigure}%
\begin{subfigure}[b]{0.33\linewidth}
\centering
\includegraphics[width=1\textwidth]{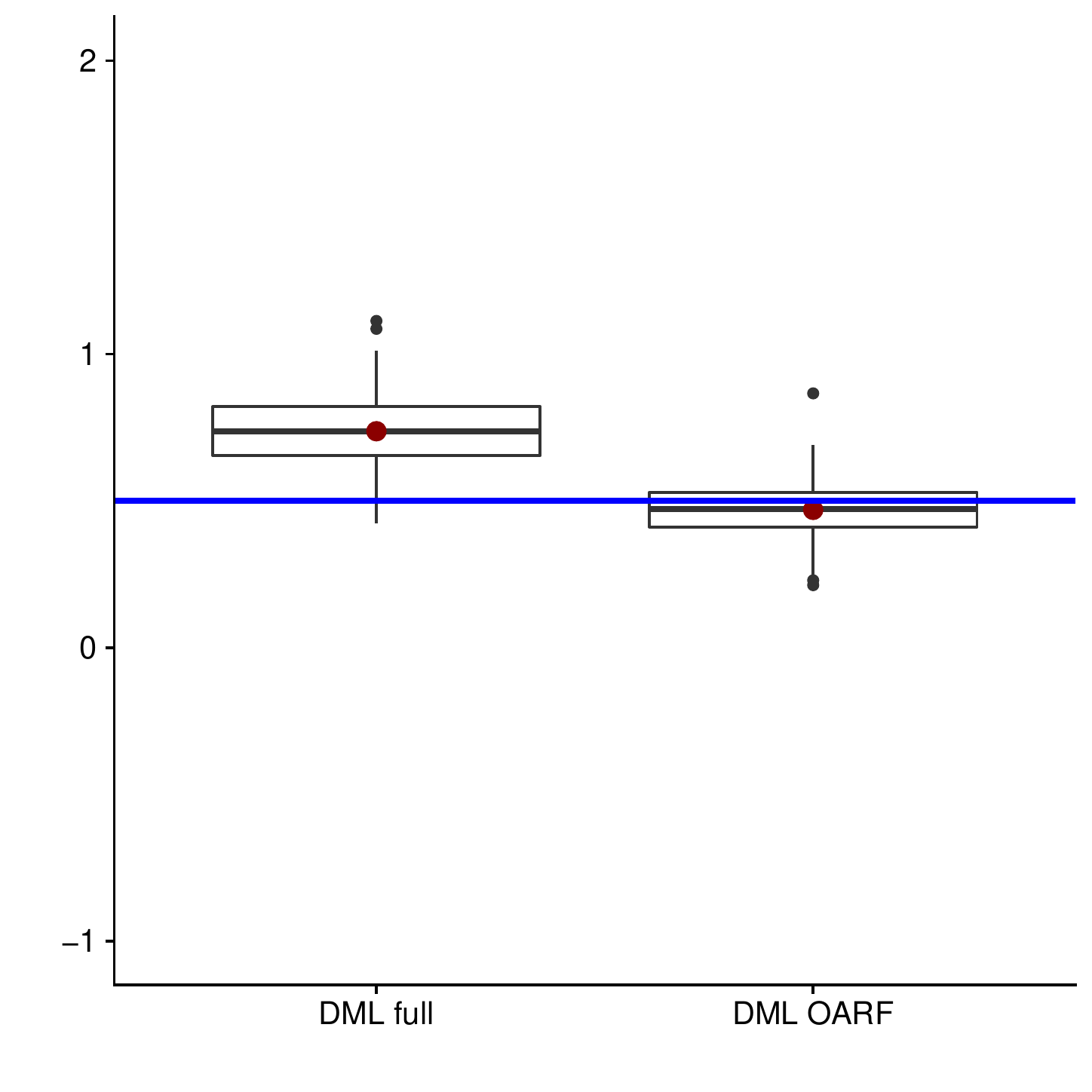}
\captionof{figure}{Setting 2}
\end{subfigure}%
\begin{subfigure}[b]{0.33\linewidth}
\centering
\includegraphics[width=1\textwidth]{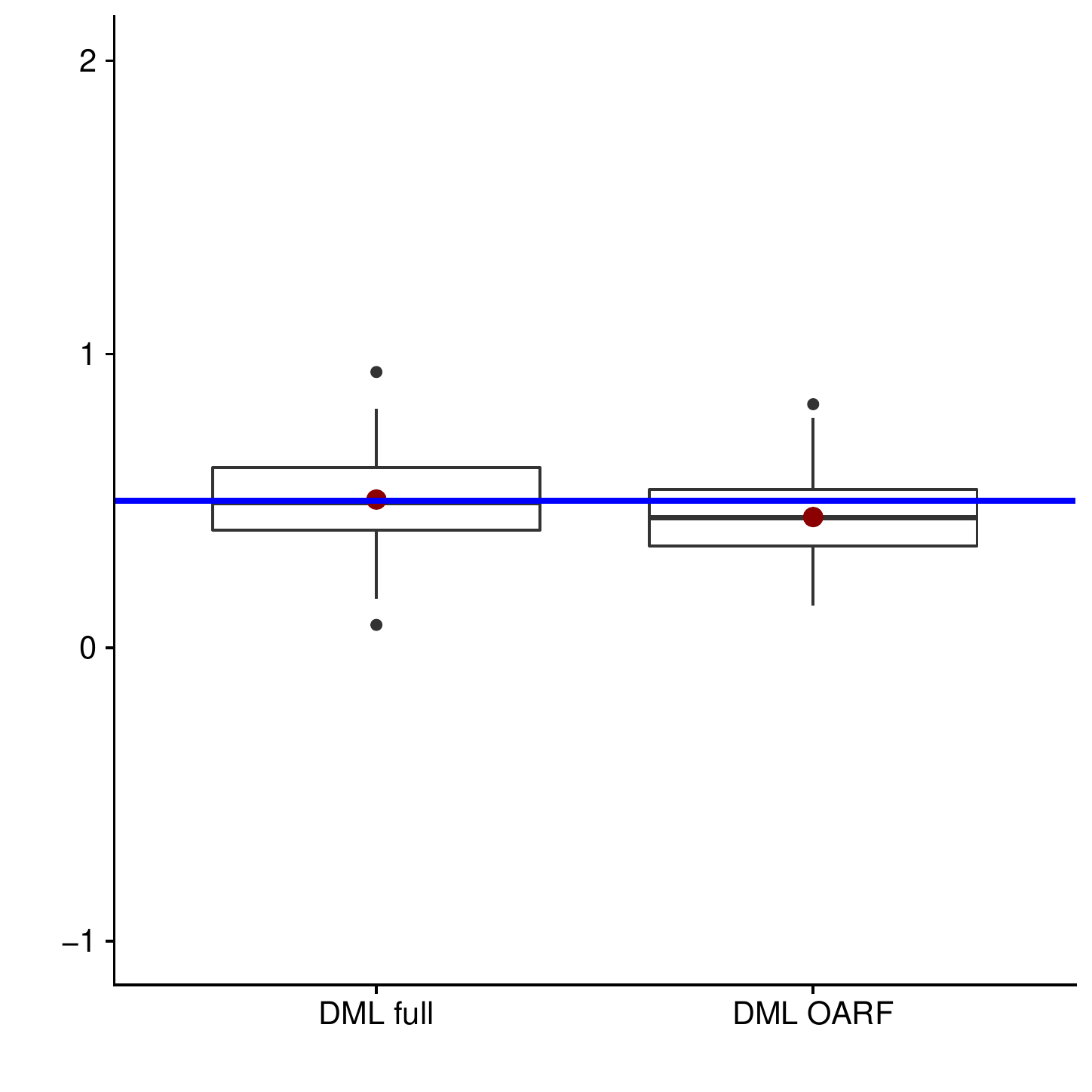}
\captionof{figure}{Setting 3}
\end{subfigure}

\begin{subfigure}[b]{0.33\linewidth}
\centering
\includegraphics[width=1\textwidth]{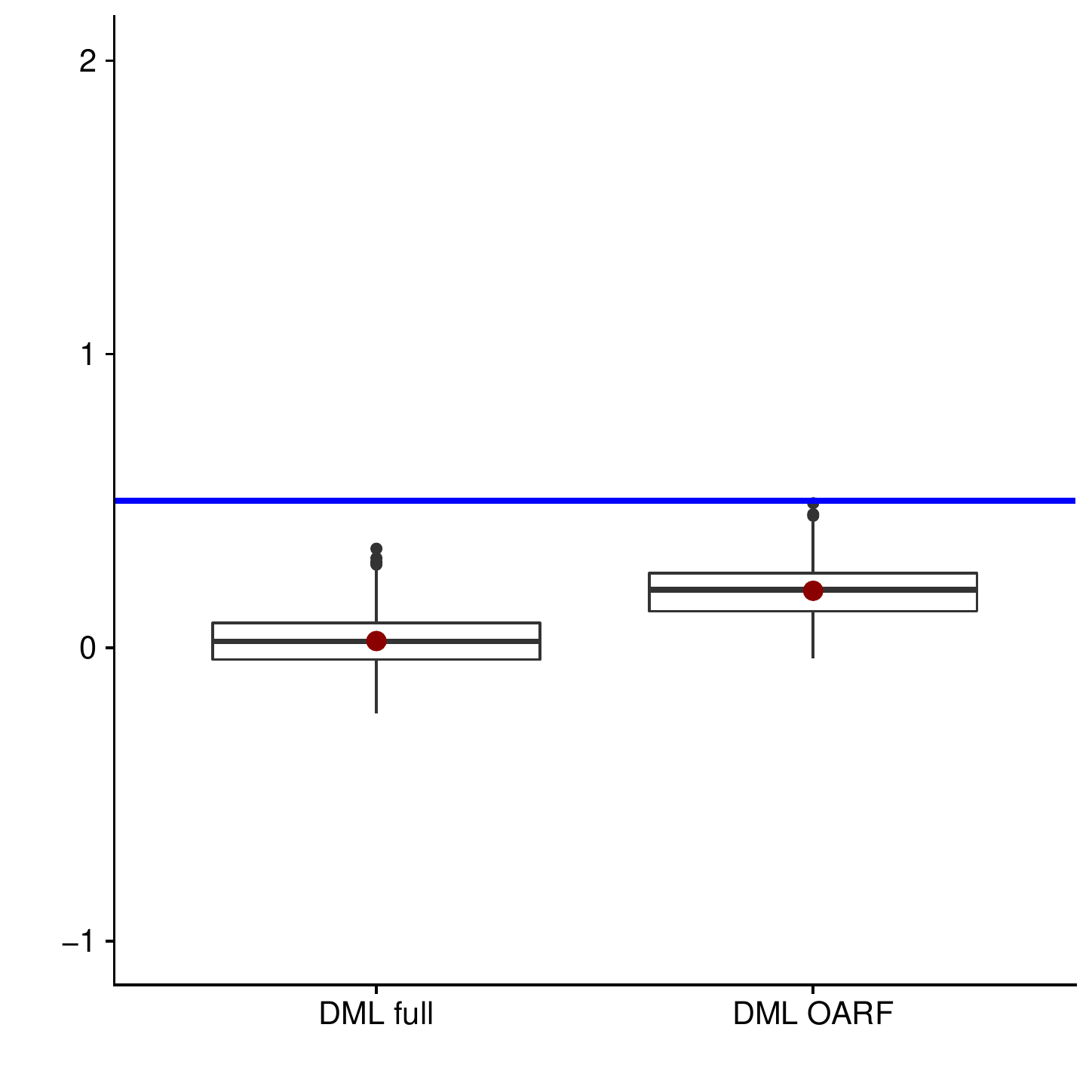}
\captionof{figure}{Setting 4}
\end{subfigure}%
\begin{subfigure}[b]{0.33\linewidth}
\centering
\includegraphics[width=1\textwidth]{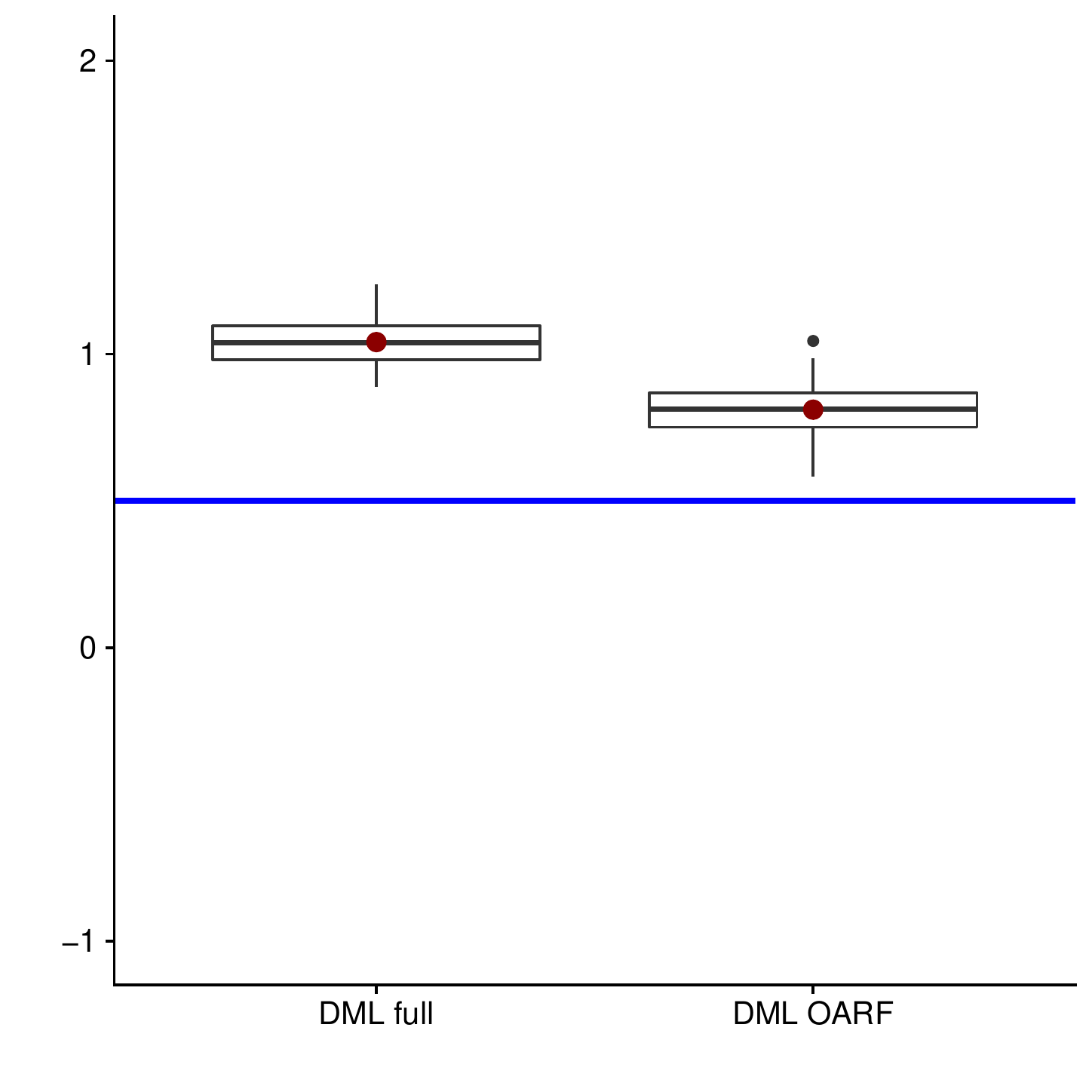}
\captionof{figure}{Setting 5}
\end{subfigure}%
\begin{subfigure}[b]{0.33\linewidth}
\centering
\includegraphics[width=1\textwidth]{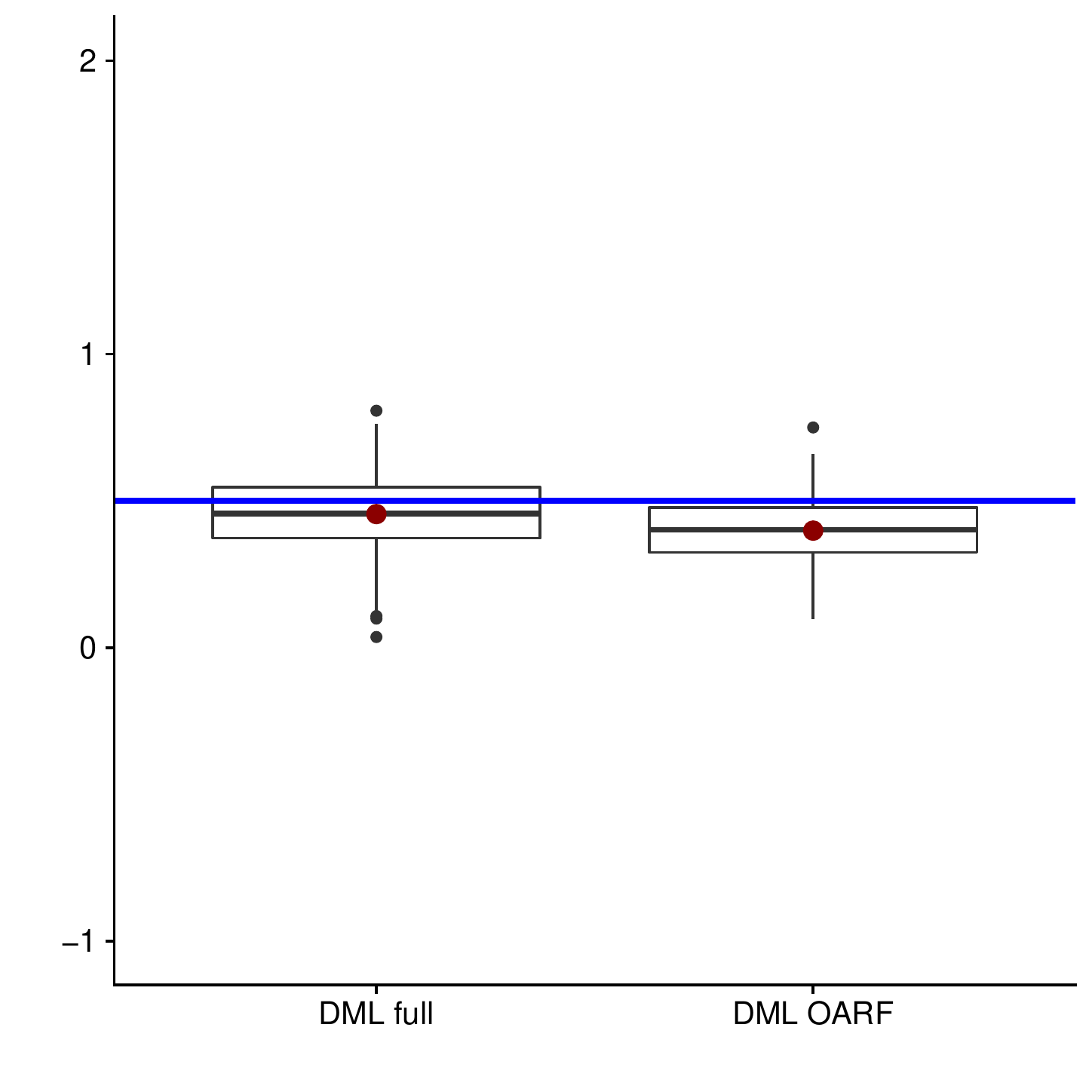}
\captionof{figure}{Setting 6}
\end{subfigure}
\caption{Illustrations of DML using all covariates and OARF based propensity score.}
\label{fig:boxplots_DML_OARF}
\end{figure}

\begin{figure}[ht]
\begin{subfigure}[b]{0.5\linewidth}
\centering
\includegraphics[width=0.8\textwidth]{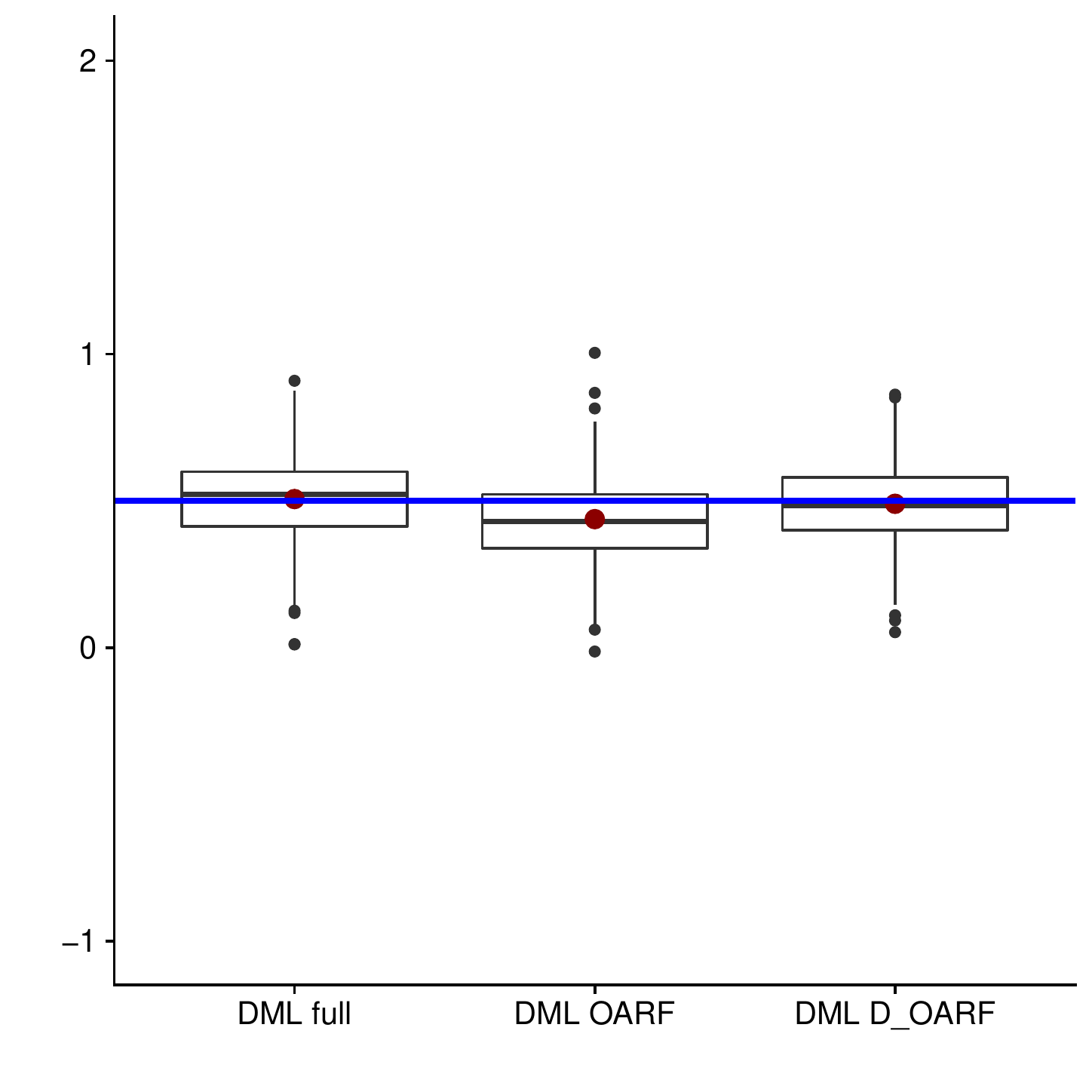}
\captionof{figure}{Setting 3}
\end{subfigure}%
\begin{subfigure}[b]{0.5\linewidth}
\centering
\includegraphics[width=0.8\textwidth]{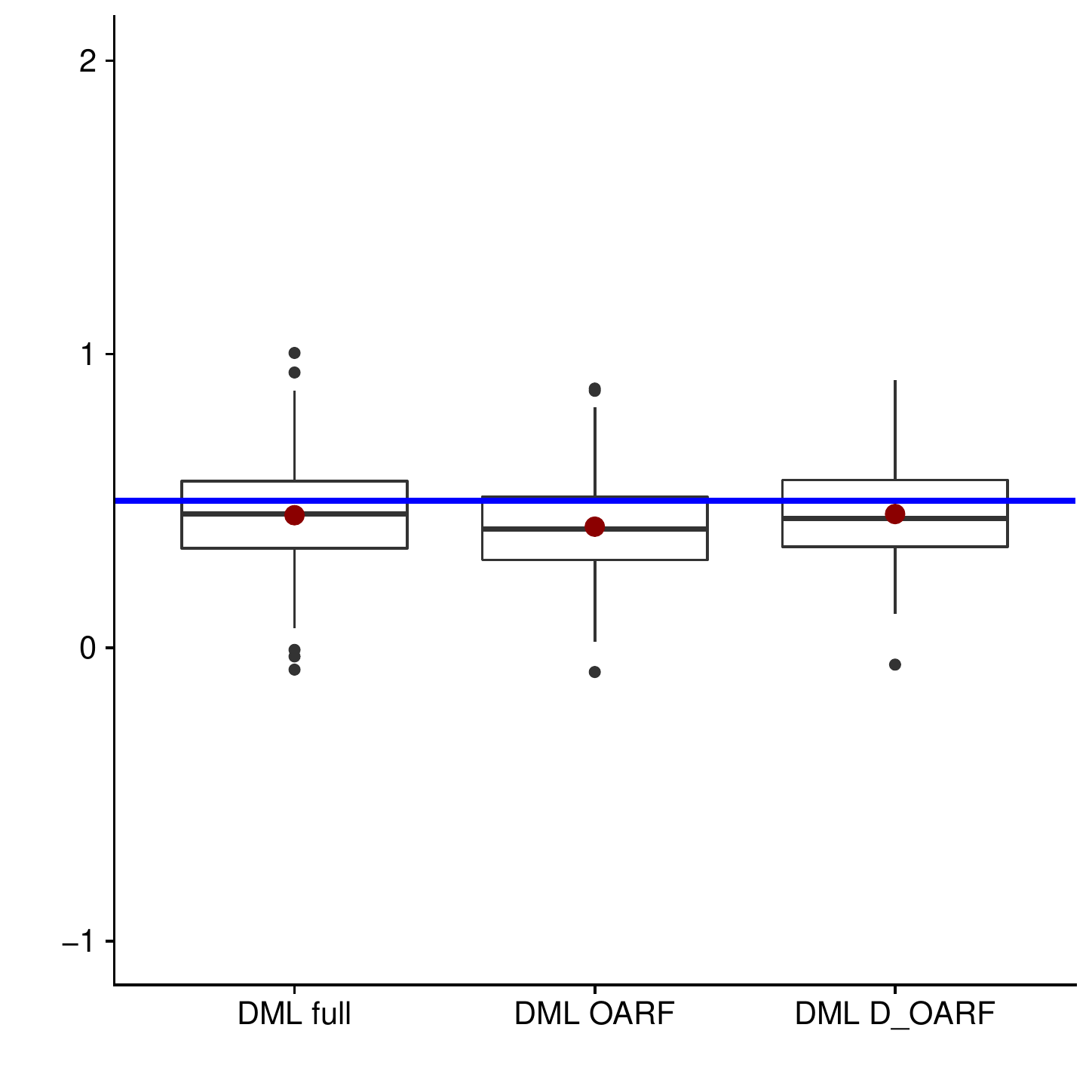}
\captionof{figure}{Setting 6}
\end{subfigure}
\caption{D\_OARF uses the OARF for both functions $\hat{\ell}(X)$ and $\hat{e}(X)$}
\label{fig:boxplots_DML_D_OARF }
\end{figure}

As in the previous settings we use sample splitting and cross-fitting to estimate the final parameter. Figure \ref{fig:boxplots_DML_OARF} shows the ATE estimated using the full RF and the OARF to estimate the propensity score. We find that using the OARF, the ATE is less biased and has a smaller variance in most settings. In settings 3 and 6, however, we find that both methods have the same variance but using the full RF might be less biased. The reason might be that if both functions, $\mathsf{E}[Y|X]$ and $\mathsf{E}[D|X]$ are quite complicated it might be desirable to also regularize the first function. This means that we only use the variables that are confounders and predictive on $Y$ to estimate the outcome function. In Figure \ref{fig:boxplots_DML_D_OARF } we show that using the regularized RF to estimate the outcome model, we can decrease the bias. This is as expected since we want to exclude variables that have no association on $Y$. We call the additional approach where we regularize both functions based on outcome variables, ``double OARF'' (DOARF). 

\end{document}